\documentclass[sigplan,screen]{acmart}
\pdfoutput=1

\setcopyright{rightsretained}
\acmPrice{}
\acmDOI{10.1145/3486607.3486748}
\acmYear{2021}
\copyrightyear{2021}
\acmSubmissionID{onward21papers-p4-p}
\acmISBN{978-1-4503-9110-8/21/10}
\acmConference[Onward! '21]{Proceedings of the 2021 ACM SIGPLAN International Symposium on New Ideas, New Paradigms, and Reflections on Programming and Software}{October 20--22, 2021}{Chicago, IL, USA}
\acmBooktitle{Proceedings of the 2021 ACM SIGPLAN International Symposium on New Ideas, New Paradigms, and Reflections on Programming and Software (Onward! '21), October 20--22, 2021, Chicago, IL, USA}

\startPage{1}

\usepackage{booktabs}   %
\usepackage{subcaption} %
\usepackage{balance}

\usepackage{xspace}
\usepackage{microtype}
\usepackage{tabularx}
\usepackage[inline]{enumitem}
\usepackage{amsfonts}
\usepackage{mleftright}
\usepackage{xcolor}
\usepackage[group-separator={,}]{siunitx}
\usepackage[framemethod=tikz]{mdframed}
\usepackage{amsmath}
\usepackage{soul}
\usepackage{colortbl}
\usepackage[normalem]{ulem}
\usepackage{booktabs}
\usepackage{xr}
\usepackage{changepage}
\usepackage{placeins}
\usepackage{xurl}
\usepackage{nowidow}

\usepackage[setmargin=true,marginparwidth=0.4in,hide=true]{marginalia}
\usepackage{cleveref}

\newcommand{\substitution}{surrogate compilation}
\newcommand{\Substitution}{Surrogate compilation}
\newcommand{\SUBSTITUTION}{Surrogate Compilation}

\newcommand{\relaxation}{surrogate optimization}
\newcommand{\Relaxation}{Surrogate optimization}
\newcommand{\RELAXATION}{Surrogate Optimization}

\newcommand{\adaptation}{surrogate adaptation}
\newcommand{\Adaptation}{Surrogate adaptation}
\newcommand{\ADAPTATION}{Surrogate Adaptation}

\newcounter{questioncounter}
\newcommand{\resetquestioncounter}{\setcounter{questioncounter}{0}}

\newcommand{\question}[2]{\refstepcounter{questioncounter}\vspace{2pt}%
  \begin{adjustwidth}{0pt}{0pt}%
    \begin{tabularx}{\columnwidth}{Xl}
      {\bf Question~\thequestioncounter:} & {\it #2}%
      \end{tabularx}
    \end{adjustwidth}%
    \nopagebreak[4]%
    \par%
  }

\renewcommand\tabularxcolumn[1]{m{#1}} %
\newcommand{\tlquestion}[2]{\refstepcounter{questioncounter}\vspace{2pt}%
  \begin{adjustwidth}{0pt}{0pt} %
    \begin{tabularx}{\columnwidth}{m{2cm}X}
      {\bf Question~\thequestioncounter:}
      & {\it #2}%
      \end{tabularx}
  \end{adjustwidth}%
  \nopagebreak[4]%
  \par%
}

\newcommand{\simcolor}{black}
\newcommand{\dualcolor}{blue}

\newcommand{\scratchcolor}{orange}

\DeclareMathOperator*{\argmax}{arg\,max}
\DeclareMathOperator*{\argmin}{arg\,min}

\newcommand{\mca}{llvm-mca}

\newcommand{\mprogramtype}{\mathcal{P}}
\newcommand{\mdataset}{\mathcal{D}}
\newcommand{\mreal}{\mathbb{R}}
\newcommand{\mobserver}{\omega}

\newcommand{\mprogram}{p}
\newcommand{\mdual}{s}
\newcommand{\minput}{x}
\newcommand{\moutput}{y}
\newcommand{\minputtype}{\mathcal{X}}
\newcommand{\moutputtype}{\mathcal{Y}}
\newcommand{\mobjective}{o}
\newcommand{\mconstraint}{c}
\newcommand{\mdistance}{d}
\newcommand{\mexecutioncost}{e}
\newcommand{\mrelaxationobjective}{\ell}

\newcommand{\mprime}{}

\makeatletter
\newcommand*{\rom}[1]{\expandafter\@slowromancap\romannumeral #1@}
\makeatother

\newcommand{\questionone}{\tlquestion{Of programs}{What neural network architecture topology does the surrogate use?}}
\newcommand{\questiontwo}{\tlquestion{Both}{How do you scale the surrogate's capacity to represent the original program?}}
\newcommand{\questionthree}{\question{Of programs}{What training data does the surrogate use?}}
\newcommand{\questionfour}{\question{Programming}{What loss function does the surrogate use?}}
\newcommand{\questionfive}{\question{Programming}{How long do you train the surrogate?}}
\newcommand{\questionsix}{\question{Programming}{What hardware does the surrogate use?}}
\newcommand{\questionseven}{\tlquestion{Programming}{What software execution environment does the~surrogate~use?}}

\makeatletter
\newcommand\cellwidth{\TX@col@width}
\makeatother

\newcommand{\Exp}{\mathop{\vcenter{\hbox{\Large$\mathbb{E}$}}}}

\usepackage[math]{cellspace}

\begin{document}

\title{Programming with Neural Surrogates of Programs}
\author{Alex Renda}
\affiliation{
  \institution{MIT CSAIL}            %
  \city{Cambridge}
  \state{MA}
  \postcode{0221}
  \country{USA}
}
\email{renda@csail.mit.edu}          %

\author{Yi Ding}
\affiliation{
  \institution{MIT CSAIL}           %
  \city{Cambridge}
  \state{MA}
  \postcode{0221}
  \country{USA}
}
\email{ding1@csail.mit.edu}         %

\author{Michael Carbin}
\affiliation{
  \institution{MIT CSAIL}           %
  \city{Cambridge}
  \state{MA}
  \postcode{0221}
  \country{USA}
}
\email{mcarbin@csail.mit.edu}         %

\begin{abstract}
  \emph{Surrogates}, models that mimic the behavior of programs, form the basis of a variety of development workflows.
We study three surrogate-based design patterns, evaluating each in case studies on a large-scale CPU simulator.

With \emph{\substitution{}}, programmers develop a surrogate that mimics the behavior of a program to deploy to end-users in place of the original program.
\Substitution{} accelerates the CPU simulator under study by $1.6\times$.
With \emph{\adaptation{}}, programmers develop a surrogate of a program then retrain that surrogate on a different task.
\Adaptation{} decreases the simulator's error by up to $50\%$.
With \emph{\relaxation{}}, programmers develop a surrogate of a program, optimize input parameters of the surrogate, then plug the optimized input parameters back into the original program.
\Relaxation{} finds simulation parameters that decrease the simulator's error by $5\%$ compared to the error induced by expert-set parameters.

In this paper we formalize this taxonomy of surrogate-based design patterns.
We further describe the programming methodology common to all three design patterns.
Our work builds a foundation for the emerging class of workflows based on programming with surrogates of programs.

\end{abstract}


\maketitle

\section{Introduction}
\label{sec:introduction}

Programmers and researchers are increasingly developing \emph{surrogates} of programs, models of a subset of the observable behavior of a given program, to solve a variety of software development challenges~\citep{ipek_exploring_2006,esmaeilzadeh_neural_2012,tercan_transfer_2018,she_neuzz_2019,tseng_hyperparameter_2019,renda_difftune_2020,munk_deep_2019,pestourie_active_2020,kustowski_transfer_2019,kwon_transfer_2020}.

Programmers train surrogates from measurements of the behavior of a program on a dataset of input examples~\citep{santner_design_2018,goodfellow_deep_2016,myers_response_2009,gramacy_surrogates_2020}.
Typical examples of surrogates include neural networks~\citep{goodfellow_deep_2016,renda_difftune_2020}, Gaussian processes~\citep{rasmussen_gaussian_2005,cherrypick}, linear models~\citep{GelmanHill:2007,ding_generalizable_2021}, and random forests~\citep{ho_random_1995,nardi_hypermapper_2019}.
Of these model architectures, \emph{neural surrogates}
have emerged as a popular design for surrogates in the literature~\citep{ipek_exploring_2006,esmaeilzadeh_neural_2012,tercan_transfer_2018,she_neuzz_2019,tseng_hyperparameter_2019,renda_difftune_2020} because for many tasks neural networks are state-of-the-art models that lead to high accuracy~\citep{krizhevsky_imagenet_2012,devlin_bert_2018}.

Programmers use surrogates for a variety of tasks including accelerating computational kernels in numerical programs~\citep{esmaeilzadeh_neural_2012}, replacing physical simulators with more accurate versions~\citep{tercan_transfer_2018}, and tuning parameters of complex simulators~\citep{renda_difftune_2020,tseng_hyperparameter_2019}.
Compared to standard development workflows, programming with surrogates requires lower development costs~\citep{renda_difftune_2020,tseng_hyperparameter_2019,she_neuzz_2019,kwon_transfer_2020,kaya_using_2019} and results in programs with lower execution~cost~\citep{esmaeilzadeh_neural_2012,mendis_thesis_2020,munk_deep_2019,pestourie_active_2020} or higher result quality~\citep{tercan_transfer_2018,kustowski_transfer_2019,renda_difftune_2020,tseng_hyperparameter_2019}.
However, the approaches in the literature for both applying and developing surrogates are disparate, with no unifying taxonomy or development methodology.

\subsection{Surrogate-Based Design Patterns}
In this paper we contribute a taxonomy that classifies the workflows above into three different design patterns:  \emph{\substitution{}}, \emph{\adaptation{}}, and \emph{\relaxation{}}.
We concretize these design patterns by demonstrating how to use each to solve one of three development tasks for llvm-mca~\citep{llvm-mca}, a 10,000 line-of-code CPU simulator that predicts the execution time of code snippets.

\paragraph{\Substitution{}}
With \substitution{}, programmers develop a surrogate that replicates the behavior~of a program to deploy to end-users in place of that program.
Key benefits of this approach include the ability to execute the surrogate on different hardware and the ability to bound or to accelerate the execution time of the surrogate~\citep{esmaeilzadeh_neural_2012,mendis_thesis_2020}.

For llvm-mca, we train a neural network to replicate \mca{}'s prediction of the execution time for a given input code snippet.
The resulting neural network executes $1.6\times$ faster than llvm-mca on the same hardware, with less than a $10\%$ deviation from~llvm\nobreakdash-mca's~predictions.

\paragraph{\Adaptation{}.}
With \adaptation{}, programmers first develop a surrogate of a program then further train that surrogate on data from a different task.
Key benefits of this approach include that \adaptation{} makes it possible to alter the semantics of the program to perform a different task of interest and that it may be more data-efficient or result in higher accuracy than training a model from scratch for the task~\citep{tercan_transfer_2018,kustowski_transfer_2019}.

We train a neural network to replicate \mca{}'s predictions then fine-tune that network on measurements of code timing on a physical CPU.
This network has as low as $50\%$ of the error of llvm\nobreakdash-mca at predicting the ground-truth timings.

\paragraph{\Relaxation{}.}
With \relaxation{}, programmers develop a surrogate of a program, optimize input parameters of that surrogate, then plug the optimized parameters back into the original program.
The key benefit of this approach is that \relaxation{} can optimize inputs faster than optimizing inputs directly against the program, due to the potential for faster execution speed of the surrogate and the potential for the surrogate to be differentiable even when the original program is not (allowing for optimizing inputs with gradient descent)~\citep{renda_difftune_2020,tseng_hyperparameter_2019,she_neuzz_2019}.

We train a neural network to replicate \mca{}'s prediction when \mca{} is parameterized with different sets of simulation parameters, then optimize against that network to find parameters that lead the network to accurately predict ground-truth timings.
We then plug these parameters back into \mca{}.
These parameters improve \mca{}'s accuracy by $5\%$ relative to expert-selected parameters.

\subsection{Programming Methodology}
The development methodologies common to these surrogate-based design patterns when instantiated with neural networks induce what we term the \emph{neural surrogate programming methodology}, consisting of the \emph{specification} of the task, the \emph{design} of the neural network architecture, the \emph{training} process for the network, and the \emph{deployment} of the system.
We present the programming methodology as a set of questions that guide development of the surrogate.
A complete set of answers to these questions constitutes a concrete plan for the development and deployment of~a~neural~surrogate.

Surrogates are constructed from input-output examples, meaning that their development methodology is the same as that of any other machine learning technique.
We present key insights related to the fact that we study surrogates of programs with known structure and behavior (e.g., how to select a neural network architecture that can represent the original program with high accuracy).
We also present insights that arise from the fact that surrogate development is itself a form of programming, constructing a function to meet a correctness specification while trading off among other objectives (e.g., how to minimize execution costs of the surrogate while satisfying an~accuracy~constraint).

\begin{figure*}
  \begin{center}
    \includegraphics[width=0.9\textwidth]{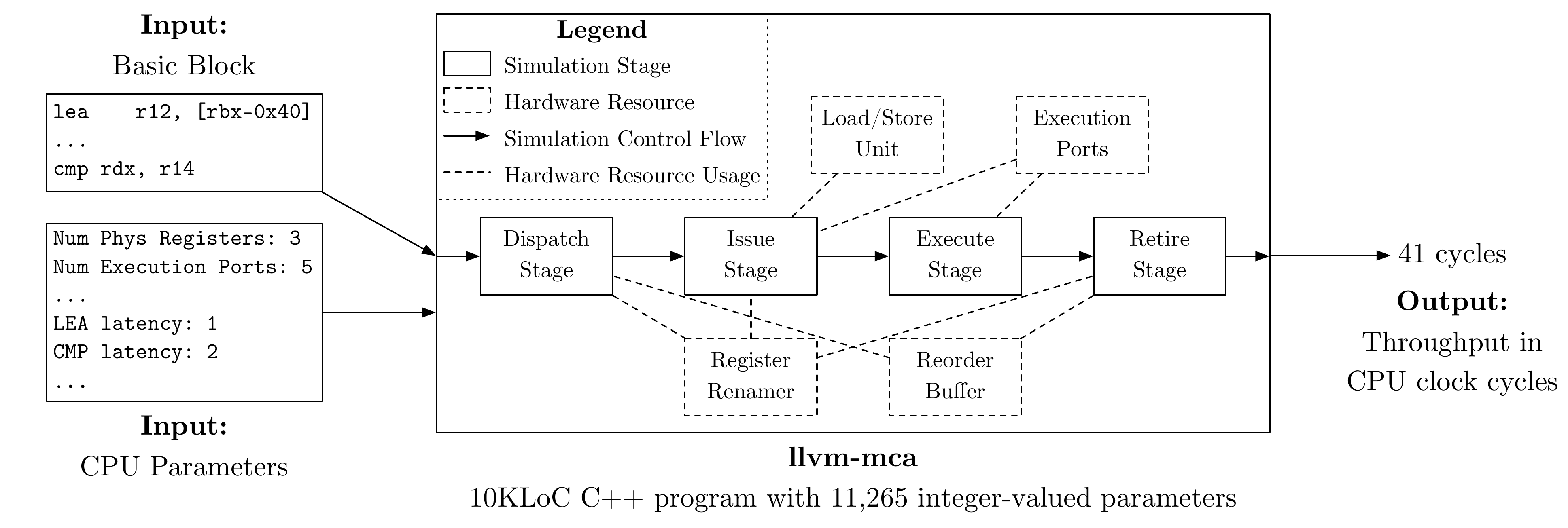}
  \end{center}
  \caption{Input-output specification and design of \mca{}.}
  \label{fig:mca}
\end{figure*}

\subsection{Contributions}
In this paper we present the following contributions:

\begin{itemize}
\item We provide three detailed case studies of programming with surrogates on a large-scale CPU simulator.
\item We formally define three design patterns that use surrogates of programs: \substitution{}, \adaptation{}, and \relaxation{}.
  We demonstrate that this taxonomy captures examples of surrogate programming from the literature.
\item We identify elements of the neural surrogate programming methodology in the form of specifications and design questions that unify these surrogate-based design patterns.
  We discuss answers to each of these design questions, showing the trade-offs that programmers must consider when developing neural surrogates.
\item We lay out future directions towards the goal of further systematizing the programming methodology underlying surrogate programming.
\end{itemize}

Surrogates are an important emerging frontier of programming with a wealth of use cases for developing complex programs.
By identifying the three surrogate-based design patterns and describing the programming methodology used to develop neural surrogates, our work provides a taxonomy for reasoning about and developing surrogates.
Our work offers a foundation on which the programming languages community can build new tools that aid in the construction and analysis of surrogates~of~programs.

\section{Case Study: Overview}
\label{sec:example}

We first demonstrate how developing surrogates of a CPU simulator makes it possible to solve three development tasks:
\begin{enumerate*}[label=(\arabic*)]
\item increasing the speed of the simulation,
\item simulating the execution behavior of a real-world processor that is not well-modeled by the simulator, and
\item finding simulation parameters that lead the simulator to accurate simulation of the behavior of a real-world processor.
\end{enumerate*}
We present a \relaxation{} case study from our prior work~\citep{renda_difftune_2020} and present two new case studies of \substitution{} and \adaptation{} on the same simulator~under~study.

\paragraph{Program under study.}
We study \mca{}~\citep{llvm-mca}, a CPU simulator included in the LLVM compiler infrastructure~\citep{llvm}.

\Cref{fig:mca} presents \mca{}'s input-output specification and design.
As input, \mca{} takes a \emph{basic block}, a sequence of assembly instructions with no jumps or loops, and a set of \emph{CPU parameters}, integers that describe properties of the CPU being modeled.
It then outputs a prediction of the \emph{throughput} of the basic block on the CPU, a prediction of the number of CPU clock cycles taken to execute the block when repeated for a fixed number of iterations.

\setnowidow[4]
\setnoclub[4]
Rather than precisely emulating the behavior of the CPU under study, \mca{} makes several modeling assumptions about the behavior of the CPU, and simulates basic blocks using an abstract execution model of that CPU.
The \mca{} system simulates a processor in four main simulation stages: \emph{dispatch}, \emph{issue}, \emph{execute}, and \emph{retire}.
Instructions pass through each of these four stages in turn.
Each stage is bottlenecked by the availability of \emph{hardware resources} in the simulation model.
The input CPU parameters specify what resources are available on the hardware and what resources to reserve for each instruction.
When all instructions of the basic block have passed through the full simulation pipeline, the simulation terminates and the final throughput prediction is the number of simulated CPU clock cycles.
\setnowidow[2]
\setnoclub[2]

Instructions first enter into the \emph{dispatch} stage.
The dispatch stage reserves the hardware resources needed to track the execution of the instruction in the~simulation~model. %

Once dispatched, instructions wait in the \emph{issue} stage until they are ready to be executed.
The issue stage holds instructions until all of their input operands and all of the hardware resources required to execute the instructions~are~available.

Instructions then enter the \emph{execute} stage, which reserves the hardware resources required to execute the instruction and holds them for the number of clock cycles specified by the CPU parameters for~the~instruction.

Finally, once instructions have executed for their duration, they enter the \emph{retire} stage, which frees the resources that were acquired for each instruction in the dispatch phase.

\paragraph{Implementation.}
The \mca{} system is a C++ program implemented as part of the LLVM compiler infrastructure, comprised of around \num{10000} lines of code.
The CPU parameters are comprised of \num{11265} integer-valued parameters, inducing a configuration space with $10^{\num{19336}}$ possible configurations.
LLVM contains expert-set CPU parameter settings for \mca{} that target common x86 hardware architectures.

\paragraph{Validation and accuracy.}

In our prior work~\citep{chen_bhive_2019}, we validate the accuracy of \mca{} by collecting BHive, a dataset of x86 basic blocks from a variety of end-user programs.
For each basic block in BHive we also collect ground-truth throughput measurements of the block by timing them on real CPUs.
We calculate the mean absolute percentage error (MAPE) of \mca{}'s throughput predictions, which is the normalized difference between \mca{}'s output $\moutput{}_{\rm pred}$ and the ground-truth measured~throughput~$\moutput{}_{\rm true}$: \[
  {\rm err}\mleft(\moutput{}_{\rm pred}, \moutput{}_{\rm true}\mright) \triangleq \frac{|\moutput{}_{\rm pred} - \moutput{}_{\rm true}|}{\moutput{}_{\rm true}}
\]

Across basic blocks in the BHive dataset and the CPU platforms that \mca{} has expert-set parameters for, \mca{} has a mean absolute percentage error of around $25\%$.

\resetquestioncounter{}

\section{Case Study: \SUBSTITUTION{}}
\label{sec:substitution}
To quickly generate throughput predictions for basic blocks, programmers must develop fast CPU simulation models.

The standard approach, used by \mca{}, is to manually implement a fast and sufficiently accurate simulation model, then use compiler optimizations to accelerate the \emph{execution speed} of the simulation code.
We define \mca{}'s execution speed as the number of basic blocks per second that \mca{} is able to generate throughput predictions for.

Other approaches in the literature for accelerating \mca{}'s execution speed include rewriting the simulation software to be faster~\citep{hager_introduction_2010} and applying compiler optimizations not included in \mca{}'s default compiler's optimization set, such as superoptimization~\citep{massalin_superoptimizer_1987,schkufza_stochastic_2013}.

\paragraph{\Substitution{}.}
An alternative approach for accelerating \mca{}'s execution speed is \substitution{}.
With \substitution{}, programmers develop a surrogate that replicates the behavior~of a program to deploy to end-users in place of the original program.

\paragraph{Results.}
When we instantiate \mca{} with its default set of Haswell CPU parameters, \mca{}'s execution speed on an Intel Xeon Skylake CPU at 3.1GHz is 1742 blocks per second.%
\footnote{Full methodological details on this evaluation are presented in \Cref{app:substitution-methodology}.}
Using \substitution{} we learn a neural surrogate of \mca{} that has an execution speed of $2820$ blocks per second on the same hardware, a speedup of $1.6\times$ over \mca{}.
This surrogate has a mean absolute percentage error (MAPE) of $9.1\%$ compared to \mca{}'s predictions.
Against BHive's ground-truth measured data on a real Haswell CPU, the surrogate has an error rate of $27.1\%$, compared to an error rate of $25.0\%$~for~\mca{}.

\subsection{Programming Methodology}

Developing the neural surrogate for \substitution{} requires thinking about the \emph{specification} of the task, the \emph{design} of the neural network architecture, the \emph{training} process for the neural network, and the \emph{deployment} considerations of the system.
We collect these concerns into what we term the neural surrogate programming methodology.

\subsubsection{Specification}
The primary concern with any programming task is its specification.
In the surrogate programming methodology, the specification comes in the form of an optimization problem with an objective and constraints.

The specification for the surrogate in this example is to maximize the execution speed of the surrogate while also constraining the error of the surrogate compared to \mca{} to be less than $10\%$ as measured by the MAPE:
\begin{gather*}
  \mdual{}^* = \argmax_{\mdual{}}\ \text{execution-speed}(\mdual{}) \\
  \text{such that}\;
  \Exp_{\minput{} \sim \mdataset{}} \mleft[ \frac{\mleft|\mdual{}\mleft(\minput{}\mright) - \mprogram{}\mleft(\minput{}, \text{haswell-params}\mright)\mright|}{\mprogram{}\mleft(\minput{}, \text{haswell-params}\mright)} \mright] \leq 10\%
\end{gather*}%
where $\mdual{}$ is the surrogate, $\mdataset{}$ is the dataset of basic blocks $\minput{}$ from BHive, $\mprogram{}$ is \mca{}, and haswell-params is LLVM's default set of Haswell~CPU~parameters.

The remainder of this case study walks through the neural surrogate programming methodology, presented as a set of design questions that guide the design, training, and deployment process of the neural surrogate.

\subsubsection{Design}
When developing a neural surrogate for a given task, the programmer must choose an architecture for the neural network underlying the surrogate, as well as scale the network's capacity appropriately.
These choices must be informed by the specification of the surrogate and by the semantics of the program that the surrogate models.

In this example the neural network architecture and capacity must be the network with the highest execution speed that meets the accuracy constraint.

\pagebreak[4]

\questionone{}
The neural network architecture topology is the connection pattern of the neurons in the neural network~\citep{goodfellow_deep_2016}.
The topology determines the types of inputs that the network can process (e.g., fixed-size inputs or arbitrary length sequences) and the \emph{inductive biases} of the network, the assumptions about the task that are baked into the neural network.

We use a BERT encoder~\citep{devlin_bert_2018}, a type of Transformer~\citep{vaswani_attention_2017}, as the neural network topology for \substitution{} of \mca{}.
Though many architectures could provide an acceptable solution to the task, we select and evaluate BERT due to its popularity~\citep{rogers_bertology_2020}, expressive power~\citep{yun_transformers_2020}, and relative ease of use~\citep{wolf_huggingface_2020} for arbitrary sequence modeling tasks (though programmers should in general choose the most appropriate neural network architecture to model the program depending on the domain).
Our BERT architecture processes raw Intel-syntax x86 basic blocks as input and predicts \mca{}'s throughput prediction as output.

\questiontwo{}

The \emph{capacity} of the surrogate is the complexity of functions that the surrogate can represent.
Higher capacity neural networks better fit the training data~\citep{belkin_double_2019}, but have higher execution cost~\citep{tan_efficientnet_2019}.
Scaling the capacity involves adding more layers or increasing the width of~each~layer.

We search among candidate capacities of the surrogate to find the smallest-capacity BERT architecture that meets the accuracy specification.
We present more details on this hyperparameter search in \Cref{app:hyperparameters-training-width}.

\subsubsection{Training}
\label{sec:substitution-training}
With the architecture in hand, the programmer must determine how to train the surrogate model.

\questionthree{}
The training data distribution is the distribution of inputs on which the surrogate is expected to~perform~well. %

For \substitution{} in general, any dataset of inputs can suffice to train the neural surrogate, as long as they constitute a sufficiently large set of representative examples of the distribution of inputs that the programmer wishes to accurately generate predictions for.
We use basic blocks from the BHive dataset~\citep{chen_bhive_2019} to train the surrogate for consistency with the case studies in Sections~4 and~5.

\questionfour{}
The \emph{loss function}, the objective in a neural network's optimization process, %
is a differentiable, continuous relaxation of the objective and constraints from the specification (which may not themselves be differentiable), with different relaxations having~different~properties~\citep[pp.\,337\nobreakdash--338]{bishop_pattern_2006}.

Because the objective of maximizing execution speed is handled in the capacity search process, the loss function for training the neural surrogate for this \substitution{} example is just the MAPE between the surrogate's prediction and \mca{}'s prediction of throughput.

\pagebreak[4]
\questionfive{}
The number of training iterations for the neural surrogate determines the trade-off between the training cost of the surrogate and the accuracy of the surrogate.
In general, the cost of training is limited either by an acceptability threshold on the error or by a fixed training budget.
Because the training procedure may be run multiple times when designing the surrogate, the threshold or budget should be set appropriately to account for the full cost of design~and~training.

We train the BERT model for 500 passes over the training set (500 epochs), recording the loss over a validation set after each epoch.
At the end of training, we select the model with the best validation loss as the final model from training.
We present more details on the training in \Cref{app:hyperparameters-training}.

\subsubsection{Deployment}
\label{sec:example-deployment}
Once the surrogate has been designed and trained, it must be deployed for its downstream task.
This takes different forms depending on the use case of the surrogate: whether the downstream task requires low-latency or high-throughput execution, whether the surrogate is distributed to end-users, what the expected hardware and software platform for the deployment is, or any other considerations related to the downstream use case of the surrogate.

\questionsix{}

For fairness of comparison with \mca{} the surrogate is deployed on identical hardware to \mca{}, which in this case is a single Intel Xeon Skylake CPU at 3.1GHz.

\questionseven{}

The BERT-based surrogate does not require any preprocessing of the input assembly.
To execute the surrogate we use the ONNX runtime~\citep{onnxruntime}, a runtime environment that accelerates neural network execution while also being portable across devices and programming languages.

\resetquestioncounter{}

\section{Case Study: \ADAPTATION{}}
\label{sec:adaptation}
Beyond just being fast, CPU simulators must be accurate.
To accurately model behaviors observed in real-world processors, a programmer must develop a model that matches the behavior of that processor.

The standard approach, exemplified by \mca{}, is to manually design, implement, and tune an abstract execution model of the processor.
This approach takes significant development effort, and can still result in inaccurate simulation, in part due to simplifying modeling assumptions that programmers must make that do not~accurately~reflect~real~CPUs.

Alternatively to hand-tuning a model, programmers can train a machine learning model from scratch based on observations of the ground-truth behavior of the processor.
Though it requires less development effort, this approach requires a significant amount of data~to~train~an~accurate~model.

\pagebreak[4]

\paragraph{\Adaptation{}.}
Another approach for developing an accurate simulation model is \adaptation{}.
With \adaptation{}, programmers first develop a surrogate of a program then further train that surrogate on data from a different task.
Key benefits of this approach include that \adaptation{} makes it possible to alter the semantics of the program to perform a different task of interest and that it may be more data-efficient or result in higher accuracy than training a model from scratch for the task~\citep{tercan_transfer_2018,kustowski_transfer_2019}.

\begin{figure}
  \includegraphics[width=0.95\columnwidth]{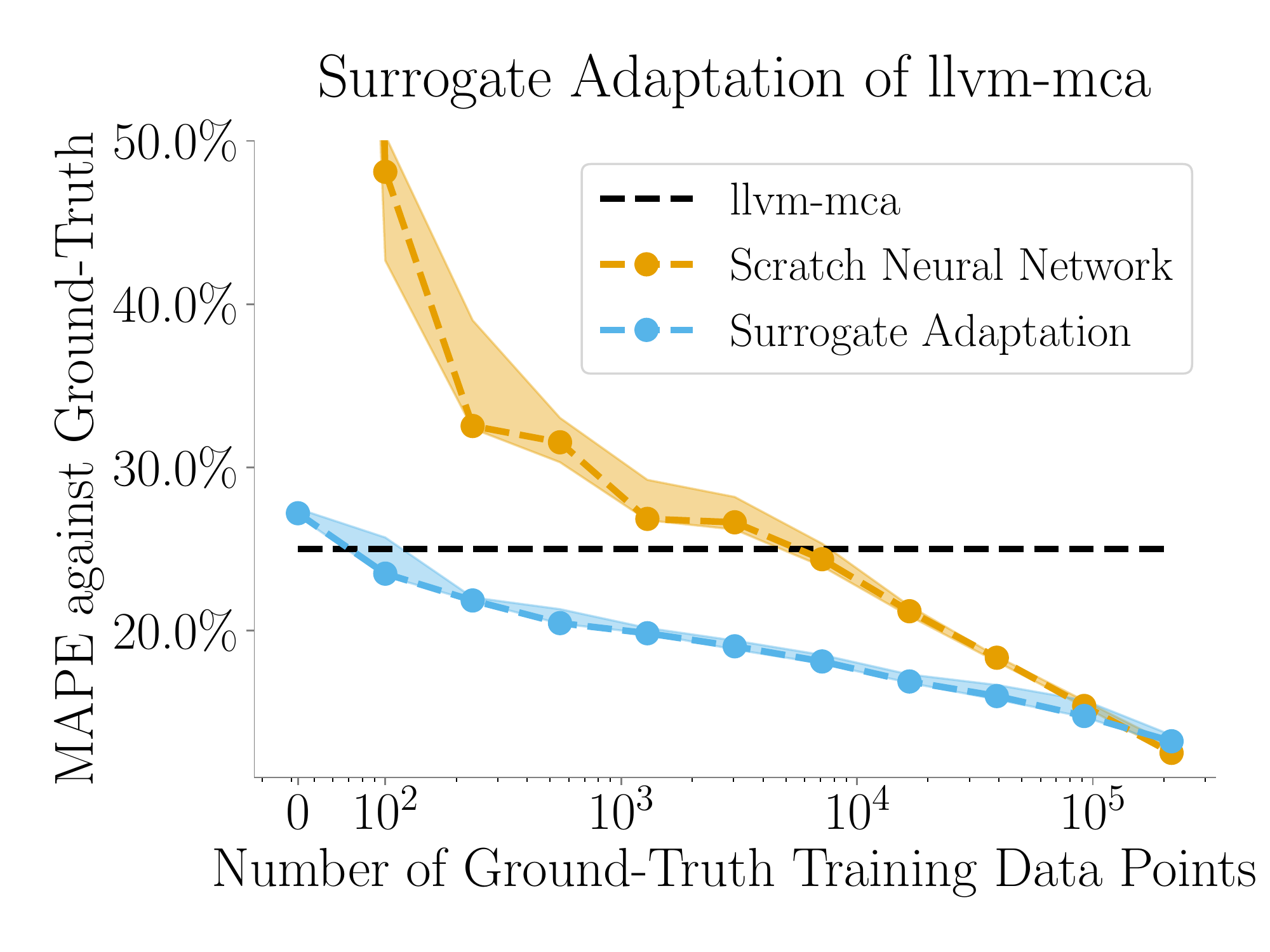}
  \vspace*{-0.1em}
  \caption{
    Error on ground-truth data of \mca{} (\simcolor{}), a neural network trained from scratch (\scratchcolor{}), and \adaptation{} of \mca{} (\dualcolor{}).
    The rightmost point corresponds to training on the entire BHive dataset.
  }
  \label{fig:mca-adaptation}
\end{figure}

\paragraph{Results.}
\Cref{fig:mca-adaptation} presents the MAPE of several approaches to predicting ground-truth basic block throughputs, as a function of the size of the training dataset of the approach.
The \simcolor{} dashed line shows \mca{}'s error rate, which is not a function of the amount of ground-truth training data available, and is constant at $25.0\%$.
The \dualcolor{} dotted line shows \adaptation{}'s error rate, which is upper bounded by \mca{}'s, as \adaptation{} is first trained to mimic \mca{}, then decreases with more training data.
The \scratchcolor{} dots show the error of a neural network trained from scratch, which results in a large error rate when trained with a small number of examples, only matching \adaptation{} when it is trained on the entire BHive training~data~set.

These results show that \adaptation{} leads to more accurate simulation than training a neural network from scratch when ground-truth data is not readily available (e.g., in cases where collecting ground-truth data is expensive), but provides no benefit when ground-truth data~is~plentiful.

\subsection{Programming Methodology}

As with \substitution{}, developing the surrogate for \adaptation{} requires a problem specification, a design for the neural network, a training procedure for the network, and a deployment configuration.

\subsubsection{Specification}
\Adaptation{} requires two steps, finding the original surrogate then adapting the surrogate to the downstream task.
This is represented as two sequential optimization problems.

In the first optimization problem for this \adaptation{} example, finding a surrogate that mimics \mca{}, we find a surrogate that minimizes the error against \mca{} without any other constraints:
\begin{align*}
  \mdual{}_1^* = \argmin_{\mdual{}}\;
  \Exp_{\minput{} \sim \mdataset{}} \mleft[ \frac{\mleft|\mdual{}\mleft(\minput{}\mright) - \mprogram{}\mleft(\minput{}, \text{haswell-params}\mright)\mright|}{\mprogram{}\mleft(\minput{}, \text{haswell-params}\mright)} \mright]
\end{align*}%
where $\mdual{}$ is the surrogate, $\mdataset{}$ is the dataset of basic blocks $\minput{}$ from BHive, $\mprogram{}$ is \mca{}, and haswell-params is LLVM's default set of Haswell~CPU~parameters.

In the second optimization problem, we optimize for accuracy on the ground-truth data: %
\begin{align*}
  \mdual{}^* &= \argmin_{\mdual{}}\;
               \Exp_{\minput{} \sim \mdataset{}} \mleft[ \frac{\mleft|\mdual{}\mleft(\minput{}\mright) - \mrelaxationobjective{}\mleft(\minput{}\mright)\mright|}{\mrelaxationobjective{}\mleft(\minput{}\mright)} \mright]%
\end{align*}%
where $\mdataset{}$ is the dataset of basic blocks $\minput{}$ from BHive, and $\mrelaxationobjective{}$ is the ground-truth measured timing of the basic block on a Haswell CPU~from~BHive. %

In \adaptation{}, the second optimization problem is seeded with the surrogate resulting from the first.

\subsubsection{Design}

In this example the neural network architecture and capacity must maximize accuracy first against \mca{} then against the ground-truth measurements.
There are no other objectives or constraints on the surrogate design.

\questionone{}

As with the \substitution{} example, we use a BERT Transformer architecture.
In general, \adaptation{} can use the same architecture as \substitution{}, though it may not have the same execution time constraints and may require an architecture that is tailored for the downstream objective.
In this \adaptation{} example the downstream objective is similar to the original program's objective, allowing us to use the same architecture.

\questiontwo{}

To minimize hyperparameter search cost, we reuse the capacity for the neural surrogate from \Cref{sec:substitution}, which has less than $10\%$ error against~llvm\nobreakdash-mca.

\subsubsection{Training}
\label{sec:adaptation-training}

\questionthree{}

As in Section 3, we use the BHive dataset to train the surrogate.
The BHive dataset is the only dataset of basic blocks with timings that correspond to the assumptions made by \mca{}, making the ground-truth errors pre- and post- \adaptation{} comparable (though for \adaptation{} in general the downstream task need not be identical to the task performed by the original program).

In the first optimization problem, the labels for training are \mca{}'s predictions on these basic blocks.
In the second optimization problem, the labels are the ground-truth measured timings on a Haswell~CPU~from~BHive.

\questionfour{}

The loss function for training the surrogate in both optimization problems is the MAPE, as specified in the specification.
In general, the loss functions for the two optimization problems do not have to be the same, if the programmer is adapting the surrogate to a substantially different problem.

\questionfive{}

In the first optimization problem of \adaptation{}, minimizing or constraining training time is not a part of the specification; we therefore reuse the neural surrogate trained in \Cref{sec:substitution}, which is the surrogate with minimum validation loss within 500 epochs of training.
In the second optimization problem, the surrogate resulting from the first step is used as a warm starting point for optimization.
We again use the minimum-validation-loss surrogate within 500 epochs of training, which constrains the surrogate in the second problem to not deviate too much from the original surrogate.
The plots of training and validation loss over the course of training are presented in \Cref{app:hyperparameters-training-telemetry}.

\subsubsection{Deployment}
Once the programmer has designed and trained the surrogate, the programmer must deploy it for its downstream task.
The specification for this \adaptation{} example does not specify objectives or constraints on the deployment for the surrogate.

\questionsix{}

The neural surrogate is trained on an NVIDIA V100 GPU, which provides sufficient throughput (over \num{512} training examples per second) to train the surrogate for each~optimization~problem.
Since the specification does not impose constraints on the deployment of the surrogate, we also deploy it on the same GPU for simplicity.

\questionseven{}

The neural surrogate is trained in PyTorch~\citep{paszke_pytorch_2019}, which automatically calculates the gradient of the surrogate for both optimization problems.
Since the specification does not impose deployment constraints, we also deploy~it~in~PyTorch.

\resetquestioncounter{}

\section{Case Study: \RELAXATION{}}
\label{sec:relaxation}
\Adaptation{} changes the semantics of the entire simulation to more accurately model ground-truth data, resulting in behavior distinct from that of the original simulation.
Such distinct behavior is not always desirable, since it leads to predictions that programmers cannot reason about with the hand-coded simulation model.
Programmers may instead want the best version of the hand-coded simulation that is possible with proper choice of~parameters~for~the~simulation.

To use \mca{} to accurately model ground-truth data, programmers must find simulation parameters that lead \mca{} to accurate simulation of the physical CPU.
The Haswell parameters in \mca{} are comprised of \num{11265} integer-valued parameters, inducing a configuration space with $10^{\num{19336}}$ possible configurations.
Each of these \num{11265} parameters be set for each different CPU that~llvm\nobreakdash-mca~targets.

The standard approach is to have experts manually set the parameters based on documentation, measurement, and intuition.
This approach again requires significant developer effort and can still result in high simulation error, due in part to the difficulties of setting \mca{}'s CPU parameters to values that lead \mca{} to low~prediction~error.

Alternatively, the parameters may be set by automatic approaches based entirely on measurement.
One class of automatic approaches for setting \mca{}'s parameters is to gather measurements of each parameter's realization in the CPU architecture that \mca{} targets~\citep{renda_difftune_2020,agner,abel_uops_2019}.

Another class of approaches is to gather coarse-grained measurements of entire basic blocks then optimize \mca{}'s parameters to best fit the timings of the basic blocks.
Due to the size of the parameter space, this is an optimization problem for which gradient-free optimization techniques~\citep{ansel_opentuner_2014} are intractable.
Gradient descent converges to local minima more quickly than gradient-free optimization with the original program~\citep{hicken_lecture}.
However, since \mca{} is not written in a differentiable programming language~\citep{baydin_automatic_2017} and operates over discrete values, it is also not possible to calculate its gradient or optimize its parameters with~gradient~descent.

\paragraph{\Relaxation{}.}
An alternative approach for optimizing parameters of the program using coarse-grained measurements is to use \relaxation{}.
We present a case study drawn from our prior work~\citep{renda_difftune_2020} of using \relaxation{} to optimize \mca{}'s parameters.

With \relaxation{}, programmers develop a surrogate of a program, optimize input parameters of that surrogate, then finally plug the optimized input parameters back into the original program.
The key benefit of this approach is that \relaxation{} can optimize inputs faster than optimizing inputs directly against the program, due to the potential for faster execution speed of the surrogate and the potential for the surrogate to be differentiable even when the original program is not (allowing for optimizing inputs with gradient descent)~\citep{renda_difftune_2020,tseng_hyperparameter_2019,she_neuzz_2019}.

\paragraph{Results.}
Using \relaxation{}, we find parameters that lead \mca{} to an average error of $23.7\%$ on the Haswell basic blocks in BHive~\citep{chen_bhive_2019}.
In contrast, the expert-tuned default Haswell parameters lead \mca{} to an average error of $25.0\%$.
OpenTuner~\citep{ansel_opentuner_2014}, a gradient-free optimization technique, is not able to find parameters that lead \mca{} to lower than $100\%$ error when given a computational budget equivalent~to~that~of~surrogate~optimization.

\subsection{Programming Methodology}

Again, developing the surrogate for \relaxation{} involves a specification, design, training, and deployment.

\subsubsection{Specification}
\Relaxation{} requires two steps, finding the original surrogate then optimizing inputs to the surrogate.
As with \adaptation{}, this is represented as two sequential optimization problems.

In the first optimization problem for \relaxation{}, the objective is to find a surrogate that minimizes the error against \mca{}'s predicted throughput for any given input basic block and set of CPU parameters: %
\begin{align*}
  \mdual{}_1^* = &\argmin_{\mdual{}}
                   \hspace*{-1.5em}
    \Exp_{\shortstack{\scriptsize $\minput{}_{\rm block}{\sim}\mdataset{}_{\rm block}$ \\ \scriptsize $\minput_{\rm params} {\sim} \mdataset{}_{\rm params}$}}
  \hspace*{-0,25em}
  \mleft[%
  \frac{\mleft|\mdual{}\mleft(\minput{}_{\rm block}, \minput{}_{\rm params}\mright) - \mprogram{}\mleft(\minput_{\rm block}, \minput{}_{\rm params}\mright)\mright|}{\mprogram{}\mleft(\minput_{\rm block}, \minput{}_{\rm params}\mright)} \mright]%
\end{align*}%
where $\mdual{}$ is the surrogate, $\mdataset{}_{\rm block}$ is the dataset of basic blocks $\minput{}_{\rm block}$ from BHive, $\mdataset{}_{\rm params}$ is a uniform distribution over parameter values $\minput{}_{\rm params}$, and $\mprogram{}$ is \mca{}.

In the second optimization problem, the objective is to find input parameters that optimize predictive accuracy against the ground-truth data: %
\begin{align*}
  \minput{}_{\rm params}^* = &\argmin_{\minput{}_{\rm params}}
                               \Exp_{\minput{}_{\rm block} \sim \mdataset{}_{\rm block}}
                               \mleft[
                               \frac{\mleft|\mdual{}_1^*(\minput{}_{\rm block}, \minput{}_{\rm params}) - \mrelaxationobjective{}\mleft(\minput{}_{\rm block}\mright)\mright|}{\mrelaxationobjective{}\mleft(\minput{}_{\rm block}\mright)}
                               \mright]
\end{align*}%
where $\mdataset{}_{\rm block}$ is the dataset of basic blocks $\minput{}_{\rm block}$ from BHive, and $\mrelaxationobjective{}$ is the ground-truth measured timing of the basic block on a Haswell CPU~from~BHive. %

\subsubsection{Design}

In this \relaxation{} example, the architecture and capacity must maximize accuracy, with no other objectives or constraints on~the~design.

\questionone{}

Due to including  $\minput{}_{\rm params}$ as input to the surrogate, the BERT architecture in Sections~3 and~4, which expects just basic blocks as input, is not sufficient for this task.
We use the neural network architecture proposed by our prior work~\citep{mendis_ithemal_2019}.

The architecture consists of a stacked pair of \NA{LSTMs}~\citep{hochreiter_lstm_1997}.
The bottommost LSTM generates a vector representation of each instruction independently.
We then concatenate each of these instruction vector representations with the relevant parameters in $\minput{}_{\rm params}$ that affect simulation of the instruction.
The topmost LSTM then processes each of these vector representations to generate a final prediction~for~the~basic~block.

We validate that this model learns to predict the throughput of basic blocks on physical Intel CPUs with low error~\citep{mendis_ithemal_2019}, a similar problem to developing a surrogate of \mca{}.

\questiontwo{}

We use a stack of 4 LSTMs in place of each original LSTM, each with a width of 256 neurons.
Stacking LSTMs increases their capacity, which is needed due to the complexity induced by adding the CPU parameters as input~to~the~surrogate.

\subsubsection{Training}
\label{sec:relaxation-training}

\questionthree{}

In both optimization problems, we use basic blocks from the BHive dataset as input basic blocks $\minput{}_{\rm block}$~\citep{chen_bhive_2019}.
In the first optimization problem, we also use a bounded uniform distribution over parameter values (informed by the range of parameter values for other CPU architectures) as input parameters $\minput{}_{\rm params}$.
As with the \adaptation{} example, in the first optimization problem the throughputs to predict are \mca{}'s predictions on these basic blocks, and in the second they are the measured~timings~from~BHive.

\questionfour{}

The loss function for training the surrogate in both optimization problems of this \relaxation{} example is the MAPE of the surrogate's prediction of \mca{}'s prediction of throughput, as specified in the specification.
As with \adaptation{}, the loss functions for both optimization problems do not have to be the same in general.

\questionfive{}

We train the surrogate and the input parameters until convergence on a validation set.
This results in training for $60$ epochs in the first training phase and $1$ epoch in the second training phase.

\subsubsection{Deployment}
Once the surrogate has been designed and trained, it is deployed for its downstream task.
Unlike \substitution{} and \adaptation{}, in \relaxation{} the surrogate is never directly deployed to end-users, instead being used entirely as an intermediate artifact in the parameter optimization process.

\questionsix{}

The surrogate itself is executed on a GPU, which provides \NA{sufficient throughput} for optimizing the input parameters.
Once found, the input parameters $\minput{}_{\rm params}^*$ are plugged back into \mca{}, which is executed on a CPU.

\questionseven{}

The surrogate and input parameters are trained in PyTorch~\citep{paszke_pytorch_2019}, which calculates the gradients for both the surrogate and the input optimization.
Once found, the input parameters $\minput{}_{\rm params}^*$ are then plugged back into \mca{}.

\section{Surrogate-Based Design Patterns}
\label{app:opt-motivation}
\label{sec:techniques}

We now present the taxonomy of surrogate-based design patterns.
We first formalize the definition and specification of a surrogate of a program.
We then present the algorithm sketches that define each design pattern, justifying these sketches with concrete examples of each design pattern from the literature.
We finally describe and provide examples of the key benefits of each design pattern.

\subsection{Surrogates of Programs}

Let $p \in \mprogramtype{}$ denote a program under study.
Let $\mobserver{} \in \mprogramtype{} \to \minputtype{} \to \moutputtype{}$ denote an \emph{interpreter}, which takes the program $\mprogram{}$ and an input $\minput{} \in \minputtype{}$ and produces an output $\moutput{} \in \moutputtype{}$.
Let $\mobserver{}^*$ denote the \emph{standard interpreter}, corresponding to the standard input-output relationship of the program according to the denotational semantics of the programming language~\citep[Chapter~5]{winksel_formal_1993}.
Other interpreters may output other aspects of the execution of the program, such as its execution time, memory usage, control flow trace, or any other aspect of its denotational or operational semantics.
Finally, let $\mdual{} \in \mprogramtype{}$ denote a surrogate~of~the~program.

The ideal surrogate $\mdual{}$ of a given interpretation $\mobserver{}_{\mprogram{}}$ of a program $\mprogram{}$ is a surrogate such that for all inputs, the standard interpretation $\mobserver{}^*$ of the surrogate has the same output as the interpretation of the program:
\[
  \forall \minput \in \minputtype{}.\, \mobserver{}^*\mleft(\mdual{}\mright)\mleft(\minput{}\mright) = \mobserver{}_{\mprogram{}}\mleft(\mprogram{}\mright)\mleft(\minput\mright)
\]

\subsection{Surrogate-Based Design Patterns}

We now formalize each of the surrogate-based design patterns.
The definitions are in the form of generic optimization problem specifications, showing the set of possible objectives and constraints on the solutions. %
These generic optimization problem specifications constitute an algorithm sketch for each surrogate-based design pattern.

Let $\mdistance{} : \moutputtype{} \times \moutputtype{} \to \mreal{}$ measure the error between two outputs.
Let $\mexecutioncost{} : \mleft(\minputtype{} \to \moutputtype{}\mright) \times \minputtype{} \to \mreal{}$ measure the cost of executing a given interpretation of a program on a given input (measured in latency, execution cost, energy, etc.).
Let $\mrelaxationobjective{} : \moutputtype{} \times \minputtype{} \to \mreal{}$ measure the error on a downstream task induced by a given prediction of a given input.

Let $\mdataset{}\mleft(\minputtype{}\mright)$ represent a distribution of program inputs that the surrogates are trained on.
Let $\mobjective{}$ and $\mconstraint{}$ denote generic objective and constraint functions for the optimization problems, which operate as reductions over the distribution of inputs $\mdataset{}\mleft(\minputtype{}\mright)$ (e.g., taking the expectation, supremum, infinimum, or other reduction over the distribution).

All together, the set of free variables for the design patterns include the choice of interpreter $\mobserver{}$ for the program, the error metric $\mdistance{}$, the execution cost metric $\mexecutioncost{}$, the downstream error metric $\mrelaxationobjective{}$, the training distribution $\mdataset{}\mleft(\minputtype{}\mright)$, the objective function $\mobjective{}$, and the constraint function $\mconstraint{}$.
The choices for each of these variables select which criteria to consider and how to weigh these criteria when training the surrogate.
In the optimization problems presented in the remainder of this section, the choice for any free variable may differ from that of any other repetition of that variable.

\paragraph{Surrogate Construction}
\label{sec:surrogate-construction}
\begin{figure*}
  \centering
  \begin{minipage}{\textwidth}
    \[
      \mdual{}_1^* = \argmin_{\mdual{}}
      \mathop{\vphantom{g}\mobjective{}}_{\minput{} \sim \mdataset{}\mleft(\minputtype{}\mright)}
      \mleft(
      \begin{matrix}
        \mdistance{} \mleft( \mobserver{}^* \mleft(\mdual{}\mright) \mleft(\minput{}\mright), \mobserver{}_{\mprogram} \mleft(\mprogram{}\mright) \mleft(\minput{}\mright) \mright), \\
        \mexecutioncost{} \mleft( \mobserver{}^* \mleft(\mdual{}\mright), \minput{} \mright)
      \end{matrix}
      \mright)
      \;\text{subject to}\;
      \mathop{\vphantom{g}\mconstraint{}}_{\minput{} \sim \mdataset{}\mleft(\minputtype{}\mright)}
      \mleft(
      \begin{matrix}
        \mdistance{} \mleft( \mobserver{}^* \mleft(\mdual{}\mright) \mleft(\minput{}\mright), \mobserver{}_{\mprogram} \mleft(\mprogram{}\mright) \mleft(\minput{}\mright) \mright), \\
        \mexecutioncost{} \mleft( \mobserver{}^* \mleft(\mdual{}\mright), \minput{} \mright)
      \end{matrix}
      \mright)
  \]
\end{minipage}
  \caption{Optimization problem for learning a surrogate $\mdual{}_1^*$ of the original program $\mprogram{}$. This optimization problem is the first step of all three surrogate-based design patterns.}
  \label{fig:surrogate-optimization}
\end{figure*}

\begin{figure*}
  \centering
  \begin{minipage}{\textwidth}
    \begin{gather*}
      \mdual{}^* = \argmin_{\mdual{}}
      \mathop{\vphantom{g}\mobjective{}^{\mprime}}_{\minput{} \sim \mdataset{}^{\mprime{}}\mleft(\minputtype{}\mright)}
      \mleft(
      \begin{matrix}
        \mrelaxationobjective{} \mleft( \mobserver{}^* \mleft(\mdual{}\mright) \mleft(\minput{}\mright), \minput{} \mright), \\
        \mdistance{} \mleft( \mobserver{}^* \mleft(\mdual{}\mright) \mleft(\minput{}\mright), \mobserver{}^* \mleft(\mdual{}^*_1\mright) \mleft(\minput{}\mright) \mright), \\
        \mdistance{} \mleft( \mobserver{}^* \mleft(\mdual{}\mright) \mleft(\minput{}\mright), \mobserver{}^{\mprime{}}_{\mprogram} \mleft(\mprogram{}\mright) \mleft(\minput{}\mright) \mright), \\
        \mexecutioncost{} \mleft( \mobserver{}^* \mleft(\mdual{}\mright), \minput{} \mright)
      \end{matrix}
      \mright)
      \;\text{subject to}\;
      \mathop{\vphantom{g}\mconstraint{}^{\mprime{}}}_{\minput{} \sim \mdataset{}^{\mprime{}}\mleft(\minputtype{}\mright)}
      \mleft(
      \begin{matrix}
        \mrelaxationobjective{} \mleft( \mobserver{}^* \mleft(\mdual{}\mright) \mleft(\minput{}\mright), \minput{} \mright), \\
        \mdistance{} \mleft( \mobserver{}^* \mleft(\mdual{}\mright) \mleft(\minput{}\mright), \mobserver{}^* \mleft(\mdual{}^*_1\mright) \mleft(\minput{}\mright) \mright), \\
        \mdistance{} \mleft( \mobserver{}^* \mleft(\mdual{}\mright) \mleft(\minput{}\mright), \mobserver{}^{\mprime{}}_{\mprogram} \mleft(\mprogram{}\mright) \mleft(\minput{}\mright) \mright), \\
        \mexecutioncost{} \mleft( \mobserver{}^* \mleft(\mdual{}\mright), \minput{} \mright)
      \end{matrix}
      \mright)
  \end{gather*}
\end{minipage}
\vspace*{-12pt}
  \caption{Second optimization problem for \adaptation{}, which re-trains a surrogate $\mdual{}^*_1$ to find another surrogate $\mdual{}^*$ with higher accuracy against a different objective. The surrogate $\mdual{}^*_1$ is used as a warm start for this problem.}
  \label{fig:adaptation-optimization}
\end{figure*}

\begin{figure*}%
  \begin{minipage}{\textwidth}
    \begin{gather*}
      \minput{}^* = \argmin_{\minput{}}
      \mobjective{}^{\mprime{}}
      \mleft(
      \begin{matrix}
        \mrelaxationobjective{} \mleft( \mobserver{}^* \mleft(\mdual{}_1^*\mright) \mleft(\minput{}\mright), \minput{} \mright) %
      \end{matrix}
      \mright)
      \;\text{subject to}\;
      \mconstraint{}^{\mprime{}}
      \mleft(
      \begin{matrix}
        \mrelaxationobjective{} \mleft( \mobserver{}^* \mleft(\mdual{}_1^*\mright) \mleft(\minput{}\mright), \minput{} \mright) %
      \end{matrix}
      \mright)
    \end{gather*}
  \end{minipage}
  \vspace*{-1em}
  \caption{Second optimization problem for \relaxation{}, which optimizes inputs $\minput{}$ of a surrogate $\mdual{}^*_1$ to minimize a different objective function on the surrogate.}
    \label{fig:relaxation-optimization}
  \end{figure*}

The first step of each surrogate-based design pattern is to train a surrogate of the original program.
\Cref{fig:surrogate-optimization} presents the generic optimization problem that defines this step.
Surrogate construction is defined by an optimization problem that finds a surrogate $\mdual{}_1^*$ that
minimizes a task-dependent objective function $\mobjective{}$
over a distribution of inputs $\minput{} \sim \mdataset{}\mleft(\minputtype{}\mright)$
of
the error $\mdistance{}$ between
the standard interpretation $\mobserver{}^*$ of the surrogate $\mdual{}$ on that input $\minput{}$
and
an interpretation $\mobserver{}_{\mprogram{}}$ of the original program $\mprogram{}$ on the input $\minput{}$,
and of the execution cost $\mexecutioncost{}$ of the standard interpretation $\mobserver{}^*$ of the surrogate $\mdual{}$ on the input $\minput{}$,
subject to a constraint function $\mconstraint{}$ of the same terms.

\subsubsection{\SUBSTITUTION{}}

In \substitution{}, the programmer simply deploys the surrogate found in the surrogate construction step to the end-user: $\mdual{}^* = \mdual_1^*$.

\subsubsection{\ADAPTATION{}}

The first step of \adaptation{} is the initial surrogate construction step.
The second step is to continue to train the surrogate to optimize a different downstream objective.

\Cref{fig:adaptation-optimization} shows the generic optimization problem that defines the second step of \adaptation{}.
This second optimization problem finds a surrogate $\mdual{}^*$ that minimizes a task-dependent objective function $\mobjective{}^{\mprime{}}$
over a distribution of inputs $\minput{} \sim \mdataset{}^{\mprime{}}\mleft(\minputtype{}\mright)$
of the downstream error $\mrelaxationobjective{}$ of the standard interpretation $\mobserver{}^*$ of the surrogate $\mdual{}$ on an input $\minput{}$,
the error $\mdistance{}$ between
the standard interpretation $\mobserver{}^*$ of the surrogate $\mdual{}$ on the input $\minput{}$ and
the standard interpretation $\mobserver{}^*$ of the surrogate $\mdual{}^*_1$ from the first optimization problem on that input $\minput{}$,
the error $\mdistance{}$ between
the standard interpretation $\mobserver{}^*$ of the surrogate $\mdual{}$ on the input $\minput{}$ and
an interpretation $\mobserver{}^{\mprime{}}_p$ of the program $\mprogram{}$ on that input $\minput{}$,
and the execution cost $\mexecutioncost{}$ of the standard interpretation $\mobserver{}^*$ of the surrogate $\mdual{}$ on that input $\minput{}$,
subject to a constraint function $\mconstraint{}^{\mprime{}}$ of the~same~terms.

In \adaptation{}, the surrogate from the first optimization problem is used as a warm starting point for the second optimization problem.

\subsubsection{\RELAXATION{}}

The first step of \relaxation{} is the surrogate construction step.
The second is to optimize inputs to the surrogate against a different objective.
\Cref{fig:relaxation-optimization} shows the optimization problem that defines the second step of \relaxation{}.

This second optimization problem finds an input $\minput{}^*$ that minimizes a task-dependent objective function $\mobjective{}^{\mprime{}}$
of the downstream error $\mrelaxationobjective{}$ of the standard interpretation $\mobserver{}^*$ of the surrogate from the first optimization problem $\mdual{}_1^*$ on the input $\minput{}$,
subject to a constraint function $\mconstraint{}^{\mprime{}}$ of~the~same~term.

\subsubsection{Specifications in the Literature}
\label{sec:specifications}
\label{sec:literature-review}

\Cref{tab:substitution-related,tab:adaptation-related,tab:relaxation-related} respextively present surveys of \substitution{}, \adaptation{}, and \relaxation{}, showing the terms in the optimization problem solved by each piece of related work.
These optimization problem specifications correspond to concrete instantiations of interpreters $\mobserver{}$, error functions $\mdistance{}$, $\mexecutioncost{}$, and $\mrelaxationobjective{}$, and objective~functions~$\mobjective{}$~and~$\mconstraint{}$.

With examples in hand, we now discuss the design considerations and trade-offs that must be considered when specifying the optimization problem for training a surrogate.

\bgroup
\def\arraystretch{0}
\cellspacetoplimit 0pt
\cellspacebottomlimit 0pt
\renewcommand\tabularxcolumn[1]{m{#1}}

\begin{table*}[!htbp]
  \caption{Optimization problem specifications of \substitution{} from the literature.}
  \label{tab:substitution-related}

  \vspace*{-11pt}

\begin{tabularx}{\textwidth}{m{2in}X}
  \toprule
  \textbf{Citation and description} & \textbf{Optimization problem specification} \\
  \midrule
  \textbf{\citet{esmaeilzadeh_neural_2012}:} Training neural surrogates of small numerical kernels to decrease their execution latency by executing them on~a~neural~network~accelerator. & %
  $$ \mdual{}^* = \argmin_{\mdual{}} \mobjective{} \mleft( \begin{matrix} \mdistance{}\mleft(\mdual{}, \mprogram{}\mright), \\ \mexecutioncost{}\mleft(\mdual{}\mright) \end{matrix} \mright) \; \text{subj.\ to}\; \mconstraint{} \mleft( \begin{matrix} \mdistance{}\mleft(\mdual{}, \mprogram{}\mright), \\ \mexecutioncost{}\mleft(\mdual{}\mright) \end{matrix} \mright) $$
  \begin{itemize}[leftmargin=*]
  \item $\mobjective{} \mleft( \mdistance{}\mleft(\mdual{}, \mprogram{}\mright) \mright)$: The mean squared error between the outputs of the surrogate and the original kernel is minimized~\cite[Section~4]{esmaeilzadeh_neural_2012}.
  \item $\mobjective{} \mleft( \mexecutioncost{}\mleft(\mdual{}\mright) \mright)$: The size of the surrogate (measured by the number of hidden units) is minimized to reduce execution time~\cite[Section~4]{esmaeilzadeh_neural_2012}.
  \item $\mconstraint{} \mleft( \mdistance{}\mleft(\mdual{}, \mprogram{}\mright) \mright)$: The end-to-end error of the application that uses the surrogate is constrained to be less than $10\%$~\cite[Section~7.1]{esmaeilzadeh_neural_2012}.
  \item $\mconstraint{} \mleft( \mexecutioncost{}\mleft(\mdual{}\mright) \mright)$: The surrogate is constrained to have lower execution latency than the original kernel~\cite[Sections~7,~8]{esmaeilzadeh_neural_2012}.
  \end{itemize}
  \\ %

  \textbf{\citet[Chapter~4]{mendis_thesis_2020}:} Training neural surrogates of compiler auto-vectorizers, to replace the original exponential-time auto-vectorizer with a linear time surrogate. &
  $$ \mdual{}^* = \argmin_{\mdual{}} \mobjective{} \mleft( \begin{matrix} \mdistance{}\mleft(\mdual{}, \mprogram{}\mright) \end{matrix} \mright) \; \text{subj.\ to}\; \mconstraint{} \mleft( \begin{matrix} \mexecutioncost{}\mleft(\mdual{}\mright) \end{matrix} \mright) $$
  \begin{itemize}[leftmargin=*]
  \item $\mobjective{} \mleft( \mdistance{}\mleft(\mdual{}, \mprogram{}\mright) \mright)$: The cross entropy error between the outputs of the surrogate and the auto-vectorizer is minimized~\cite[Chapter~4.4]{mendis_thesis_2020}.
  \item $\mconstraint{} \mleft( \mexecutioncost{}\mleft(\mdual{}\mright) \mright)$: The surrogate has predictable (and not data-dependent) linear running time~\cite[Chapters~1.3.4,~4.8]{mendis_thesis_2020}.
  \end{itemize}
  \\ %

  \textbf{\citet{munk_deep_2019}:} Training neural surrogates of stochastic simulators to accelerate simulation and inference using the simulator. &
  $$ \mdual{}^* = \argmin_{\mdual{}} \mobjective{} \mleft( \begin{matrix} \mdistance{}\mleft(\mdual{}, \mprogram{}\mright), \\ \mexecutioncost{}\mleft(\mdual{}\mright) \end{matrix} \mright) \; \text{subj.\ to}\; \mconstraint{} \mleft( \begin{matrix} \mexecutioncost{}\mleft(\mdual{}\mright) \end{matrix} \mright) $$
  \begin{itemize}[leftmargin=*]
  \item $\mobjective{} \mleft( \mdistance{}\mleft(\mdual{}, \mprogram{}\mright) \mright)$: The KL divergence between the outputs of the surrogate and the original stochastic simulator is minimized~\cite[Section~3.1]{munk_deep_2019}.
  \item $\mobjective{} \mleft( \mexecutioncost{}\mleft(\mdual{}\mright) \mright)$: The surrogate is as fast as possible to maximize the execution throughput speedup over the original simulator~\cite[Section~3.2]{munk_deep_2019}.
  \item $\mconstraint{} \mleft( \mexecutioncost{}\mleft(\mdual{}\mright) \mright)$: The surrogate is constrained to have higher execution throughput than the original simulator~\cite[Section~3.2]{munk_deep_2019}.
  \end{itemize}
  \\ %

    \textbf{\citet{pestourie_active_2020}:} Training neural surrogates of partial differential equation (PDE) solvers to aid designing material composites, using active learning to minimize the training cost of the surrogate. &
  $$ \mdual{}^* = \argmin_{\mdual{}} \mobjective{} \mleft( \begin{matrix} \mdistance{}\mleft(\mdual{}, \mprogram{}\mright) \\ \mexecutioncost{}\mleft(\mdual{}\mright) \end{matrix} \mright) \; \text{subj.\ to}\; \mconstraint{} \mleft( \begin{matrix} \mexecutioncost{}\mleft(\mdual{}\mright) %
  \end{matrix} \mright) $$
  \begin{itemize}[leftmargin=*]
  \item $\mobjective{} \mleft( \mdistance{}\mleft(\mdual{}, \mprogram{}\mright) \mright)$: The MAPE between the outputs of the surrogate and the original PDE solver is minimized~\cite[Figure~5]{pestourie_active_2020}.
  \item $\mobjective{} \mleft( \mexecutioncost{}\mleft(\mdual{}\mright) \mright)$: The surrogate is as fast as possible to maximize execution latency speedup over the original solver~\cite[``Introduction'']{pestourie_active_2020}.
  \item $\mconstraint{} \mleft( \mexecutioncost{}\mleft(\mdual{}\mright) \mright)$: The surrogate must have higher execution throughput than the original PDE solver~\cite[``Introduction'']{pestourie_active_2020}.
  \end{itemize}
  \\

  \bottomrule
\end{tabularx}

\end{table*}

\egroup

\bgroup
\renewcommand\tabularxcolumn[1]{m{#1}}

\begin{table*}[!htbp]
  \caption{Optimization problem specifications of \adaptation{} from the literature.}
\label{tab:adaptation-related}

\begin{tabularx}{\textwidth}{m{1.55in}X}
  \toprule
  \textbf{Citation and description} & \textbf{Optimization problem specification} \\
  \midrule

  \textbf{\citet{tercan_transfer_2018}:} Training neural surrogates of computer simulations of plastic injection molding, then adapting the surrogates on real-world experiments of injection molding to close the gap between simulated and real results. & %
  {\hspace*{3em}\begin{tabularx}{5.25in}{rcl}$\displaystyle
    \mdual{}_1^* = \argmin_{\mdual{}} \mobjective{}_1 \mleft( \begin{matrix} \mdistance{}\mleft(\mdual{}, \mprogram{}\mright) \end{matrix} \mright)
    $ & \hspace*{3em} & $\displaystyle
    \mdual{}^* =   %
    \left\{
      \begin{alignedat}{1}
    &\argmin_{\mdual{}} \mobjective{}^{\mprime{}}_2 \mleft( \begin{matrix} \mrelaxationobjective{}\mleft(\mdual{}_1^*, \minput{}\mright), \\ \mexecutioncost{}\mleft(\mdual{}\mright) \end{matrix} \mright)  \\  %
    &\text{subj.\ to}\; \mconstraint{}^{\mprime{}}_2 \mleft( \begin{matrix} \mrelaxationobjective{}\mleft(\mdual{}_1^*, \minput{}\mright) %
    \end{matrix} \mright) %
    \end{alignedat}
  \right.
    $
  \end{tabularx}}
  \begin{itemize}[leftmargin=*]
  \item $\mobjective{}_1 \mleft( \mdistance{}\mleft(\mdual{}, \mprogram{}\mright) \mright)$: The Pearson correlation coefficient between the outputs of the surrogate and the original simulation is maximized~\cite[Section~5.2]{tercan_transfer_2018}.
  \item $\mobjective{}^{\mprime{}}_2 \mleft( \mrelaxationobjective{}\mleft(\mdual{}_1^*, \minput{}\mright) \mright)$: The Pearson correlation coefficient between the trained surrogate and the results of the real-world experiments is maximized~\cite[Section~5.2]{tercan_transfer_2018}.
  \item $\mobjective{}^{\mprime{}}_2 \mleft( \mexecutioncost{}\mleft(\mdual{}\mright) \mright)$: The surrogate is cheaper to execute than physical experiments~\cite[Section~1]{tercan_transfer_2018}.
  \item $\mconstraint{}^{\mprime{}}_2 \mleft( \mrelaxationobjective{}\mleft(\mdual{}_1^*, \minput{}\mright) \mright)$: The L1 loss of the surrogate is constrained to be less than $0.01$~\cite[Section~4]{tercan_transfer_2018}.
  \end{itemize}
  \\

    \textbf{\citet{kustowski_transfer_2019}:} Training neural surrogates of computer simulations of inertial confinement fusion, then adapting on a small number of results from real-world experiments to close the gap between simulated and~real~results. & %
    {\hspace*{3em}\begin{tabularx}{5.25in}{rcl}$\displaystyle
  \mdual{}_1^* = \argmin_{\mdual{}} \mobjective{}_1 \mleft( \begin{matrix} \mdistance{}\mleft(\mdual{}, \mprogram{}\mright) \end{matrix} \mright)
  $ & \hspace*{3em} & $\displaystyle
  \mdual{}^* = \argmin_{\mdual{}} \mobjective{}^{\mprime{}}_2 \mleft( \begin{matrix} \mrelaxationobjective{}\mleft(\mdual{}_1^*, \minput{}\mright), \\ \mdistance{}\mleft(\mdual{}, \mdual{}^*_1\mright), \\ \mexecutioncost{}\mleft(\mdual{}\mright) \end{matrix} \mright) %
  $
\end{tabularx}}
  \begin{itemize}[leftmargin=*]
  \item $\mobjective{}_1 \mleft( \mdistance{}\mleft(\mdual{}, \mprogram{}\mright) \mright)$: The Pearson correlation coefficient between the outputs of the surrogate and the original simulation is maximized~\cite[Section~\rom{2}]{kustowski_transfer_2019}.
  \item $\mobjective{}^{\mprime{}}_2 \mleft( \mrelaxationobjective{}\mleft(\mdual{}_1^*, \minput{}\mright) \mright)$: The Pearson correlation coefficient between the trained surrogate and the results of the real-world experiments is maximized~\cite[Section~\rom{2}]{kustowski_transfer_2019}.
  \item $\mobjective{}^{\mprime{}}_2 \mleft( \mdistance{}\mleft(\mdual{}, \mdual{}^*_1\mright) \mright)$: $\mdual{}^*$ is biased to be close to $\mdual{}_1^*$ by freezing the weights in most layers in the neural network to be equal to their values in $\mdual{}_1^*$~\cite[Section~\rom{3}.B]{kustowski_transfer_2019}.
  \item $\mobjective{}^{\mprime{}}_2 \mleft( \mexecutioncost{}\mleft(\mdual{}\mright) \mright)$: The surrogate is cheaper to run than real-world experiments~\cite[Section~\rom{1}]{kustowski_transfer_2019}.
  \end{itemize}
  \\

  \textbf{\citet{kwon_transfer_2020}:} Training neural surrogates of computer architecture simulations of programs for design space exploration of the architecture, then adapting the surrogates for accurate design space exploration when simulating other programs. &
  {\hspace*{3em}\begin{tabularx}{5.25in}{rcl}$\displaystyle
  \mdual{}_1^* = \argmin_{\mdual{}} \mobjective{}_1 \mleft( \begin{matrix} \mdistance{}\mleft(\mdual{}, \mprogram{}\mright) \end{matrix} \mright) %
  $ & \hspace*{3em} & $\displaystyle
  \mdual{}^* = \argmin_{\mdual{}} \mobjective{}^{\mprime{}}_2 \mleft( \begin{matrix} \mrelaxationobjective{}\mleft(\mdual{}_1^*, \minput{}\mright), \\ \mdistance{}\mleft(\mdual{}, \mdual{}^*_1\mright), \\ \mexecutioncost{}\mleft(\mdual{}\mright)  \end{matrix} \mright) %
  $
  \end{tabularx}}
  \begin{itemize}[leftmargin=*]
  \item $\mobjective{}_1 \mleft( \mdistance{}\mleft(\mdual{}, \mprogram{}\mright) \mright)$: The mean squared error between the outputs of the surrogate and the simulated running time for the training programs is minimized~\cite[Section~1]{kwon_transfer_2020}.
  \item $\mobjective{}^{\mprime{}}_2 \mleft( \mrelaxationobjective{}\mleft(\mdual{}_1^*, \minput{}\mright) \mright)$: The mean squared error of the surrogate on new programs not in the surrogate's original training set is minimized~\cite[Section~2]{kwon_transfer_2020}.
  \item $\mobjective{}^{\mprime{}}_2 \mleft( \mdistance{}\mleft(\mdual{}, \mdual{}^*_1\mright) \mright)$: $\mdual{}^*$ is biased to be close to $\mdual{}_1^*$ by using the weights from $\mdual{}_1^*$ as a warm starting point for the optimization problem~\cite[Section~3]{kwon_transfer_2020}.
  \item $\mobjective{}^{\mprime{}}_2 \mleft( \mexecutioncost{}\mleft(\mdual{}\mright) \mright)$: The surrogate is cheaper to run than simulation~\cite[Section~1]{kwon_transfer_2020}.
  \end{itemize}
  \\

  \textbf{\citet{kaya_using_2019}:} Training neural surrogates of physics simulations of properties of a given material for designing structures with that material, then adapting those surrogates to aid simulation-based design with~other~materials. &
  {\hspace*{3em}\begin{tabularx}{5.25in}{rcl}$\displaystyle
  \mdual{}_1^* = \argmin_{\mdual{}} \mobjective{}_1 \mleft( \begin{matrix} \mdistance{}\mleft(\mdual{}, \mprogram{}\mright) \end{matrix} \mright) %
  $ & \hspace*{3em} & $\displaystyle
  \mdual{}^* =
  \left\{
    \begin{alignedat}{1}
      &\argmin_{\mdual{}} \mobjective{}^{\mprime{}}_2 \mleft( \begin{matrix} \mrelaxationobjective{}\mleft(\mdual{}_1^*, \minput{}\mright), \\ \mdistance{}\mleft(\mdual{}, \mdual{}^*_1\mright), \\ \mexecutioncost{}\mleft(\mdual{}\mright) \end{matrix} \mright) \\
      &\text{subj.\ to}\; \mconstraint{}^{\mprime{}}_2 \mleft( \begin{matrix} \mdistance{}\mleft(\mdual{}, \mprogram{} \mright) %
    \end{matrix} \mright)
    \end{alignedat}
    \right.
  $
  \end{tabularx}}
  \begin{itemize}[leftmargin=*]
  \item $\mobjective{}_1 \mleft( \mdistance{}\mleft(\mdual{}, \mprogram{}\mright) \mright)$: The mean squared error between the outputs of the surrogate and simulation on the base material is minimized~\cite[``Results~and~Discussion~--~Base~Case'']{kaya_using_2019}.
  \item $\mobjective{}^{\mprime{}}_2 \mleft( \mrelaxationobjective{}\mleft(\mdual{}_1^*, \minput{}\mright) \mright)$: The error of the outputs of the trained surrogate on the new material is minimized~\cite[``Results and Discussion~--~Transfer~Cases'']{kaya_using_2019}.
  \item $\mobjective{}^{\mprime{}}_2 \mleft( \mdistance{}\mleft(\mdual{}, \mdual{}^*_1\mright) \mright)$: $\mdual{}^*$ is biased to be close to $\mdual{}_1^*$ by using the weights from $\mdual{}_1^*$ as a warm starting point for the optimization problem~\cite[``Introduction'']{kaya_using_2019}.
  \item $\mobjective{}^{\mprime{}}_2 \mleft( \mexecutioncost{}\mleft(\mdual{}\mright) \mright)$: The surrogate is cheaper to run than simulation~\cite[``Introduction'']{kaya_using_2019}.
  \item $\mconstraint{}^{\mprime{}}_2 \mleft( \mdistance{}\mleft(\mdual{}, \mprogram{}\mright) \mright)$: If $\mdual{}^*$ is less accurate than simulation, then the transfer learning results in low accuracy and is rejected~\cite[``Results and Discussion~--~Transfer~Cases'']{kaya_using_2019}.
  \end{itemize}
  \\

  \bottomrule
\end{tabularx}

\end{table*}

\egroup

\bgroup
\def\arraystretch{0}
\cellspacetoplimit 0pt
\cellspacebottomlimit 0pt
\renewcommand\tabularxcolumn[1]{m{#1}}

\begin{table*}

  \caption{Optimization problem specifications of \relaxation{} from the literature.}
\label{tab:relaxation-related}

\begin{tabularx}{\textwidth}{m{1.75in}X}
  \toprule
  \textbf{Citation and description} & \textbf{Optimization problem specification} \\
  \midrule

  \textbf{\citet{renda_difftune_2020}:} Training neural surrogates of CPU simulators that predict execution time of code, then optimizing parameters of the CPU simulator to more closely match ground-truth execution times measured on real hardware. &
  {\hspace*{3em}\begin{tabularx}{5in}{ccc}
  $\displaystyle \mdual{}_1^* = \argmin_{\mdual{}} \mobjective{} \mleft( \begin{matrix} \mdistance{}\mleft(\mdual{}, \mprogram{}\mright) \end{matrix} \mright) %
    $ & \hspace*{3em} & $\displaystyle
  \minput{}^* = \argmin_{\minput{}} \mobjective{}^{\mprime{}}\mleft( \mrelaxationobjective{}\mleft(\mdual{}_1^*, \minput{}\mright)  \mright) %
  $
  \end{tabularx}}
  \begin{itemize}[leftmargin=*]
  \item $\mobjective{} \mleft( \mdistance{}\mleft(\mdual{}, \mprogram{}\mright) \mright)$: The MAPE between the outputs of the surrogate and the CPU simulator on a given input code snippet is minimized~\cite[Section~\rom{3}]{renda_difftune_2020}.
  \item $\mrelaxationobjective{}\mleft(\mdual{}_1^*, \minput{}\mright) $: The MAPE of the output of the trained surrogate induced by the set of simulation parameters is minimized against the ground-truth data~\cite[Section~\rom{3}]{renda_difftune_2020}.
  \end{itemize}
  \\

  \textbf{\citet{she_neuzz_2019}:} Training neural surrogates of the branching behavior of programs to find inputs that trigger branches that cause bugs in the program. &
  {\hspace*{3em}\begin{tabularx}{5in}{ccc}
  $\displaystyle \mdual{}_1^* = \argmin_{\mdual{}} \mobjective{} \mleft( \begin{matrix} \mdistance{}\mleft(\mdual{}, \mprogram{}\mright) \end{matrix} \mright) %
  $ & \hspace*{3em} & $\displaystyle
  \minput{}^* = \argmin_{\minput{}} \mobjective{}^{\mprime{}}\mleft( \mrelaxationobjective{}\mleft(\mdual{}_1^*, \minput{}\mright)  \mright) %
  $
  \end{tabularx}}
  \begin{itemize}[leftmargin=*]
  \item $\mobjective{} \mleft( \mdistance{}\mleft(\mdual{}, \mprogram{}\mright) \mright)$: The binary cross-entropy error between the output of the surrogate and the actual branching behavior of the program is minimized~\cite[Section~\rom{4}.B]{she_neuzz_2019}.
  \item $\mrelaxationobjective{}\mleft(\mdual{}_1^*, \minput{}\mright) $: Gradient descent tries to find an input that lead to an unseen set of branches taken in the program~\cite[Section~\rom{4}.C]{she_neuzz_2019}.
  \end{itemize}
  \\
  \textbf{\citet{tseng_hyperparameter_2019}:} Training neural surrogates of camera pipelines, to find parameters for the pipelines that lead to the cameras producing the most photorealistic images. &
  {\hspace*{3em}\begin{tabularx}{5in}{ccc}
    $\displaystyle \mdual{}_1^* = \argmin_{\mdual{}} \mobjective{} \mleft( \begin{matrix} \mdistance{}\mleft(\mdual{}, \mprogram{}\mright) \end{matrix} \mright) %
    $ & \hspace*{3em} & $\displaystyle
  \minput{}^* = \argmin_{\minput{}} \mobjective{}^{\mprime{}}\mleft( \mrelaxationobjective{}\mleft(\mdual{}_1^*, \minput{}\mright)  \mright) %
      $
  \end{tabularx}}
  \begin{itemize}[leftmargin=*]
  \item $\mobjective{} \mleft( \mdistance{}\mleft(\mdual{}, \mprogram{}\mright) \mright)$: The L2 error between the predicted image from the surrogate and the image resulting from the pipeline is minimized~\cite[Section~4.2]{tseng_hyperparameter_2019}.
  \item $\mrelaxationobjective{}\mleft(\mdual{}_1^*, \minput{}\mright) $: Gradient descent tries to find parameters that lead to images being as similar as possible in L2 distance to the ground-truth~\cite[Section~4.2]{tseng_hyperparameter_2019}.
  \end{itemize}
  \\

  \textbf{\citet{shirobokov_differentiating_2020}:} Training neural surrogates of physics simulators to find inputs that lead to local optima. &
  {\hspace*{3em}\begin{tabularx}{5in}{ccc}
    $\displaystyle
    \mdual{}_1^* = \argmin_{\mdual{}} \mobjective{} \mleft( \begin{matrix} \mdistance{}\mleft(\mdual{}, \mprogram{}\mright) \end{matrix} \mright) %
    $ & \hspace*{3em} & $\displaystyle
      \minput{}^* = \argmin_{\minput{}} \mobjective{}^{\mprime{}}\mleft( \mrelaxationobjective{}\mleft(\mdual{}_1^*, \minput{}\mright)  \mright) %
      $
  \end{tabularx}}

  \begin{itemize}[leftmargin=*]
  \item $\mobjective{} \mleft( \mdistance{}\mleft(\mdual{}, \mprogram{}\mright) \mright)$: The error (as measured by a domain-specific loss function per-task) between the outputs of the surrogate and the simulation is minimized~\cite[Section~2.2]{shirobokov_differentiating_2020}.
  \item $\mrelaxationobjective{}\mleft(\mdual{}_1^*, \minput{}\mright) $: Gradient descent tries to find parameters that lead to local optima in the problem space against the same domain-specific loss function~\cite[Section~2.2]{shirobokov_differentiating_2020}.
  \end{itemize}
    \\
  \bottomrule

\end{tabularx}

\end{table*}

\egroup

\paragraph{Surrogate error.}
A surrogate must compute a similar function to that computed by its source program.
When the surrogate is deployed to end-users as in \substitution{} and \adaptation{}, the error metric for the surrogate is that of the domain~\citep{esmaeilzadeh_neural_2012}.
When the surrogate is used as an intermediate artifact as in \relaxation{}, other error metrics may help to learn a surrogate that allows for successful~downstream~optimization~\citep{tseng_hyperparameter_2019}.

In the second step of \adaptation{}, the final surrogate may also be constrained to be close to the original surrogate, another instantiation of surrogate error (treating the original surrogate as a source program)~\citep{kwon_transfer_2020,kaya_using_2019}.

\paragraph{Downstream error.}
For \adaptation{} and \relaxation{}, the second optimization problems use an error metric beyond that of mimicking the original program.
This downstream error metric may be that of the downstream task that the original program targets~\citep{renda_difftune_2020,tercan_transfer_2018}.
The downstream error metric may also be unrelated to the domain of the original program: for instance, \citet{kwon_transfer_2020} use an error metric for \adaptation{} that adapts the surrogate to inputs and outputs of a different domain.
\citet{she_neuzz_2019} use an error metric for \relaxation{} that measures the extent to which the discovered inputs trigger unseen control flow paths in the program.

\paragraph{Execution Cost}
Regardless of the intended use case, a surrogate must be efficient, not exceeding resource budgets to deploy.
The execution cost of a surrogate measures the resources required to execute the surrogate in its execution environment.
The ideal is a surrogate that is efficient to execute, with low execution latency~\citep{esmaeilzadeh_neural_2012}, high throughput~\citep{mendis_thesis_2020}, low storage cost~\citep{han_deep_2016}, and minimal energy cost~\citep{esmaeilzadeh_neural_2012}.

\subsection{Key Benefits}
We now demonstrate the key benefits of each design pattern, detailing examples beyond those of the case study.

\subsubsection{\SUBSTITUTION{}}
\Substitution{} allows for the ability to execute the surrogate on different hardware and the ability to bound or to accelerate the execution time of the surrogate~\citep{esmaeilzadeh_neural_2012,mendis_thesis_2020}.

\paragraph{Compiling to different hardware.}
\citet{esmaeilzadeh_neural_2012} develop surrogates of small computational kernels, then deploy the surrogates on a hardware accelerator that reduces the latency and energy cost of executing the surrogate.
More generally, surrogates can be deployed on any hardware that supports the surrogate architecture, resulting in different trade-offs compared to the CPU architectures that many conventional programs execute on.

\paragraph{Different algorithmic complexity.}
Algorithmic complexity can differ between a program and its surrogate: for example, while an algorithm may require an exponential number of operations in the size of the input, a surrogate of that algorithm may only require a linear number of operations to approximate the algorithm to satisfactory accuracy~\citep{mendis_vemal_2019,karpathy_software2_2017}.

\subsection{\ADAPTATION{}}
\Adaptation{} makes it possible to alter the semantics of the program to perform a different task of interest.
\Adaptation{} may be more data-efficient or result in higher accuracy than training a model from scratch~\citep{tercan_transfer_2018,kustowski_transfer_2019}.

\paragraph{Data efficiency.}
\citet{tercan_transfer_2018} develop models that accurately simulate a plastic injection molding process.
\citeauthor{tercan_transfer_2018} train surrogates of computer simulations of injection molding, then adapt the surrogates on real-world experiments of the injection molding process to close the gap between simulation and ground-truth.
\citeauthor{tercan_transfer_2018} show that the surrogate resulting from \adaptation{} requires less training data than a neural network trained~from~scratch.

\paragraph{Accuracy.}
\citet{kustowski_transfer_2019} learn a model of a physical process involved in nuclear fusion, inertial confinement fusion (ICF).
Physical simulation is critical for this area of research, but it is not accurate in part due to unknown biases and inaccuracies in the models of ICF.
\citeauthor{kustowski_transfer_2019} use \adaptation{} to increase the accuracy of simulators by training a surrogate of simulation then adapting the surrogate on data from physical experiments.

\subsubsection{\RELAXATION{}}
\Relaxation{} optimizes inputs faster than optimizing inputs directly against the program, due to the potential for faster execution speed of the surrogate and the potential for the surrogate to be differentiable even when the original program is not~\citep{renda_difftune_2020,tseng_hyperparameter_2019,she_neuzz_2019}.

\paragraph{Faster execution time.}
\citet{ipek_exploring_2006} perform design space exploration on a simulated computer architecture, finding the physical parameters (e.g., cache size, cache associativity, etc.) that lead to the best performance.
\citeauthor{ipek_exploring_2006} use surrogate optimization to optimize these parameters, exploiting the significantly faster execution of the surrogate compared to the execution of the original simulation.

\paragraph{Differentiable output domain of programs.}
\citet{she_neuzz_2019} construct neural surrogates of programs for \emph{fuzzing}, generating inputs that cause bugs in the program.
For a given input, a classical program has an \emph{execution trace}, the set of edges taken in the control flow graph, which can be represented as a bitvector where $1$ denotes that a given edge is taken, and $0$ denotes that it is not.
\citeauthor{she_neuzz_2019} construct a neural surrogate that, for a given input, predicts an approximation of the execution trace of the program with each element between $0$ and $1$ (rather than strictly set to $0$ or $1$).
This allows for a smooth output of the surrogate, which then allows \citeauthor{she_neuzz_2019} to use gradient descent to find inputs that induce a specific execution trace on the original program.

\paragraph{Relaxing the input domain of programs.}
\citet{grathwohl_backpropagation_2018} use neural surrogates to approximate the gradient of non-differentiable functions, in order to reduce the variance of gradient estimators of random variables.
Though the input variables are discrete, \citeauthor{grathwohl_backpropagation_2018}'s surrogates take continuous values as input, allowing for optimizing these inputs with gradient descent.

\resetquestioncounter{}

\section{Design}
\label{sec:design}

Given a set of optimization problems that constitute a specification for the surrogate, a programmer must then determine how to design, train, and deploy the surrogate to meet the specification.
In this and the following sections we detail the design questions driving the neural surrogate programming methodology.
We discuss possible answers to each of these design questions, showing the trade-offs that programmers must navigate when developing neural surrogates.

This section describes the neural network architecture design approaches for neural surrogates used in the literature.

\questionone{}

\paragraph{Domain-agnostic architectures.}

One design methodology is to use a domain-agnostic architecture for the surrogate, a neural network architecture designed independently of the behavior and domain of application of the program under consideration.
A common choice of domain-agnostic architectures for neural surrogates with fixed-size inputs are multilayer perceptrons (MLPs)~\citep{ipek_exploring_2006,she_neuzz_2019}.
In \Cref{sec:substitution,sec:adaptation} we use a BERT encoder~\citep{devlin_bert_2018}, a type of Transformer~\citep{vaswani_attention_2017}, which is a common architecture for sequence processing tasks.
While simple to design, such domain-agnostic architectures may have high training costs or low accuracy~\citep{urban_deep_2017,neyshabur_towards_2020}.

\paragraph{Domain-specific architectures.}
An alternative is to design the architecture based on the program and domain under study~\citep{renda_difftune_2020,tseng_hyperparameter_2019}.
However designing such architectures requires manual effort and expertise, both in the original program and in its domain.
For instance, our surrogate optimization case study~\citep{renda_difftune_2020} uses a derivative of the architecture proposed by our prior work~\citep{mendis_ithemal_2019}, a model with high accuracy on basic block throughput prediction.
This architecture also exploits input sparsity in the simulation: rather than using the entire set of CPU parameters, we only input parameters that influence simulation of instructions~in~the~basic~block.

\questiontwo{}

Determining the capacity of the neural surrogate trades off between accuracy and execution cost, core tasks in any approximate programming task~\citep{stanley-marbell_exploiting_2020}.
Possible approaches include manually selecting the architecture based on reasoning about the complexity of the program~\citep{renda_difftune_2020} and automatically searching for the capacity that leads to the optimal trade-offs among the components of the surrogate's specification~\citep{esmaeilzadeh_neural_2012}.

\section{Training}
\label{sec:training}
With the neural network architecture in hand, the programmer must determine how to train the neural surrogate.

\questionthree{}

The training data of the surrogate defines the distribution of inputs on which the surrogate is expected to perform well.
The data must be representative of inputs for the downstream task for which the surrogate is deployed.
The data must also be plentiful and diverse enough to train the surrogate model to generalize the observed behavior of the program.

\paragraph{Instrumenting the program.}
One approach is to instrument the execution of the original program and record observed inputs~\citep{chen_bhive_2019,esmaeilzadeh_neural_2012}.
This approach is prevalent in \substitution{}.
An underlying challenge is that it may not be possible to guarantee that the
training workload is reflective of the workload of the downstream task, especially when the surrogate is deployed directly~to~end\nobreakdash-users.

\paragraph{Manually-defined random sampling.}
When data reflective of the downstream task is not available, or when the downstream data distribution is not known \textit{a priori}, another common approach is to randomly sample inputs from some hand-defined sampling distribution~\citep{renda_difftune_2020,tseng_hyperparameter_2019,tercan_transfer_2018}.

\setnowidow[4]
\setnoclub[4]

\paragraph{Neural surrogate and program symmetries.}
The training data must also reflect the symmetries enforced in the program and the surrogate.
For instance, when the original program is invariant to a specific change in the input but the neural surrogate architecture is not (e.g., a program that calculates the area of a shape is invariant to translation of that shape), the training data should include augmentations on the data that reflect those symmetries, to train the surrogate to be invariant to that symmetry~\citep{shorten_survey_2019}.

\setnowidow[2]
\setnoclub[2]

\pagebreak[4]

\questionfour{}

The \emph{loss function} is the objective in a neural network's optimization process which measures how bad a neural network's prediction is compared to the ground truth. %
The loss function should reflect the downstream specification for the surrogate (such that a reduction in the loss results in a better surrogate for the task) while also being a differentiable function that is possible to optimize with~gradient~descent.

\questionfive{}

With training data and loss function in hand, the programmer must then train the surrogate.
This results in a trade-off between accuracy and training cost. %
Because the training procedure may be run multiple times during hyperparameter search, the threshold or budget should be set appropriately to account for the full cost of design and training.

There are two primary approaches in the literature for determining an appropriate training time of the surrogate.
One approach is training for a fixed training time, typically determined via experiments on a validation set~\citep{renda_difftune_2020,esmaeilzadeh_neural_2012}.
Another approach is training until an acceptable accuracy is reached, whether via a plateau of the training loss~\citep{tseng_hyperparameter_2019} or via reaching a minimum acceptable accuracy~\citep{tercan_transfer_2018}.
Such variable-length training time approaches are discussed in more depth by~\citet[Chapter~7.8]{goodfellow_deep_2016}.

  Determining the training length for \adaptation{} is especially important due to the challenges imposed by \emph{catastrophic forgetting}~\citep{catastrophic_mccloskey_1989,ratcliff_connectionist_1990}, when a neural network's performance on a task it was trained on in the past degrades when it is trained on a new task.
  There are a number of approaches in the literature for addressing catastrophic forgetting~\citep{yosinski_transferable_2014,serra_overcoming_2018,kirkpatrick_overcoming_2017,chronopoulou_embarrassingly_2019}; in the case study in \Cref{sec:adaptation} we simply select the (relatively small) training time that results in the minimum validation error on a held-out~test~set.

\section{Deployment}
\label{sec:deployment}

Once the programmer has designed and trained the surrogate, the programmer must deploy the surrogate into its execution context.
Neural networks can execute on diverse hardware and runtimes, and require different representations of the input data than those of the original program.

\questionsix{}

The hardware that the surrogate is deployed on impacts the surrogate's execution time properties, efficiency, and available optimization opportunities.
When a surrogate is deployed using different hardware than the original program, developers must also consider the costs of data and control transfer between the original program and the surrogate.

\paragraph{GPUs.}
Modern large-scale deep neural networks can be executed on GPUs~\citep{ciresan_flexible_2011}, which achieve high throughput (the number of inputs that can be processed per unit time) and low energy consumption per example at the cost of high latency (the end-to-end time to process a single input) and high energy consumption per unit time~\citep{hanhirova_latency_2018,li_evaluating_2016,han_thesis_2017}.

\paragraph{CPUs.}
Other applications use a CPU to deploy the surrogate~\citep{ipek_exploring_2006}.
CPUs typically result in lower latency and energy consumption per unit time than GPUs, at the cost of higher energy consumption per example and reduced throughput~\citep{lee_debunking_2010,hazelwood_applied_2018,han_thesis_2017,li_evaluating_2016} (though recent work challenges some of these assumptions~\citep{daghaghi_accelerating_2021}).
CPUs are also more widely available than GPUs, including on edge devices~\citep{wu_machine_2019}.

\paragraph{Machine learning accelerators.}
\citet{esmaeilzadeh_neural_2012} design and deploy a custom neural processing unit (NPU) to accelerate neural surrogates with low latency and energy cost. %
Other machine learning accelerators offer different trade-offs, such as TPUs increasing throughput even further~\citep{jouppi_tensor_2017}, or the Efficient Inference Engine decreasing energy costs while approximating the surrogate~\citep{han_efficient_2016}.

\questionseven{}

Neural networks require specialized software runtime environments.
Choosing the runtime environment requires navigating concerns about both the implementation of the program that uses the surrogate and the deployment of the surrogate across varying devices.
Software execution environments include custom frameworks and runtimes which provide bespoke trade-offs for specific applications~\citep{esmaeilzadeh_neural_2012}.

The choice of software environment can also impact the availability and performance of the surrogate across hardware platforms.
Certain software runtimes are only available for certain devices (e.g., CPUs), some devices are supported by specific software runtimes (e.g., TPUs by TensorFlow), and some runtimes are specialized for resource-constrained devices (e.g., TensorFlow Lite for edge devices).

\paragraph{Normalization.}
\emph{Data normalization}, which involves pre- and post-processing the inputs and outputs to be suitable for neural networks~\citep{LeCun2012}, induces complexity into the program that deploys the surrogate, with normalization and denormalization requiring additional code when integrating the surrogate into the original program's execution context.
Data processing bugs in such code are difficult to diagnose and lead to reduced accuracy~\citep{sculley_machine_2014}.
\citet{esmaeilzadeh_neural_2012} address these issues by integrating the normalization and denormalization steps into the custom hardware (the NPU), eliminating the opportunity for~software~bugs.

\paragraph{Batching.}
\emph{Batching}, determining the number of inputs to process at a time, induces a trade-off between latency and throughput for the surrogate.
Parrot~\citep{esmaeilzadeh_neural_2012}, a \substitution{} approach that deploys the surrogate to end-users, focuses entirely on latency and uses a single data item in each batch, sacrificing throughput for decreased latency.
DiffTune~\citep{renda_difftune_2020}, a \relaxation{} approach, has no explicit latency requirements and focuses entirely on throughput, increasing throughput by batching large numbers of training examples into single invocations of the surrogate.

\section{Future Work}
\label{sec:future}

While we have presented a programming methodology that details the questions and trade-offs that must be addressed when developing a neural surrogate, there are still several open problems related to the development and application of surrogates.
This section details open problems and future work not addressed in this paper.

\paragraph{Broadening to other surrogate models.}
Though the design patterns in \Cref{sec:techniques} are general to all types of surrogate models, the neural surrogate programming methodology in \Cref{sec:design,sec:training,sec:deployment} is specific to when using neural networks as surrogate models.
However, other surrogate models are popular in the literature, including surrogates based on Gaussian processes~\citep{rasmussen_gaussian_2005,cherrypick}, linear models~\citep{GelmanHill:2007,ding_generalizable_2021}, and random forests~\citep{ho_random_1995,nardi_hypermapper_2019}.
Future work in this direction can extend the programming methodology presented in this paper to other classes of surrogate models beyond just neural networks.

\paragraph{More mechanization and systematization.}
We have presented a programming methodology for developing neural surrogates.
However, our methodology is not mechanized: programmers still must manually navigate the trade-off space between desiderata.
Future work in this domain should mechanize the various aspects of surrogate construction, from automating the surrogate's design based on the semantics of the original program, to automatically training the surrogate based on specifications and objectives over a data distribution, to automatically integrating the surrogate into the original program's execution context.
While prior work has addressed some of these concerns~\citep{esmaeilzadeh_neural_2012}, fully mechanizing this process is an important direction for future~work.

\paragraph{Defining the scope of applicability.}
We have shown that surrogates provide state-of-the-art solutions to large-scale programming problems.
However, we have not precisely characterized what problems these surrogate-based design patterns are not suitable for.
Future work in this domain can more precisely characterize what aspects of a given task admit or preclude surrogates as a candidate solution.

\paragraph{Generalization and robustness.}
Large-scale neural networks struggle to generalize outside of their training dataset \citep{barnard_extrapolation_1992,ilyas_adversarial_2019,xu_neural_2021}.
Generalization consists of interpolation and extrapolation; while neural networks interpolate well, they struggle to extrapolate.
On the other hand, formal program reasoning techniques can prove properties about the behavior of programs on entire classes of inputs~\citep{platzer_logical_2010}.
To address situations where the neural surrogate is expected to extrapolate outside of its training data, neural surrogate programmers must develop new approaches to recognizing and addressing generalization issues.
This may be easier for surrogates of programs than for neural networks in general, because programmers still have access to the original program when developing a surrogate of that program.

\paragraph{Interpretability.}
Neural networks do not generate explanations for predictions~\citep{gilpin_explaining_2018}, leading to difficulties when reasoning about neural surrogates' predictions.
Future work can address these issues by better characterizing what interpretability means for different domains, developing interpretability tools for neural surrogates specifically (again aided by access to the original program), and characterizing when interpretability is and is not a relevant concern for neural surrogates.
For example, \relaxation{} uses surrogates as an intermediate artifact to aid another optimization process, where interpretability is less of a concern.

\section{Related Work Addressing Similar Tasks}

In this section we discuss related work that provides alternative solutions to the surrogate-based design patterns and the neural surrogate programming methodology.

\paragraph{\NA{Function approximation.}}
Surrogate construction is an instance of function approximation, which encompasses a broad set of techniques ranging from polynomial approximations like the Taylor series to machine learning approaches like Gaussian processes and neural networks~\citep{trefethen_approximation_2012,rasmussen_gaussian_2005}.
The conventional wisdom is that compared to other approaches, neural networks excel at \emph{feature~extraction}~\citep{huang_feature_2006}, converting function inputs (including discrete and structured inputs) into vectors which can then be processed by machine learning algorithms.
Neural networks also excel when given a large amount of training data~\citep{krizhevsky_imagenet_2012}.
Other function approximation approaches have different trade-offs relative to neural networks, and may be appropriate in circumstances with limited execution cost or data, or when requiring specific bounds on the behavior of the function approximation.

\paragraph{Program repair.}
Similar to \adaptation{}, program repair techniques alter the semantics of a program to meet a downstream objective~\citep{long_prophet_2016,weimer09icse,perkins_automatically_2009}.
These approaches typically make local changes to a program in response to a single identified bug.
In contrast, \adaptation{} can change the entire behavior of the program to achieve good performance on a large dataset of examples.

\paragraph{Probabilistic programming.}
Probabilistic programming is a broad set of techniques for defining probabilistic models, then fitting parameters for these probabilistic models automatically given observations of real-world data~\citep{cusumano-towner_gen_2019,goodman_church_2008}.
When fitting parameters of a probabilistic program, such techniques require the program to be explicitly specified as a probabilistic program.
The parameters are then optimized using inference techniques like Monte Carlo inference~\citep{Neal93probabilisticinference} and variational inference~\citep{blei_variational_2017}.
In contrast, when optimizing parameters with \relaxation{} the original program can be specified in any form, while the parameters are optimized with stochastic gradient descent.

\paragraph{Differentiable programming.}
Differentiable programming is a set of techniques that calculates the derivatives of programs with respect to their input parameters~\citep{baydin_automatic_2017}.
In contrast with estimating the program's gradient with \relaxation{}, differentiable programming calculates the exact derivative without requiring the design and training processes of developing neural surrogates.

While differentiable programming is an appropriate alternative to \relaxation{} in contexts with smooth and continuous original programs, \NA{it struggles} in cases where the original program is not smooth or is not continuous.
For instance, differentiating through control flow constructs like branches and loops results in a discontinuity.
Such control flow constructs can also induce a true derivative of 0 almost everywhere, which poses challenges for gradient-based optimization.
Differentiable programming also relies on implementing the program in a language amenable to differentiable programming such as Pytorch or TensorFlow~\citep{paszke_pytorch_2019,abadi_tensorflow_2016,bischof_adifor_1996}.

In contrast, \relaxation{} approximates the program regardless of the provenance of its original implementation.
This means that while some points in the original program may be non-smooth, discontinuous, or have derivative 0, those points may be better behaved in the surrogate model (which only approximates the original program)
allowing for optimizing the inputs with gradient descent despite challenges posed by the original program~\citep{renda_difftune_2020}.

\paragraph{Program smoothing.}
\citet{chaudhuri_smooth_2010} present a method to approximate numerical programs by executing the programs probabilistically.
This approach lets \citeauthor{chaudhuri_smooth_2010} apply gradient descent to optimize parameters of arbitrary numerical programs, similar to \relaxation{}.
However, the semantics presented by \citeauthor{chaudhuri_smooth_2010} only apply to a limited set of program constructs and do not easily extend to the set of program constructs exhibited by large-scale programs.
In contrast, \relaxation{} estimates the gradients of arbitrary programs regardless of the constructs used in the program's implementation.

\paragraph{Automating construction of surrogates.}
\citet{munk_deep_2019} present an approach for automatic construction of neural surrogates of stochastic simulators for \substitution{}.
\citeauthor{munk_deep_2019} propose an LSTM architecture that predicts the sequence of samples output by the original stochastic simulator.
This approach is applicable to all stochastic simulators, regardless of the number or order of samples output by the original simulator.
\citeauthor{munk_deep_2019} show that this surrogate executes faster than the original simulator.
Though this approach addresses some questions of our neural surrogate programming methodology (specifically, how to design a surrogate for a given program), it does not address questions about how to train and how to deploy~the~surrogate.

\section{Related Work Addressing Other Tasks}
\label{sec:unrelated}

This section details approaches which, while related in that they use machine learning and programs together, are not examples of surrogates of programs.
The intent is to clarify the scope of our study of surrogates of programs.

\paragraph{Surrogates of non-programs.}
Surrogates of black-box processes (beyond just programs) are used across a wide variety of domains from computer systems to physical sciences~\citep{sun_review_2019,giuseppe_machine_2019,mendis_ithemal_2019}.
For example, in our prior work~\citep{mendis_ithemal_2019} we train a surrogate of the execution behavior of Intel CPUs to predict the execution time of code.
This is not an example of a surrogate of a program because this is performed without precise knowledge of the execution behavior of the CPU. %
This paper focuses on constructing surrogates of programs for which we have an intensional representation of the semantics of the program (e.g., program source code) rather than developing surrogates of black-box~functions.

\paragraph{\NA{Residual} models.}
Another approach is training \emph{residual models} on top of programs, neural networks that add to rather than simply replacing the original program's behaviors~\citep{verma_imitation_2019,watson_applying_2019,toderici_full_2017}.
Formally, if the original program is a function $f(x)$ then the residual approach learns a neural network $g(x)$ and adds the result to that of the original program, such that the final program computes $f(x) + g(x)$.
For example, \citet{verma_imitation_2019} train neural networks that augment programmatic reinforcement learning policies~\citep{Sutton1998}.
While learning such residual models is a form of programming, the neural networks are not surrogates of programs, and are thus out of scope for this paper.

\paragraph{Programs synthesized to mimic neural networks.}
Several approaches in the literature train neural networks, taking advantage of their relative ease of training for high accuracy on downstream tasks, then synthesize a program that mimics the neural network~\citep{bastani_verifiable_2018,verma_programmatically_2018,verma_imitation_2019}.
For example, after training a residual model, \citet{verma_imitation_2019} synthesize a new program $f'$ that mimics the original program with its residual: $f'(x) \approx f(x) + g(x)$.
This class of approaches is also out of the scope of this paper due to the significant differences in programming methodologies when synthesizing a program that mimics a neural network and developing a surrogate that mimics a program.

\section{Conclusion}

Our work demonstrates the promise of using surrogates to develop complex systems, especially in contexts where programmers lack a full characterization of the system and its operating environment.
By identifying the surrogate-based design patterns and describing the methodology used to develop neural surrogates, our work provides a taxonomy for developing surrogates.
Our work builds a foundation on which the programming languages community can build new tools that aid in the development of surrogates~of~programs.

\begin{acks}
  We would like to thank Alana Marzoev, Ben Sherman, Cambridge Yang, Charith Mendis, Charles Yuan, Eric Atkinson, Jesse Michel, Saman Amarasinghe, Stella Lau, and the anonymous reviewers for their helpful comments and suggestions.
  This work was supported in part by the \grantsponsor{GS100000001}{National Science Foundation}{http://dx.doi.org/10.13039/100000001} (\grantnum{GS100000001}{CCF-1918839} and \grantnum{GS100000001}{CCF-1751011}), the \grantsponsor{GS100000002}{Defense Advanced Research Projects Agency}{http://dx.doi.org/10.13039/100000185} (\grantnum{GS100000002}{\#HR00112190046} and \grantnum{GS100000002}{\#HR0011-18-C-0059}), a \grantsponsor{GS100000003}{Google}{http://dx.doi.org/10.13039/100006785} \grantnum{GS100000003}{Faculty Research Award}, and a \grantsponsor{GS100000004}{Sloan}{https://doi.org/10.13039/100000879} \grantnum{GS100000004}{Research Fellowship}.
  Yi Ding’s work is supported by the \grantsponsor{GS100000001}{National Science Foundation}{http://dx.doi.org/10.13039/100000001} under Grant No.~\grantnum{GS100000001}{2030859} to the Computing Research Association for the CIFellows Project.
  Any opinions, findings, and conclusions or recommendations expressed in this material are those of the authors and do not necessarily reflect the views of the sponsors.

\end{acks}

\bibliography{references}

\appendix
\section{Methodology for \SUBSTITUTION{} Experiments}
\label{app:substitution-methodology}

This appendix provides more details on the performance evaluation experiments performed in \Cref{sec:substitution}.

All experiments were performed on a Google Cloud Platform \texttt{c2-standard-4} instance, using a single core of an Intel Xeon Skylake CPU at 3.1 GHz.

We compile \mca{} in release mode from version 8.0.1, using the version at \url{https://github.com/ithemal/DiffTune/tree/9992f69/llvm-mca-parametric}.
We invoke \mca{} a single time and pass it a random sample of \num{10000} basic blocks from BHive over stdin, with the following invocation:
\begin{verbatim}
  llvm-mca -parameters noop -march=x86-64 \
   -mtriple=x86_64-unknown-unknown -mcpu=haswell \
   --all-views=0 --summary-view -iterations=100
\end{verbatim}
The reported execution speed is the time from invocation to exit of the \verb|llvm-mca| command.

The neural surrogate is run on the same CPU, using a network compiled, optimized, and loaded with the ONNX runtime, version 1.7.0~\citep{onnxruntime}.
The surrogate implementation is the Hugging Face Transformers v4.6.1 BertForSequenceClassification with a hidden size of 64, 2 hidden layers, 2 attention heads, an intermediate size of 256, and dropout probability of 0.
The surrogate is compiled to ONNX using \url{https://github.com/huggingface/transformers/blob/acc3bd9/src/transformers/convert_graph_to_onnx.py}.
The surrogate is optimized using the ONNX transformer optimization script with default settings: \url{https://github.com/microsoft/onnxruntime/blob/4fd9fef9ee04c0844d679e81264779402cfa445c/onnxruntime/python/tools/transformers/optimizer.py}.

The surrogate is set to use a single thread by setting the OMP, MKL, and ONNX number of threads to 1, and is set to a single CPU affinity.
The surrogate uses a batch size of 1.
The surrogate is invoked repeatedly by a Python script, and is passed the same \num{10000} basic blocks to predict timing values for.
The reported execution speed is the time from the invocation of the Python script to its exit.

\begin{table}[H]
  \caption{
The validation error and speedup of BERT models over a range of candidate embedding widths.
The MAPE is the best MAPE observed on the validation set over the course of training. The speedup is the speedup relative to the default BERT-Tiny (W=128).
An embedding width of 64 results in the fastest BERT model that achieves less than $10\%$~validation~MAPE.
}
  \label{tab:hparam-search-results}

  \begin{center}
    \begin{tabular}{c|ccc}
      \toprule
      \textbf{Embedding Width} & \textbf{MAPE} & \textbf{Speedup over W=128} \\
      \midrule
      128 & $8.9\%$ & $1\times$ \\
      \textbf{64} & $\mathbf{9.5\%}$ & $\mathbf{1.57\times}$ \\
      32 & $10.1\%$ & $2.01\times$ \\
      16 & $10.8\%$ & $2.22\times$ \\
      \bottomrule
    \end{tabular}%
  \end{center}
\end{table}

\section{BERT Hyperparameter Selection and Training Telemetry}
\label{app:hyperparameters-training}

This appendix describes the hyperparameter selection process, the loss curves over the course of training, and the epochs with minimum validation loss for the BERT models used in \Cref{sec:substitution,sec:adaptation}.
In \Cref{app:hyperparameters-training-width} we describe the hyperparameters used for the model and show the hyperparameter search process used to find the hidden size of 64.
In \Cref{app:hyperparameters-training-telemetry}, we show the training, validation, and test loss curves of the models, along with the total amount of time taken to train all model and the epochs resulting in minimum validation loss.

Code to reproduce \Cref{sec:substitution,sec:adaptation,app:substitution-methodology,app:hyperparameters-training} is available at \url{https://github.com/psg-mit/programming-with-neural-surrogates-of-programs}.

\subsection{Hyperparameters}
\label{app:hyperparameters-training-width}

We base our BERT model on the BERT-Tiny model described by \citet{turc_well_2019}, which has an embedding width of 128, 2 layers, and 2 self-attention heads.
From this base architecture we search across alternative embedding widths that are a factor of two between 16 and 128.
The objective is to find the fastest-to-execute architecture that has a validation error of less than $10\%$ MAPE.

\Cref{tab:hparam-search-results} shows the results of the hyperparameter search, with the bolded row describing the selected model (with an embedding with of 64).
Embedding widths of both 128 and 64 achieve less than $10\%$ MAPE; because an embedding width of 64 achieves the fastest execution speed among this set, it is chosen as the final model.
Embedding widths of 32 and 16 provide increasing execution speedups, but do not satisfy the error criteria of a MAPE of less than $10\%$.

\subsection{Training Telemetry}
\label{app:hyperparameters-training-telemetry}

\begin{figure*}
  \begin{center}
    \textbf{\huge \SUBSTITUTION{} Telemetry}\\

    \begin{minipage}{0.48\textwidth}\includegraphics[width=\textwidth]{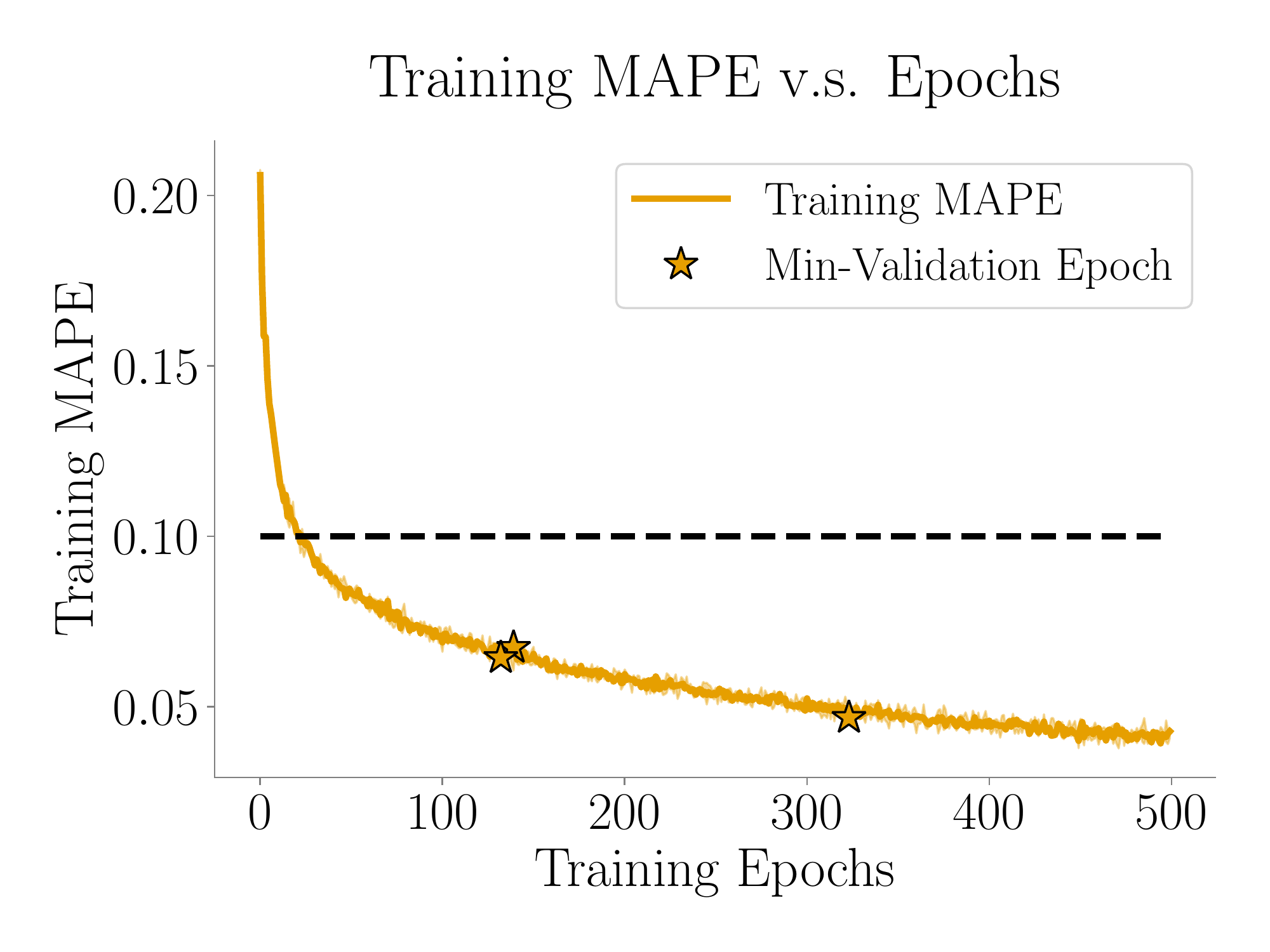}\end{minipage}
    \begin{minipage}{0.48\textwidth}\includegraphics[width=\textwidth]{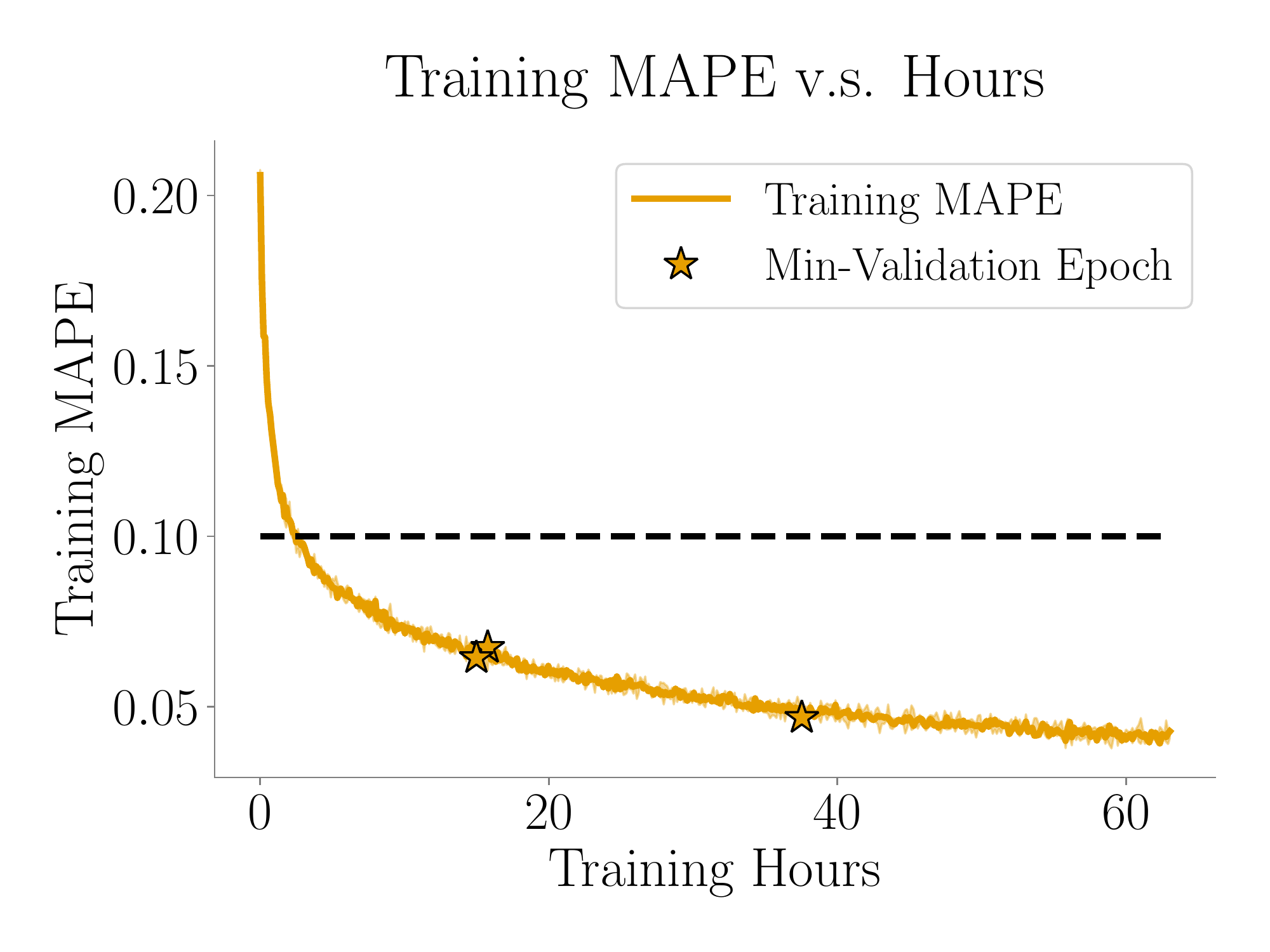}\end{minipage} \\
    \begin{minipage}{0.48\textwidth}\includegraphics[width=\textwidth]{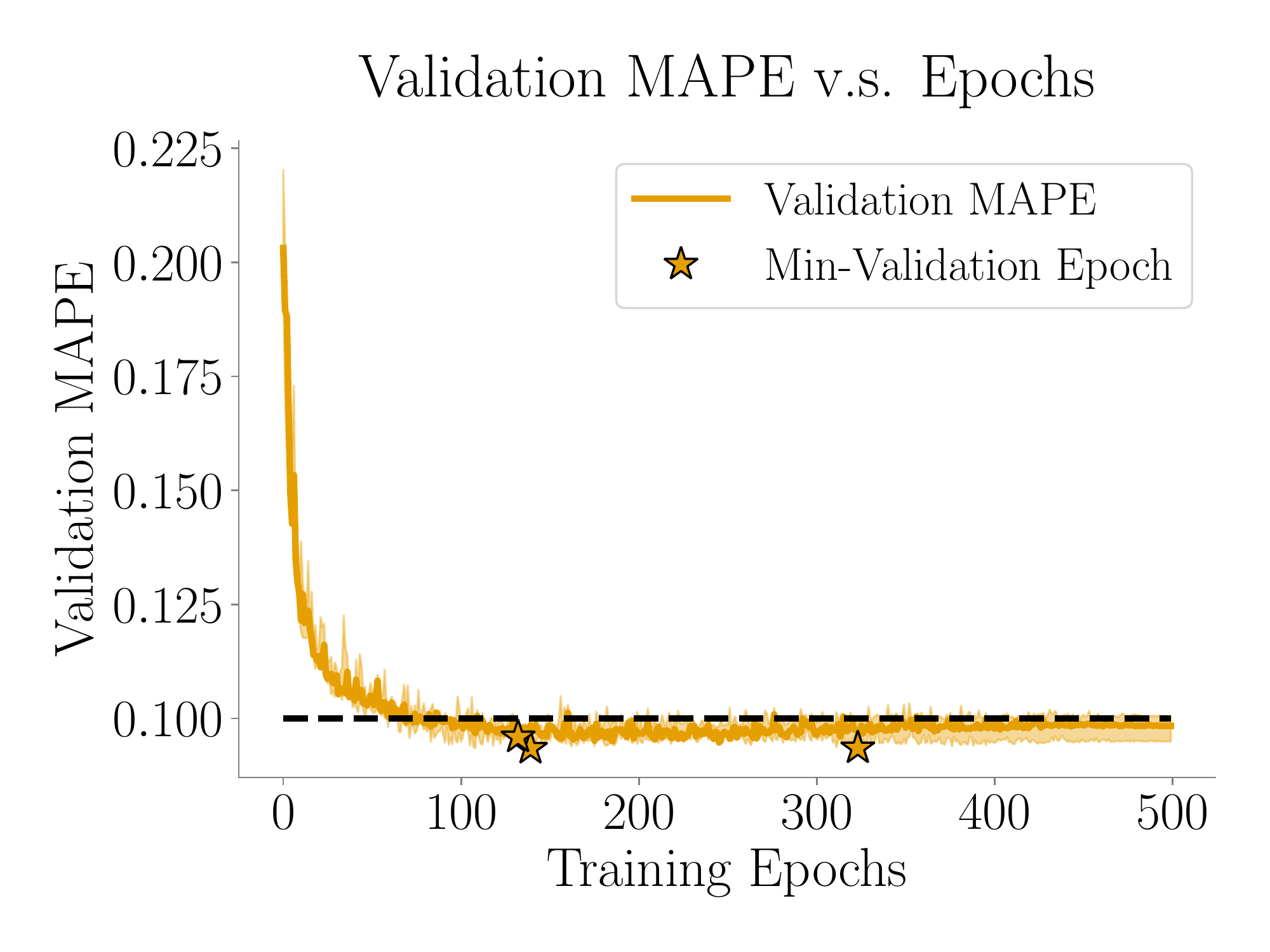}\end{minipage}
    \begin{minipage}{0.48\textwidth}\includegraphics[width=\textwidth]{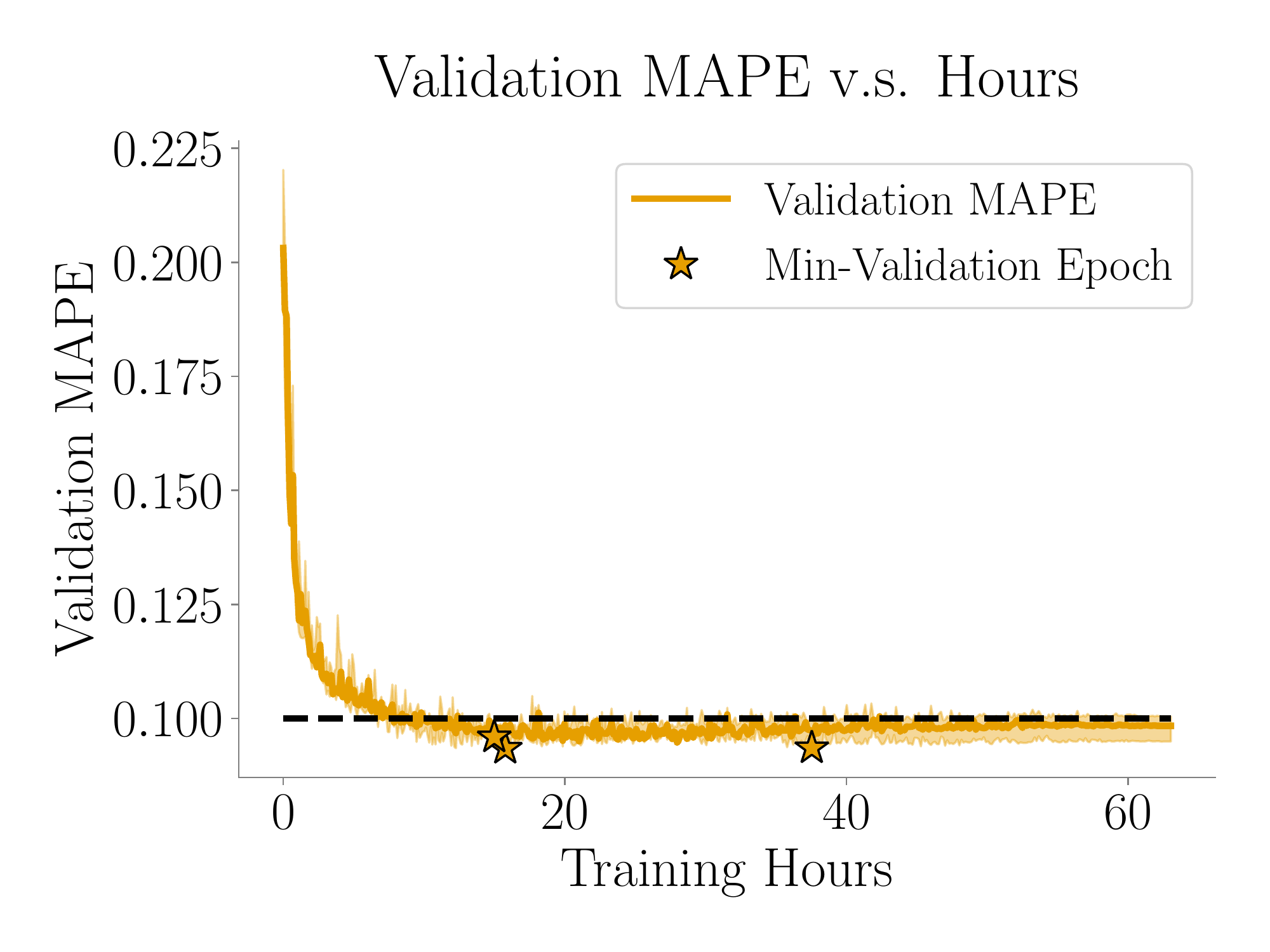}\end{minipage} \\
    \begin{minipage}{0.48\textwidth}\includegraphics[width=\textwidth]{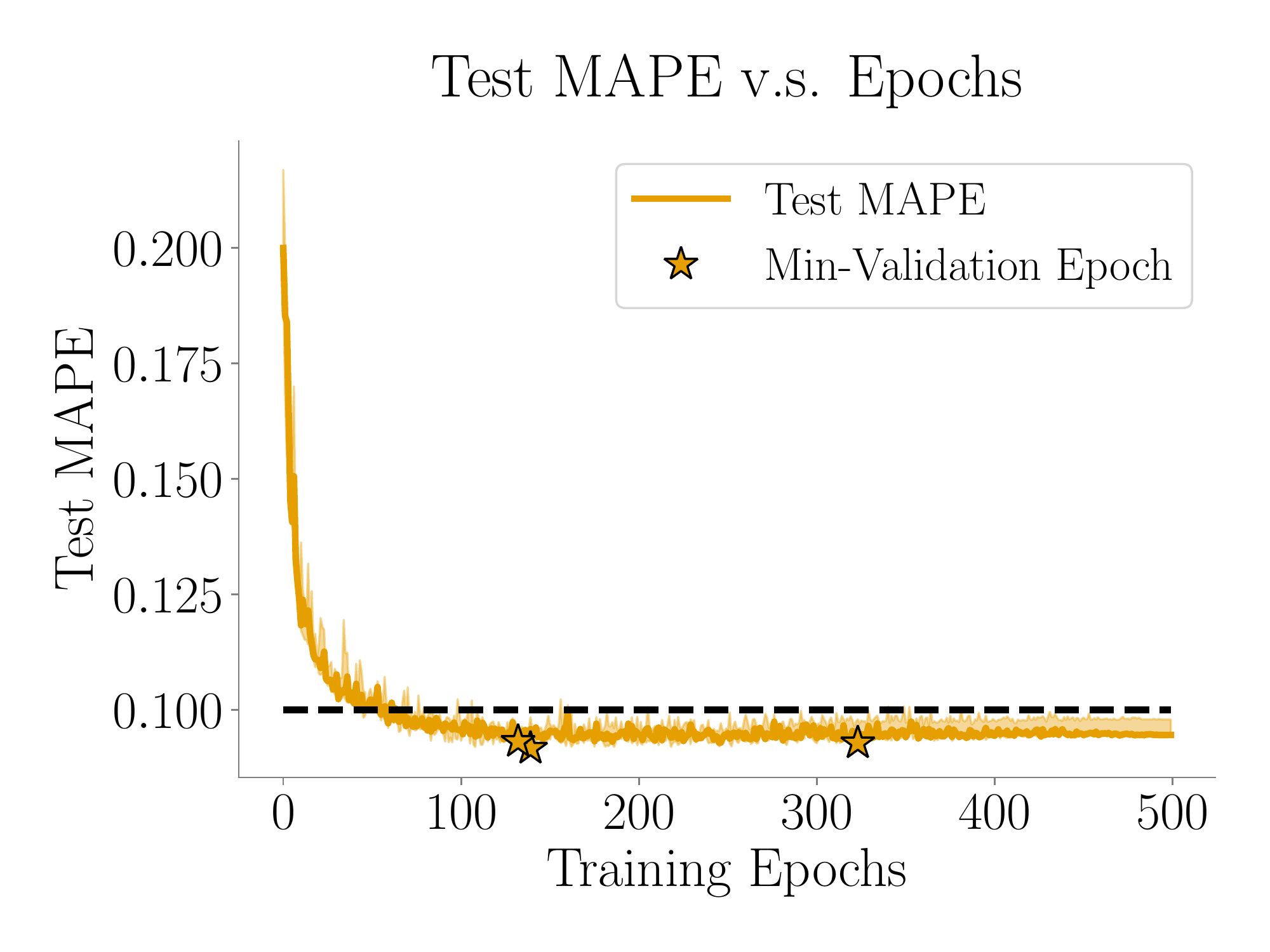}\end{minipage}
    \begin{minipage}{0.48\textwidth}\includegraphics[width=\textwidth]{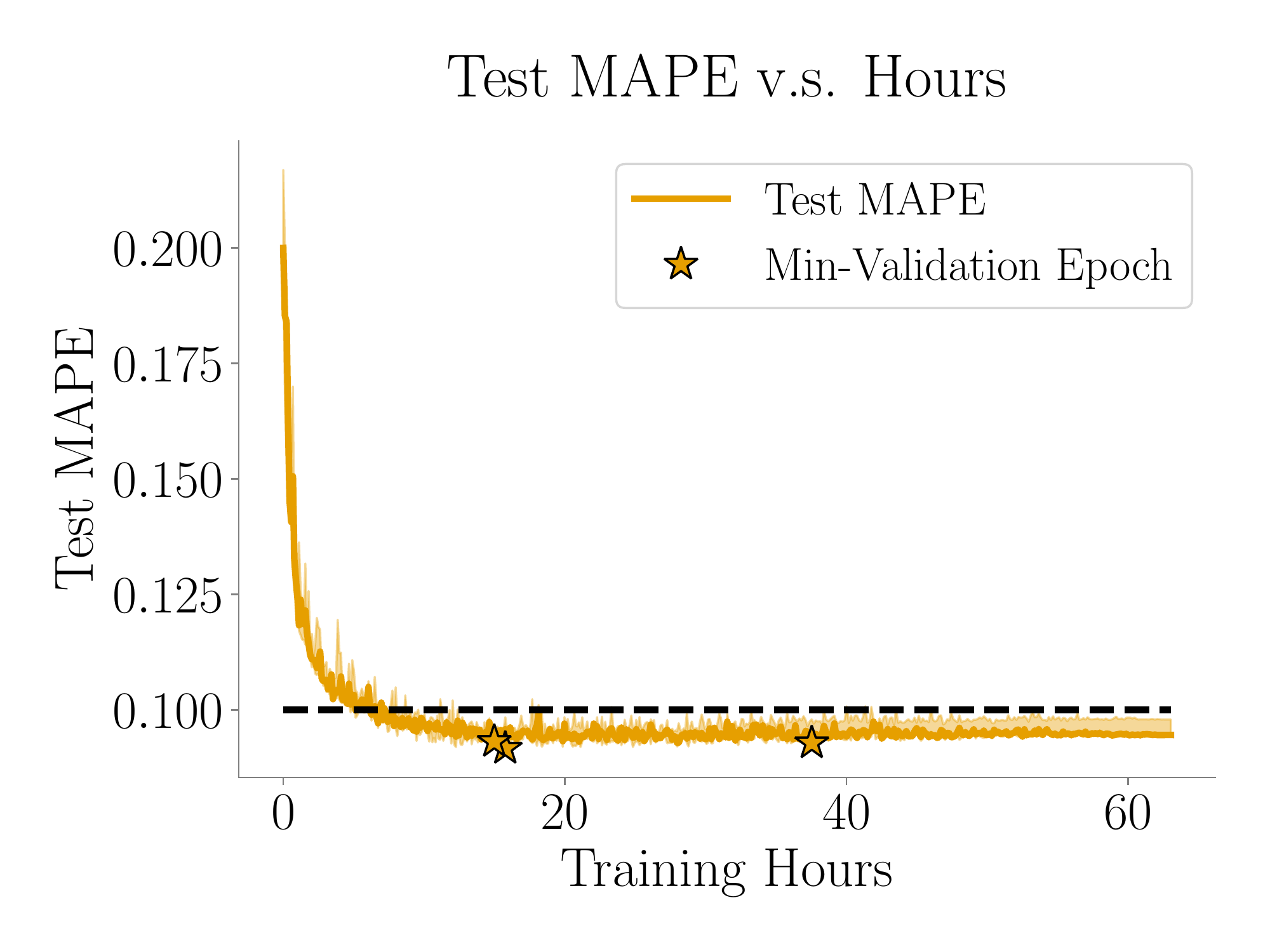}\end{minipage} \\
  \end{center}

  \caption{Training, validation, and test losses for \substitution{} over the course of training as a function of training iterations (left) and wall-clock time (right).}
  \label{fig:training-telemetry-mca}
\end{figure*}

  \Cref{fig:training-telemetry-mca} shows the training, validation, and test losses over the course of training when mimicking \mca{} (as in \Cref{sec:substitution} and the first optimization problem of \Cref{sec:adaptation}), as a function of training iteration (left column) and wall-clock time (right column).
  The plots show the median, minimum, and maximum error for each of three trials.
  The starred points on each plot represent the epoch with the minimum validation error for each trial.

  \Cref{fig:training-telemetry-fracs} shows errors over the course of training when using \adaptation{} to adapt a surrogate (as in the second optimization problem in \Cref{sec:adaptation}; in \dualcolor{}), and training a network trained from scratch (in \scratchcolor{}).
  The experiment is ran for each fraction of training data shown in \Cref{fig:mca-adaptation}.
  The left column shows error as a function of training iteration; the right column shows error as a function of wall-clock time.
  For \adaptation{}, the trials are seeded from the minimum-validation-error models from \Cref{fig:training-telemetry-mca}.
  The starred points on each plot represent the epoch with minimum validation error for each trial.

\begin{figure*}
  \begin{center}
  {\huge\bf {\ADAPTATION{} Telemetry}}\\
  {\LARGE $0.05\%$ Data} \\
    \begin{minipage}{0.48\textwidth}\includegraphics[width=\textwidth]{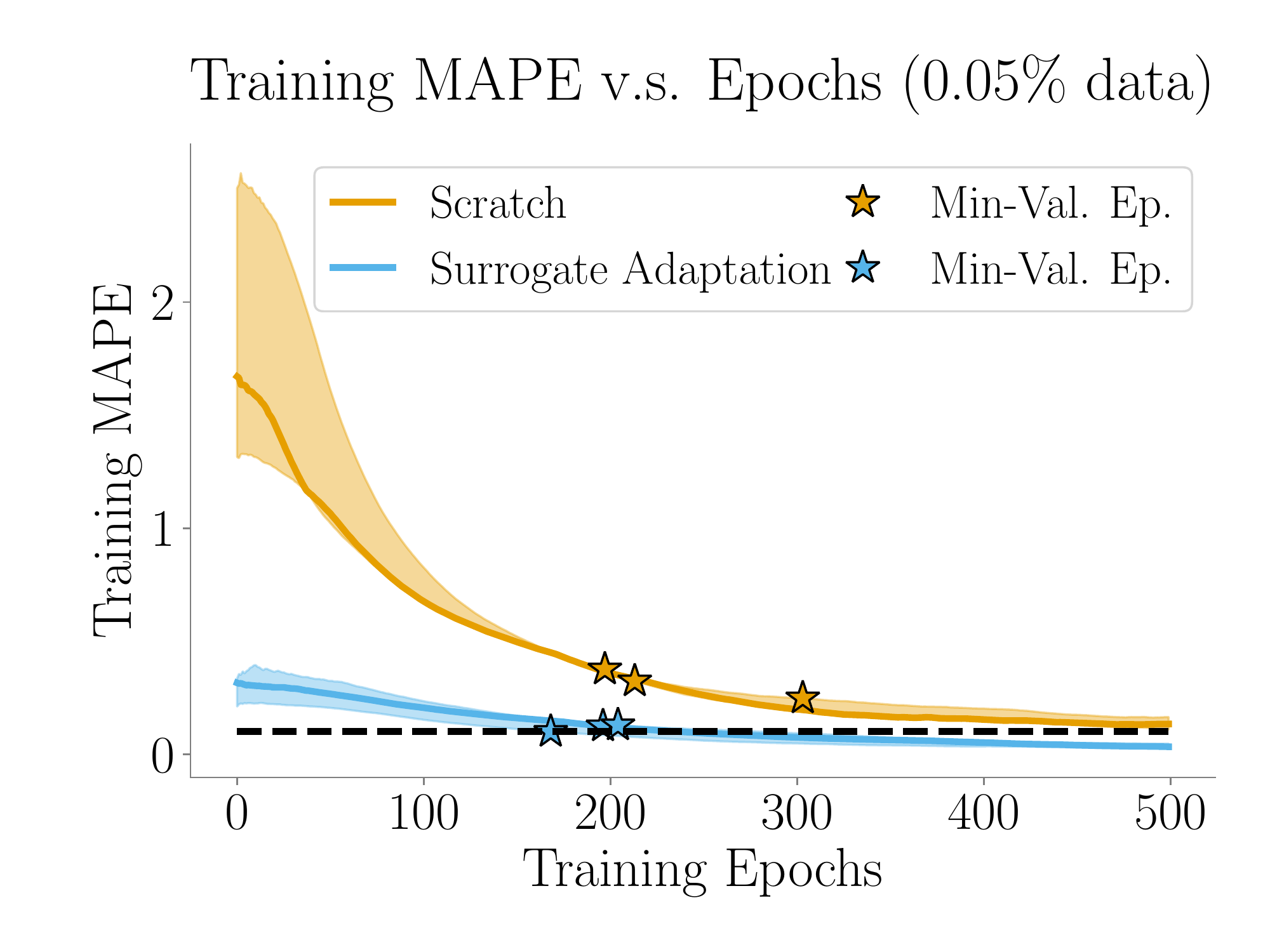}\end{minipage}
    \begin{minipage}{0.48\textwidth}\includegraphics[width=\textwidth]{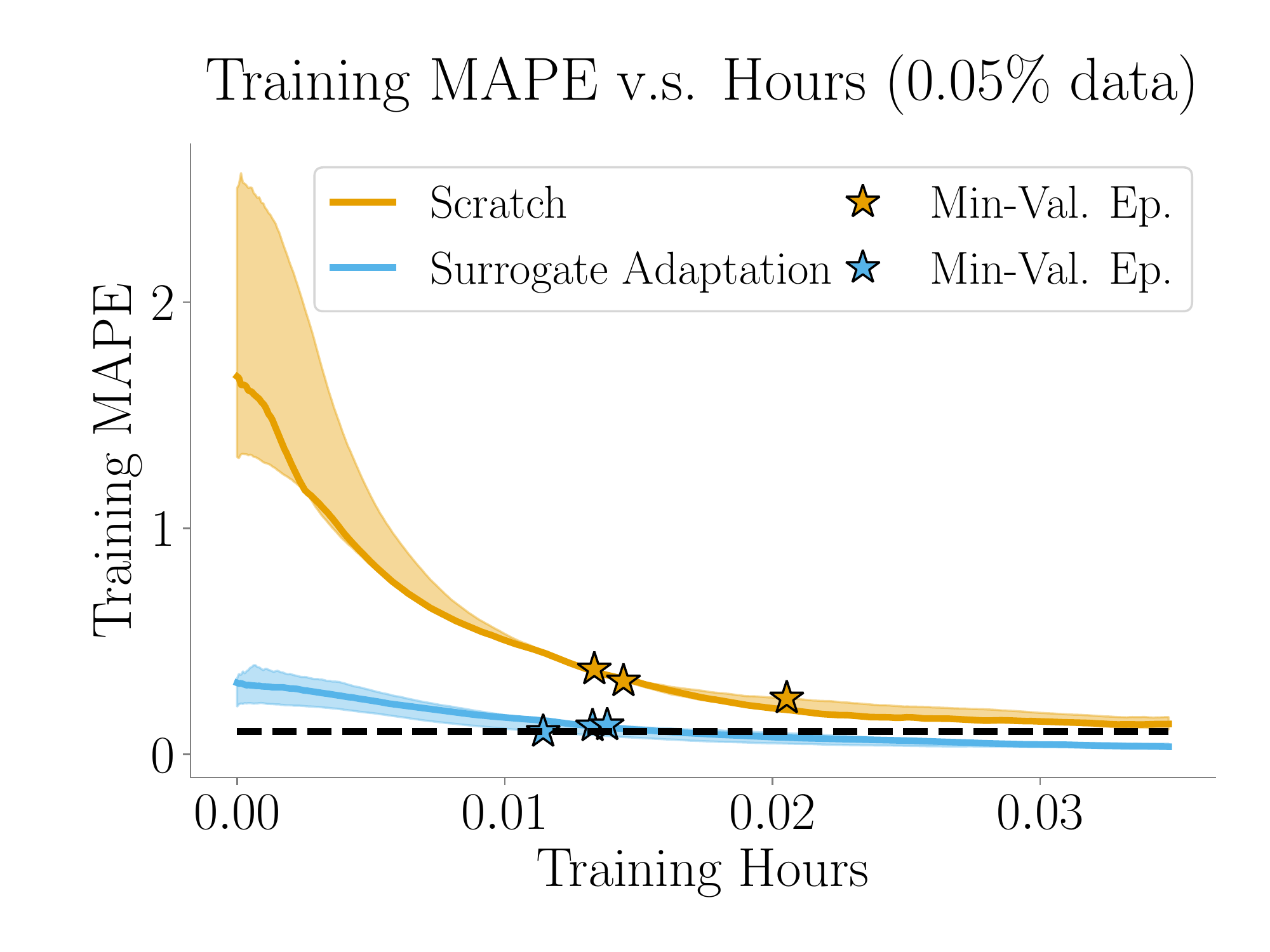}\end{minipage} \\
    \begin{minipage}{0.48\textwidth}\includegraphics[width=\textwidth]{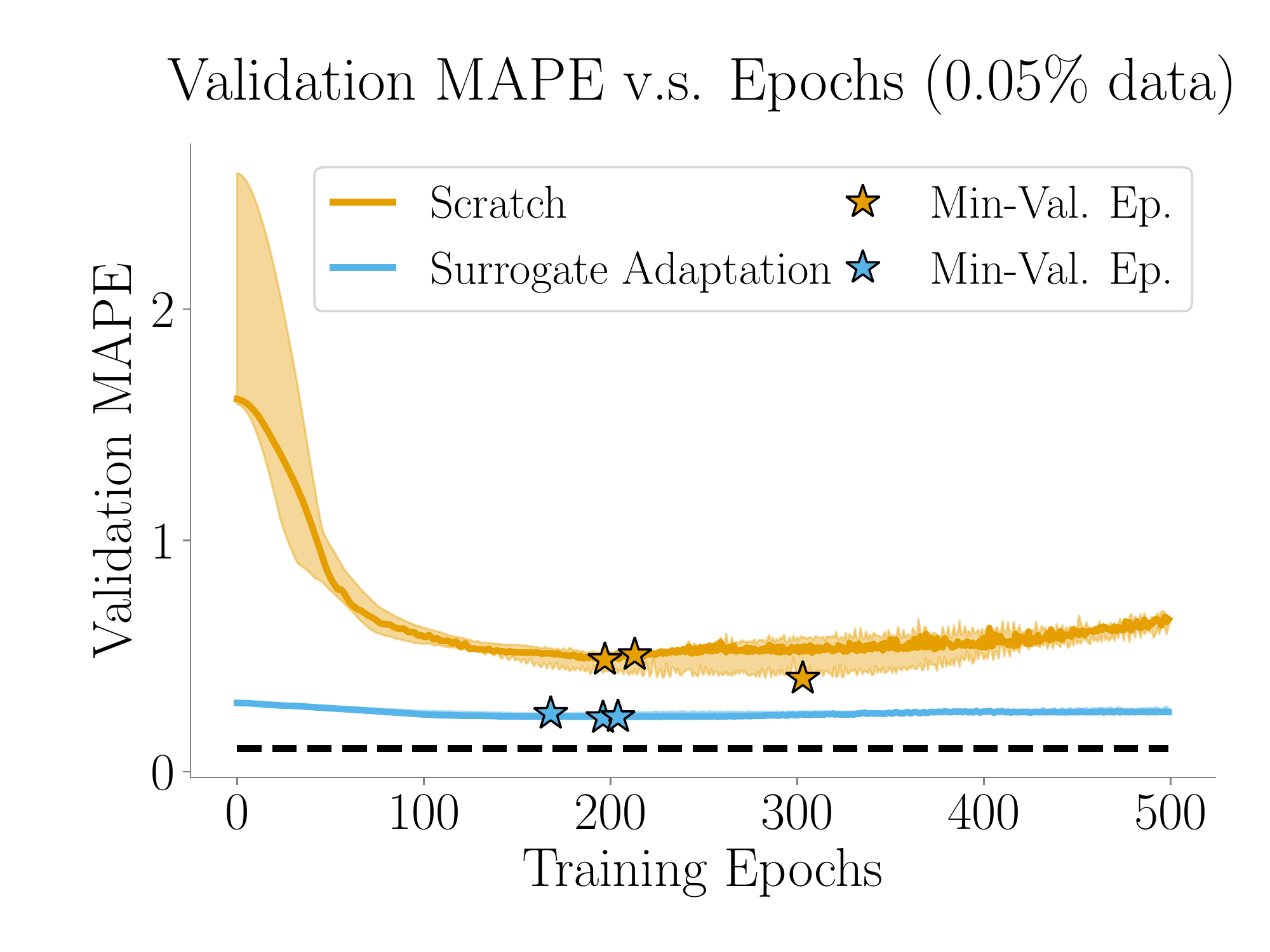}\end{minipage}
    \begin{minipage}{0.48\textwidth}\includegraphics[width=\textwidth]{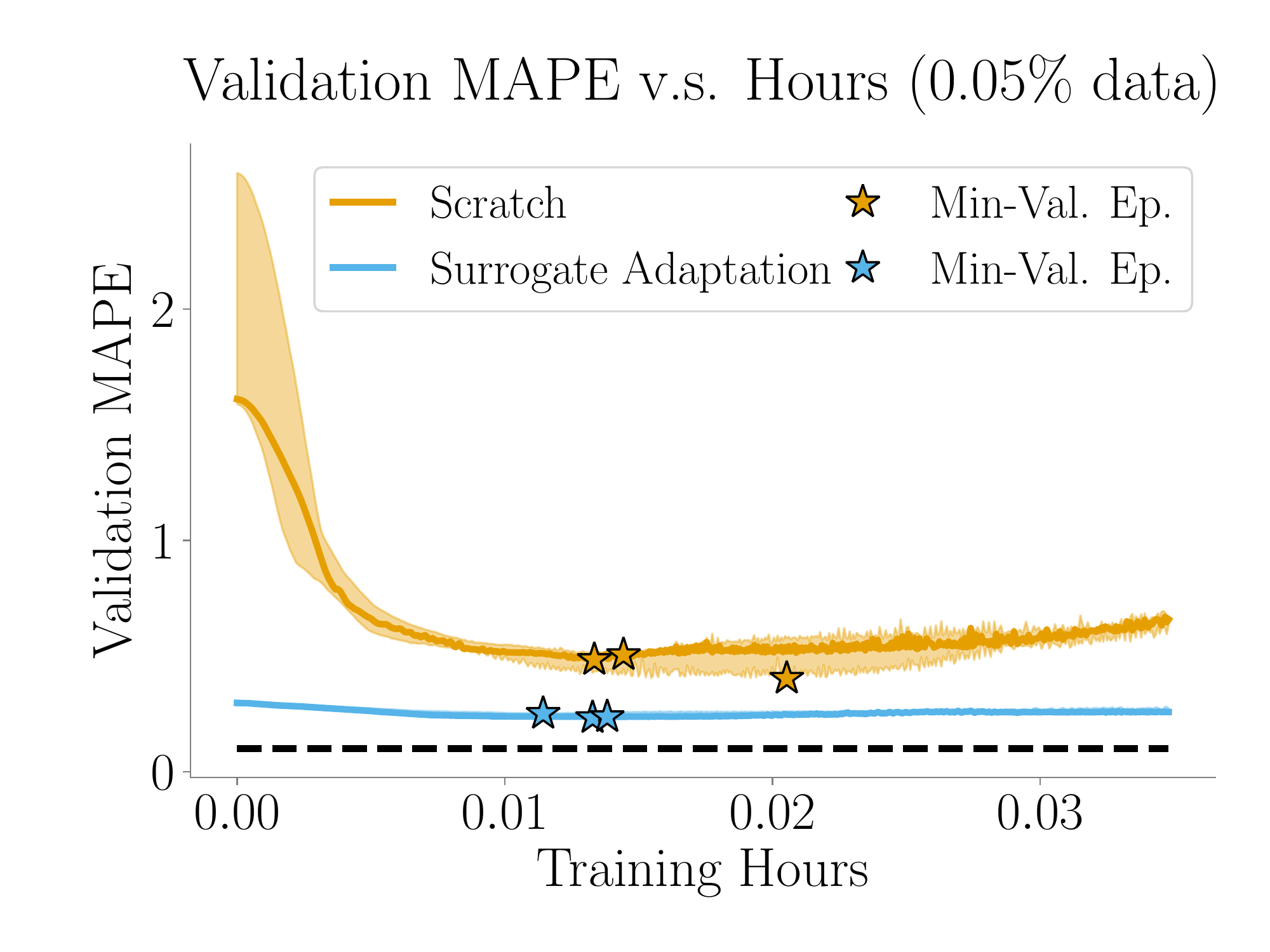}\end{minipage} \\
    \begin{minipage}{0.48\textwidth}\includegraphics[width=\textwidth]{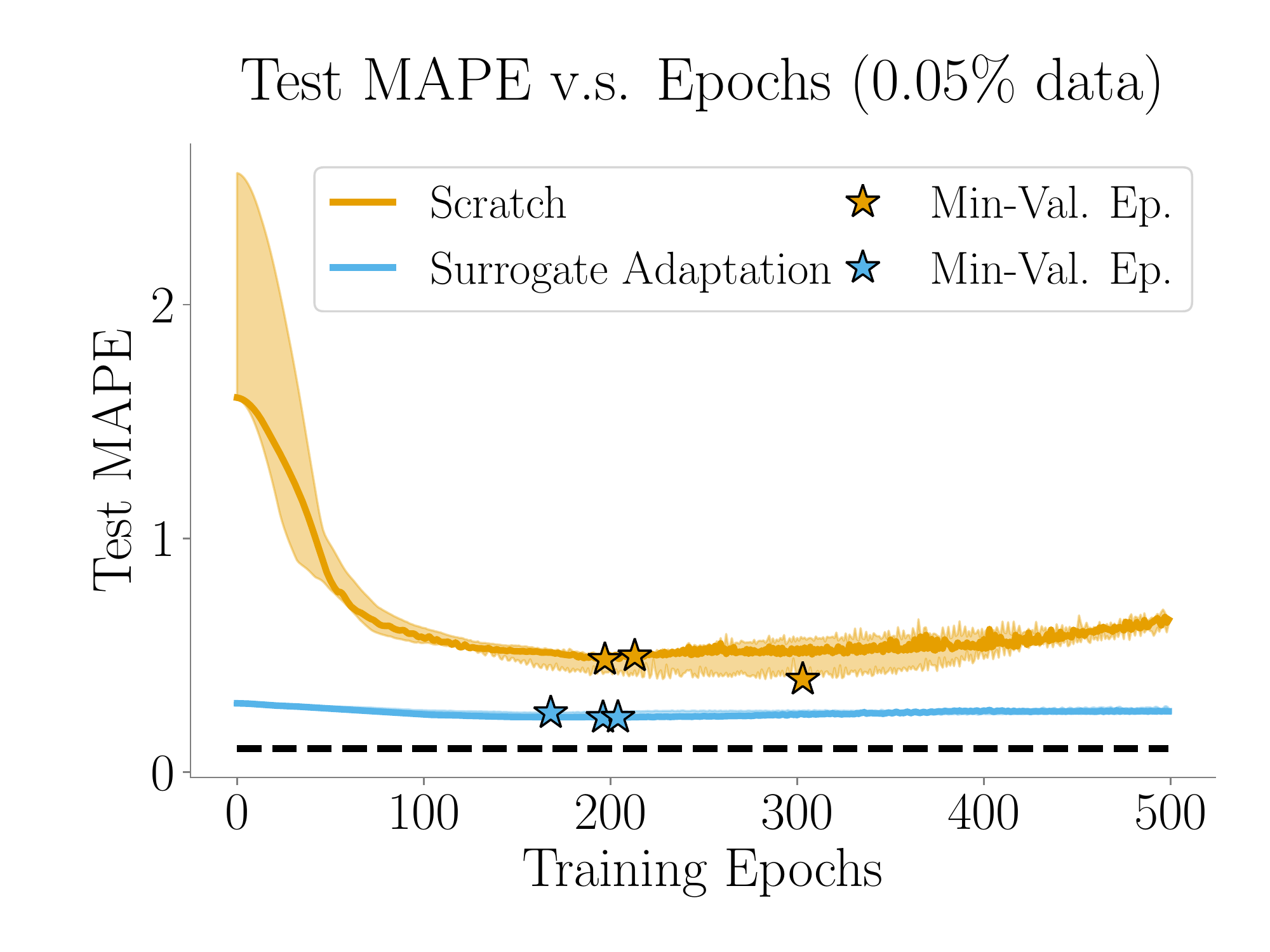}\end{minipage}
    \begin{minipage}{0.48\textwidth}\includegraphics[width=\textwidth]{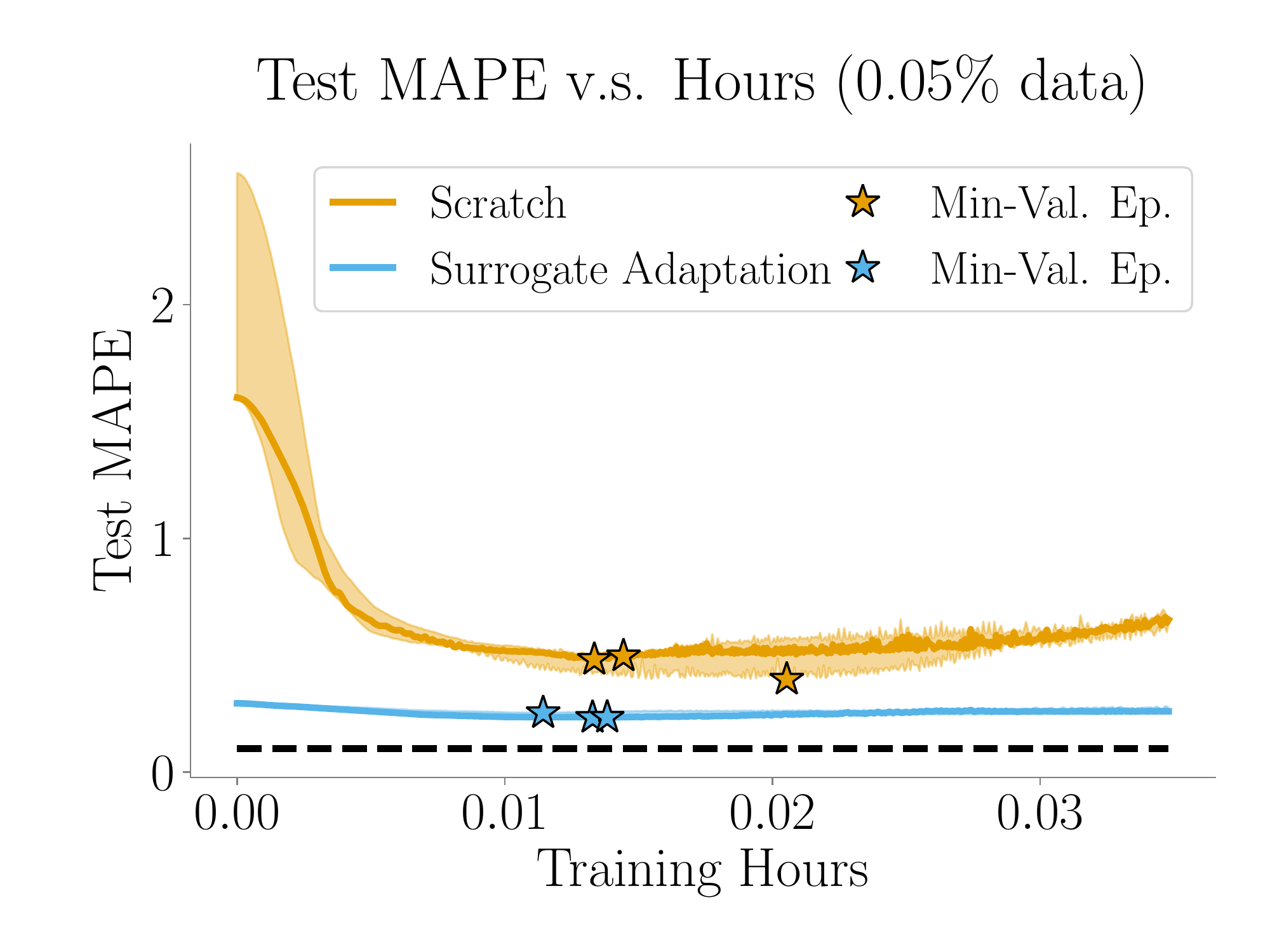}\end{minipage} \\
  \end{center}

  \caption{Training, validation, and test losses for \adaptation{} for each data fraction over the course of training as a function of training iterations (left) and wall-clock time (right).}
  \label{fig:training-telemetry-fracs}
\end{figure*}%
\begin{figure*}
  \ContinuedFloat
  \begin{center}
  {\huge\bf {\ADAPTATION{} Telemetry}}\\
  {\LARGE $0.11\%$ Data} \\
    \begin{minipage}{0.48\textwidth}\includegraphics[width=\textwidth]{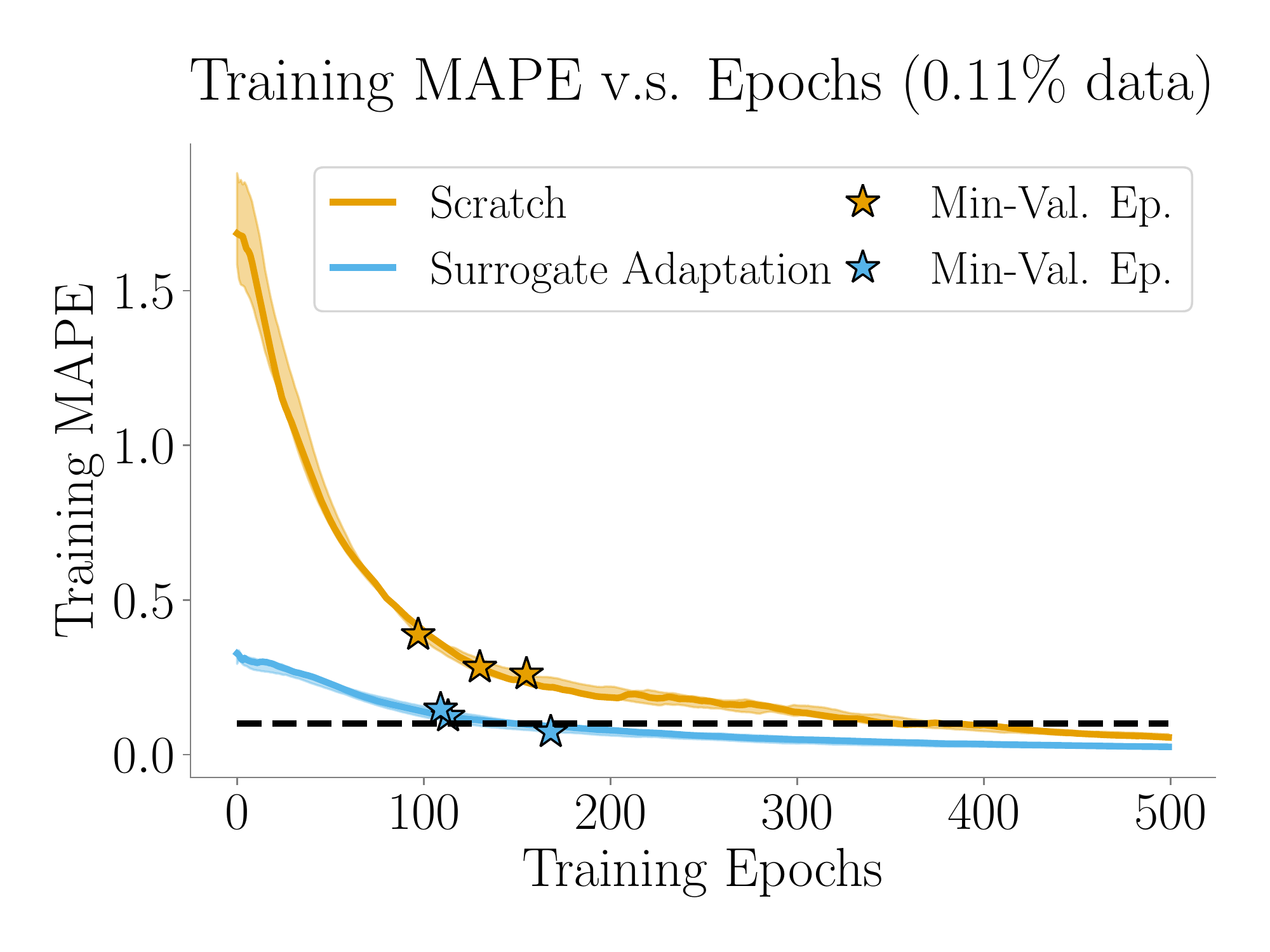}\end{minipage}
    \begin{minipage}{0.48\textwidth}\includegraphics[width=\textwidth]{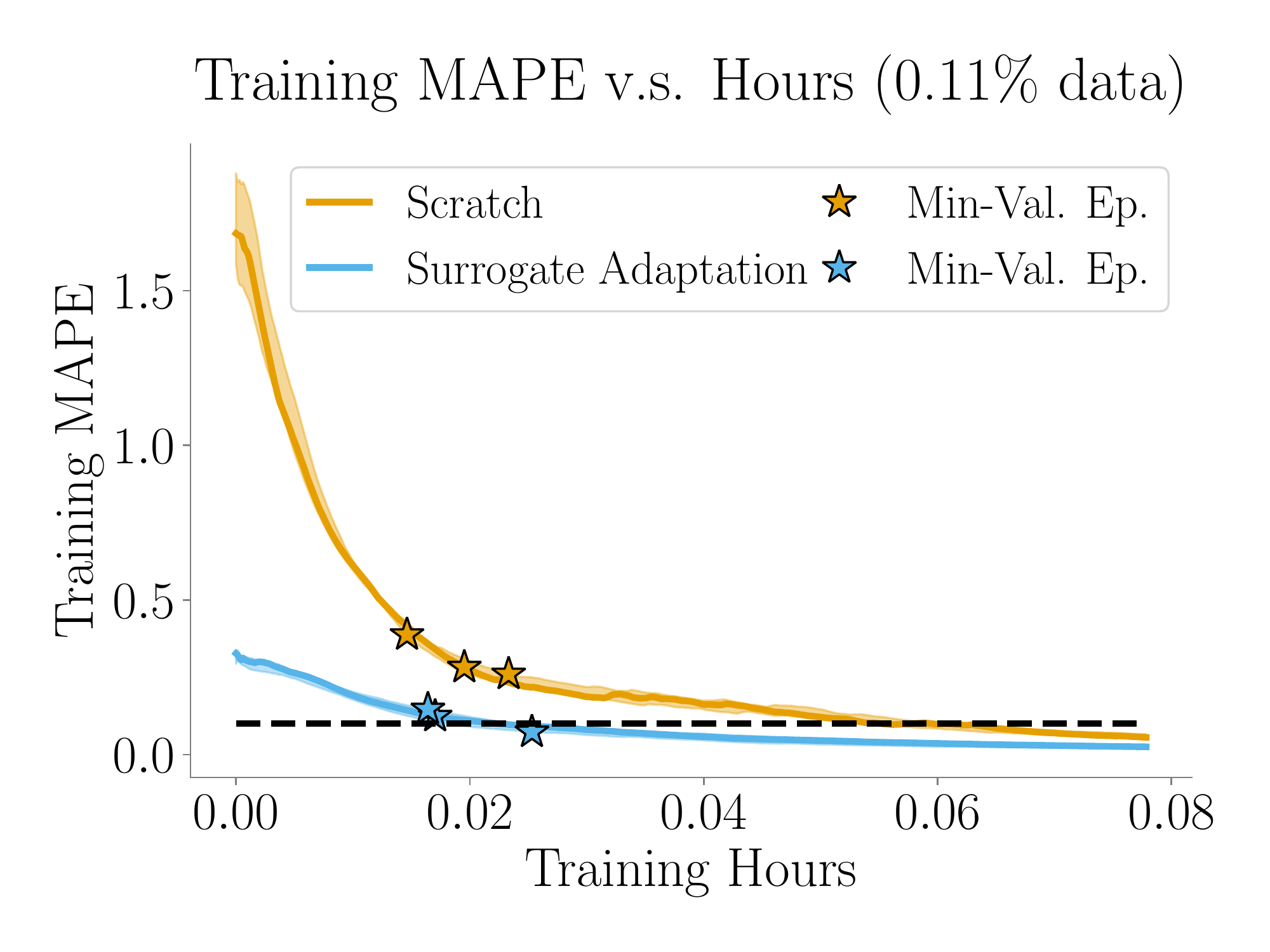}\end{minipage} \\
    \begin{minipage}{0.48\textwidth}\includegraphics[width=\textwidth]{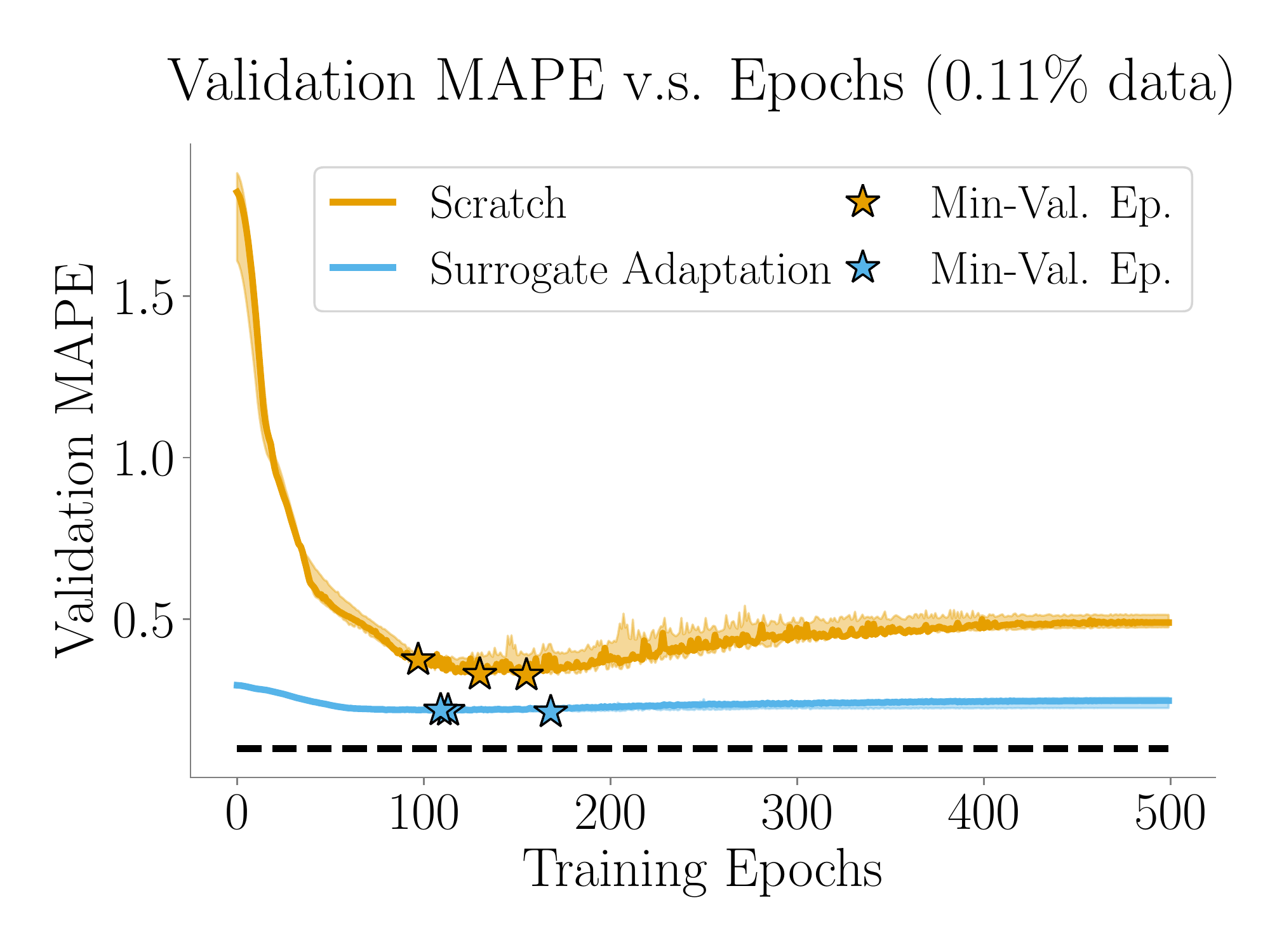}\end{minipage}
    \begin{minipage}{0.48\textwidth}\includegraphics[width=\textwidth]{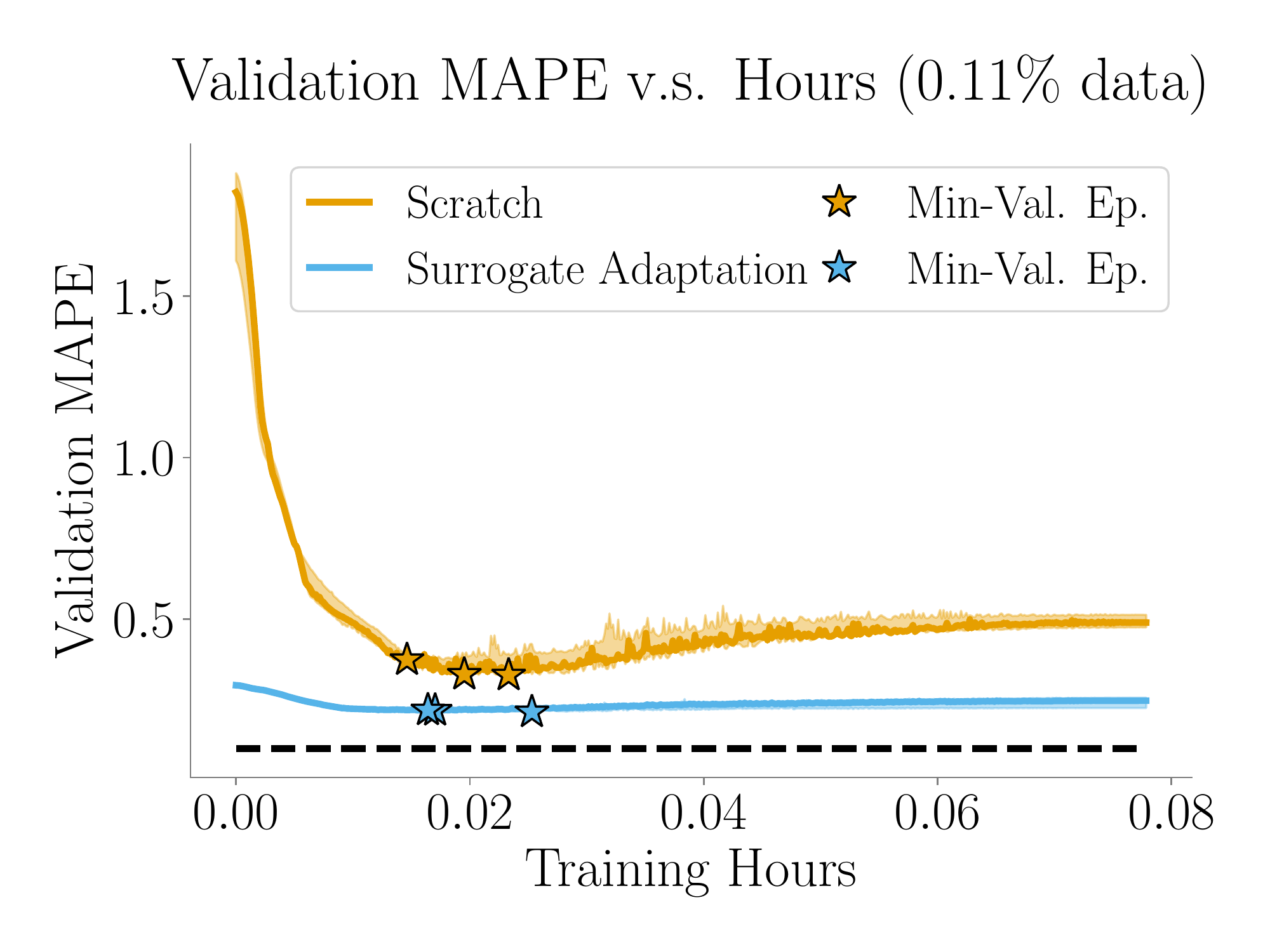}\end{minipage} \\
    \begin{minipage}{0.48\textwidth}\includegraphics[width=\textwidth]{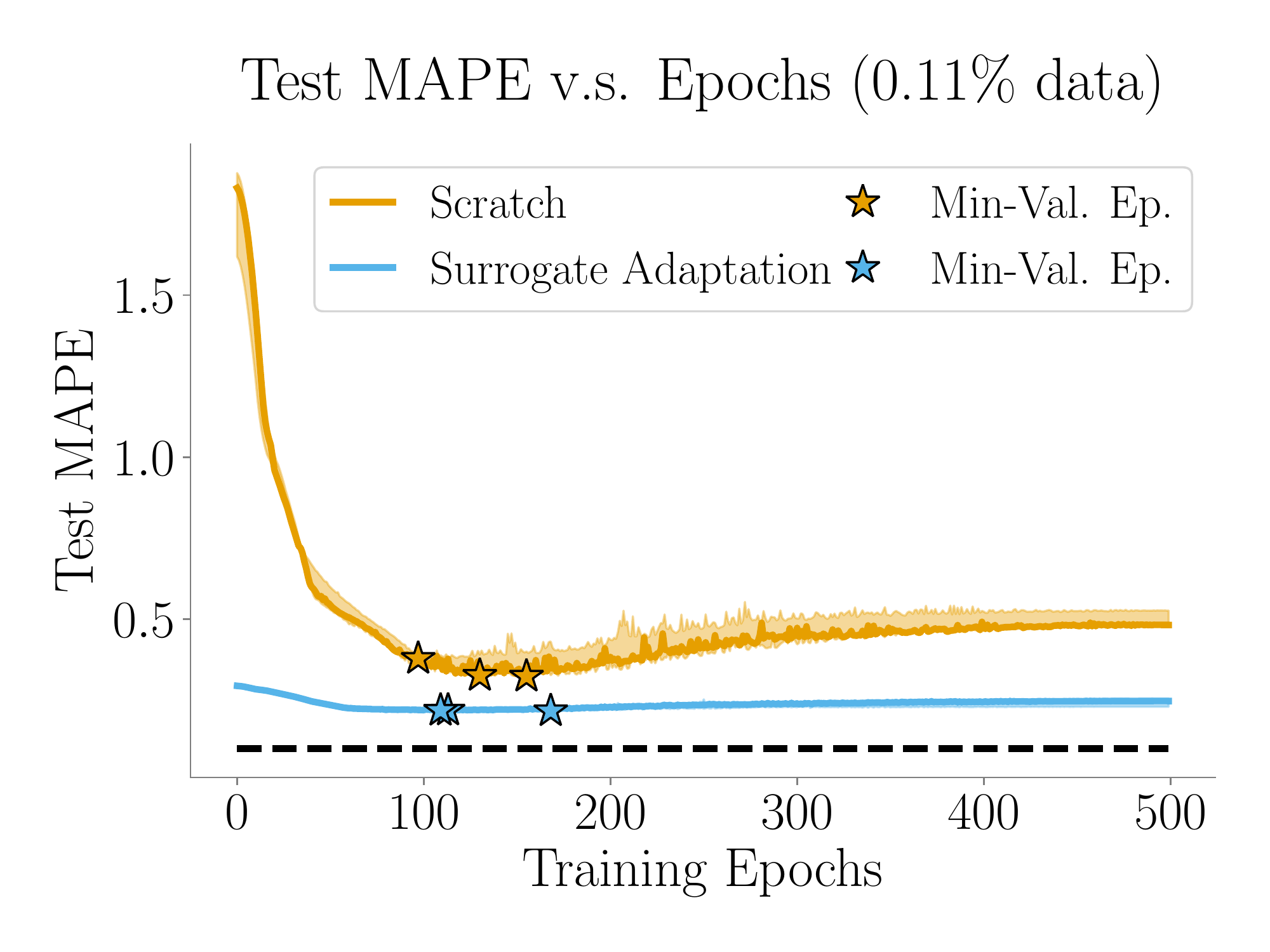}\end{minipage}
    \begin{minipage}{0.48\textwidth}\includegraphics[width=\textwidth]{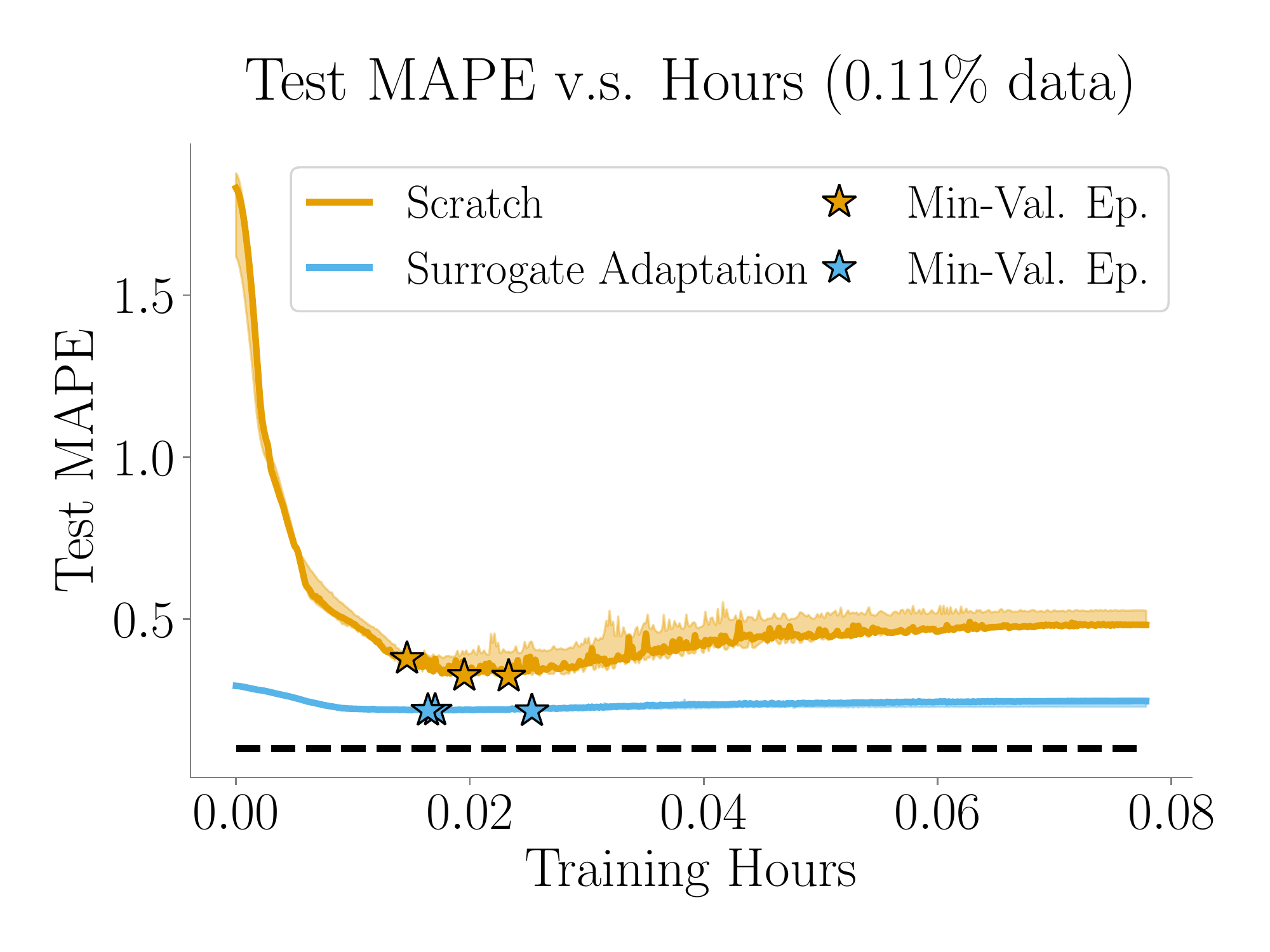}\end{minipage} \\
  \end{center}
\end{figure*}%
\begin{figure*}
  \ContinuedFloat
  \begin{center}
  {\huge\bf {\ADAPTATION{} Telemetry}}\\
  {\LARGE $0.26\%$ Data} \\
    \begin{minipage}{0.48\textwidth}\includegraphics[width=\textwidth]{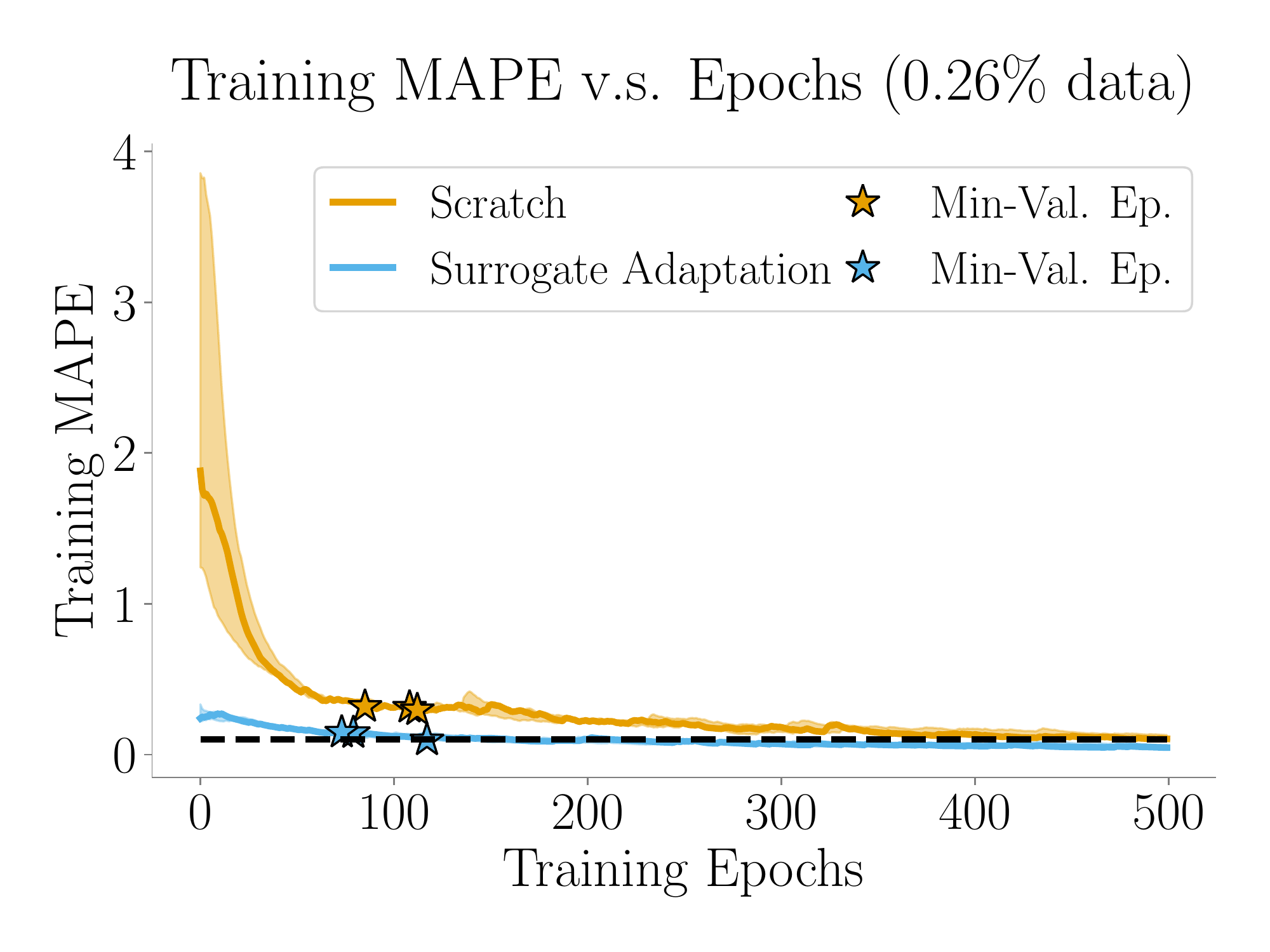}\end{minipage}
    \begin{minipage}{0.48\textwidth}\includegraphics[width=\textwidth]{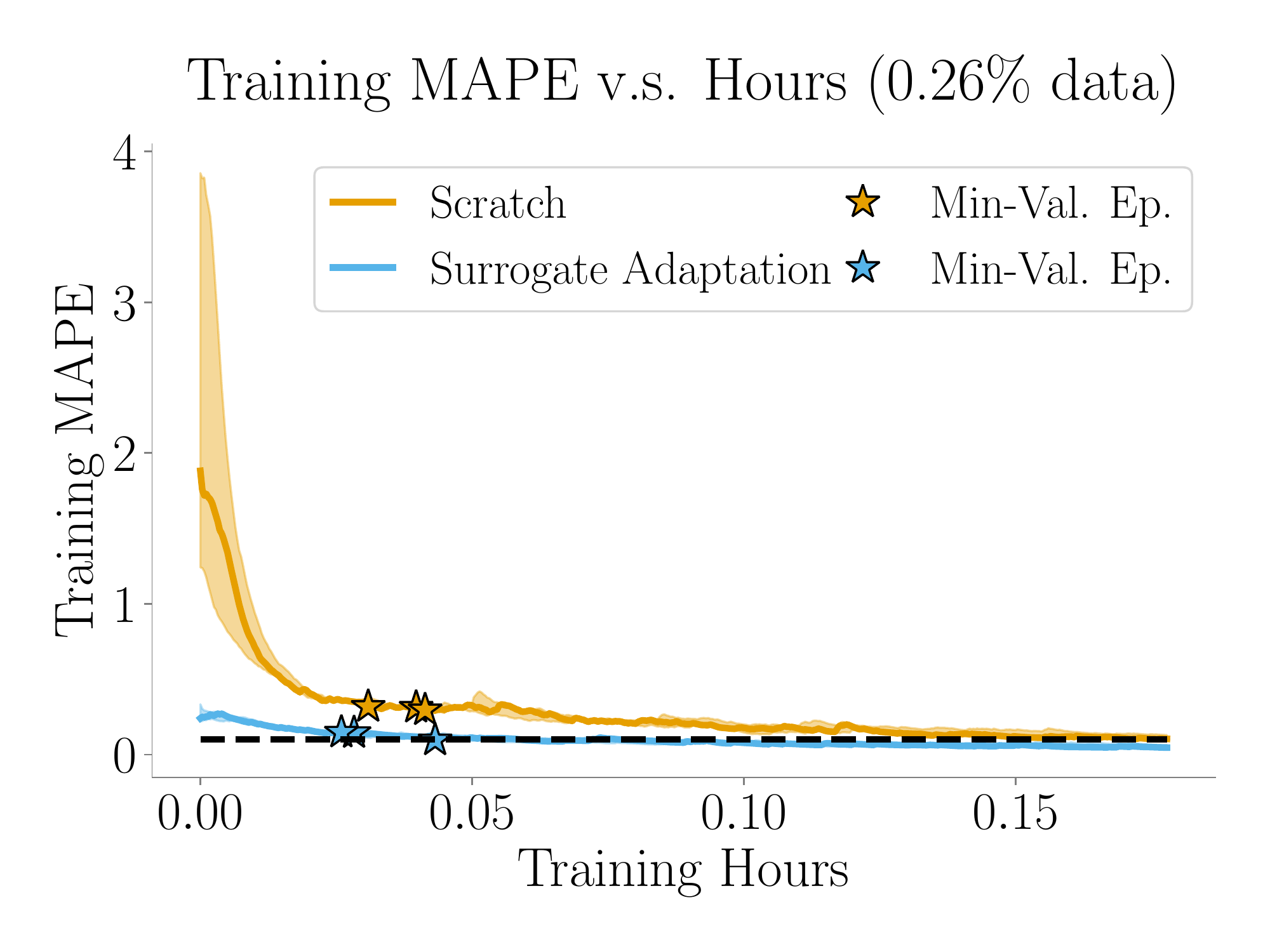}\end{minipage} \\
    \begin{minipage}{0.48\textwidth}\includegraphics[width=\textwidth]{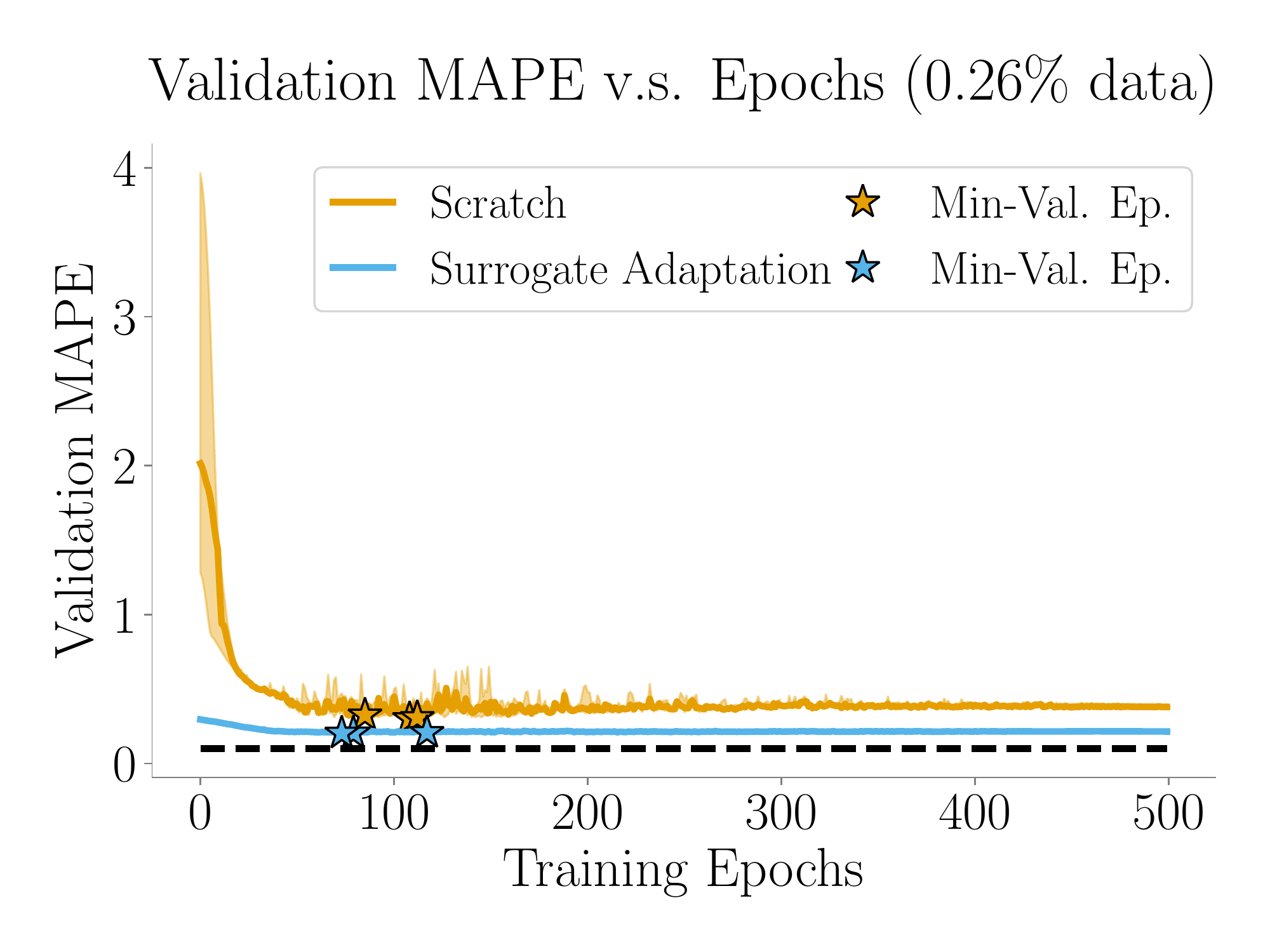}\end{minipage}
    \begin{minipage}{0.48\textwidth}\includegraphics[width=\textwidth]{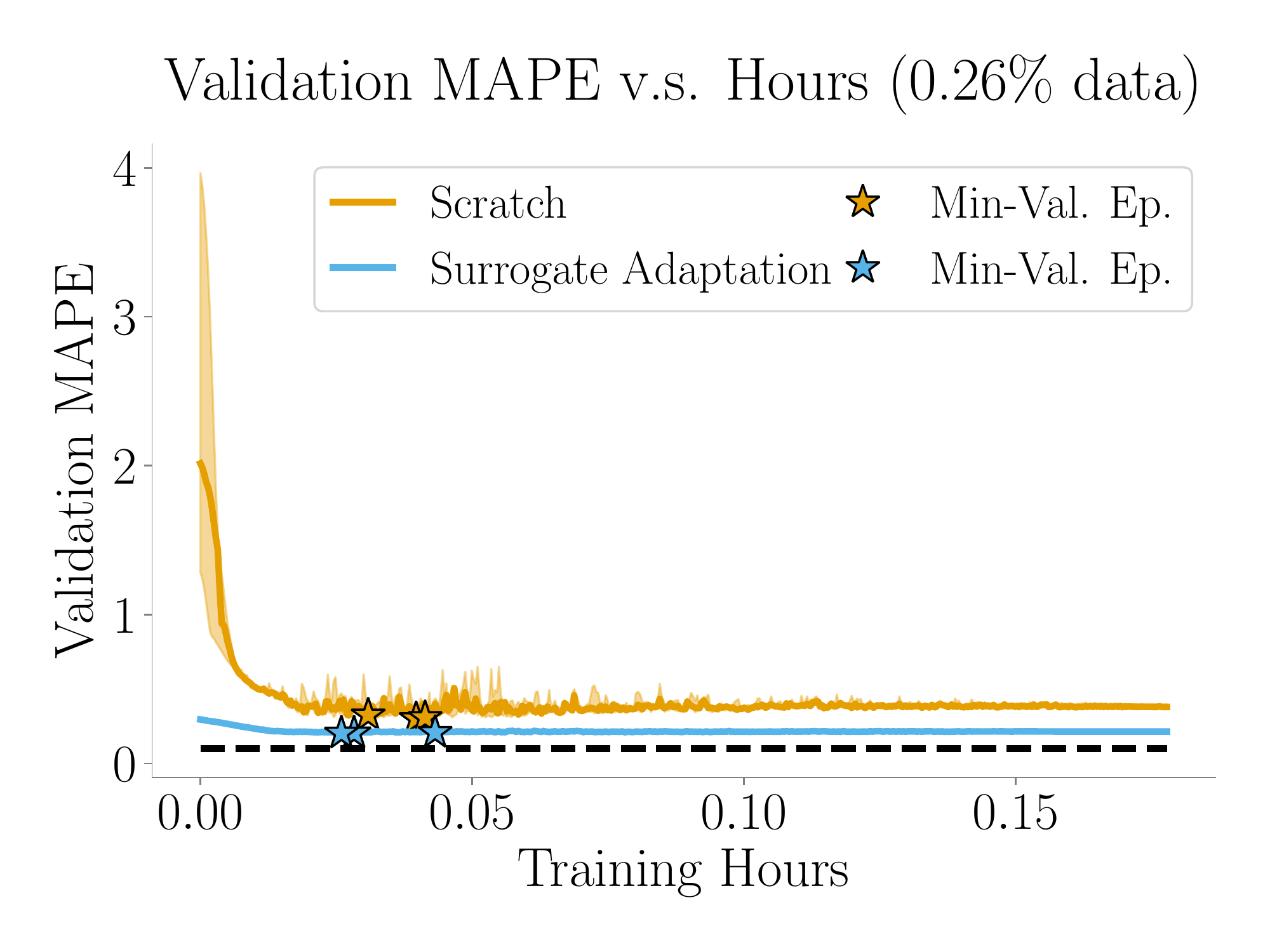}\end{minipage} \\
    \begin{minipage}{0.48\textwidth}\includegraphics[width=\textwidth]{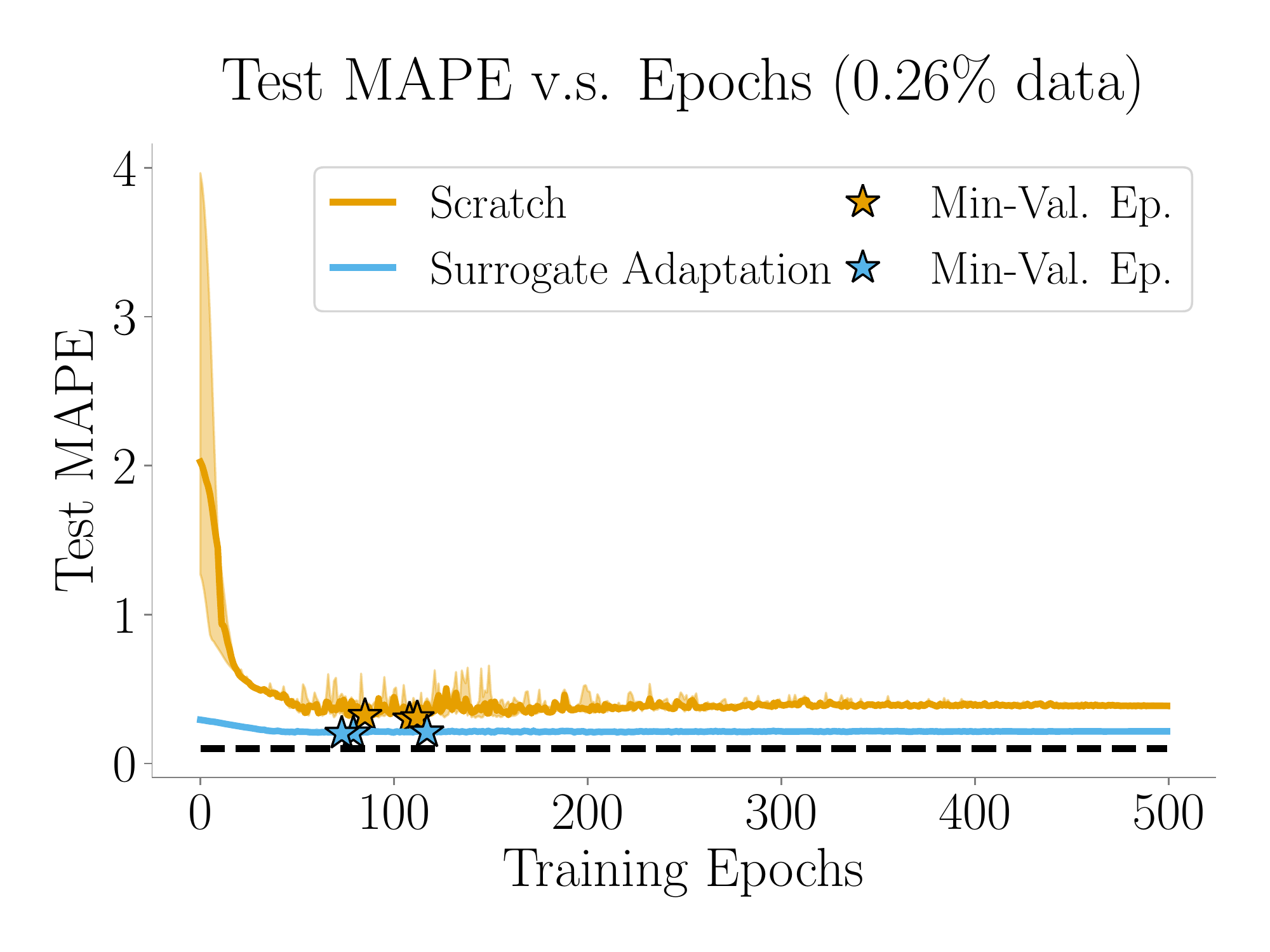}\end{minipage}
    \begin{minipage}{0.48\textwidth}\includegraphics[width=\textwidth]{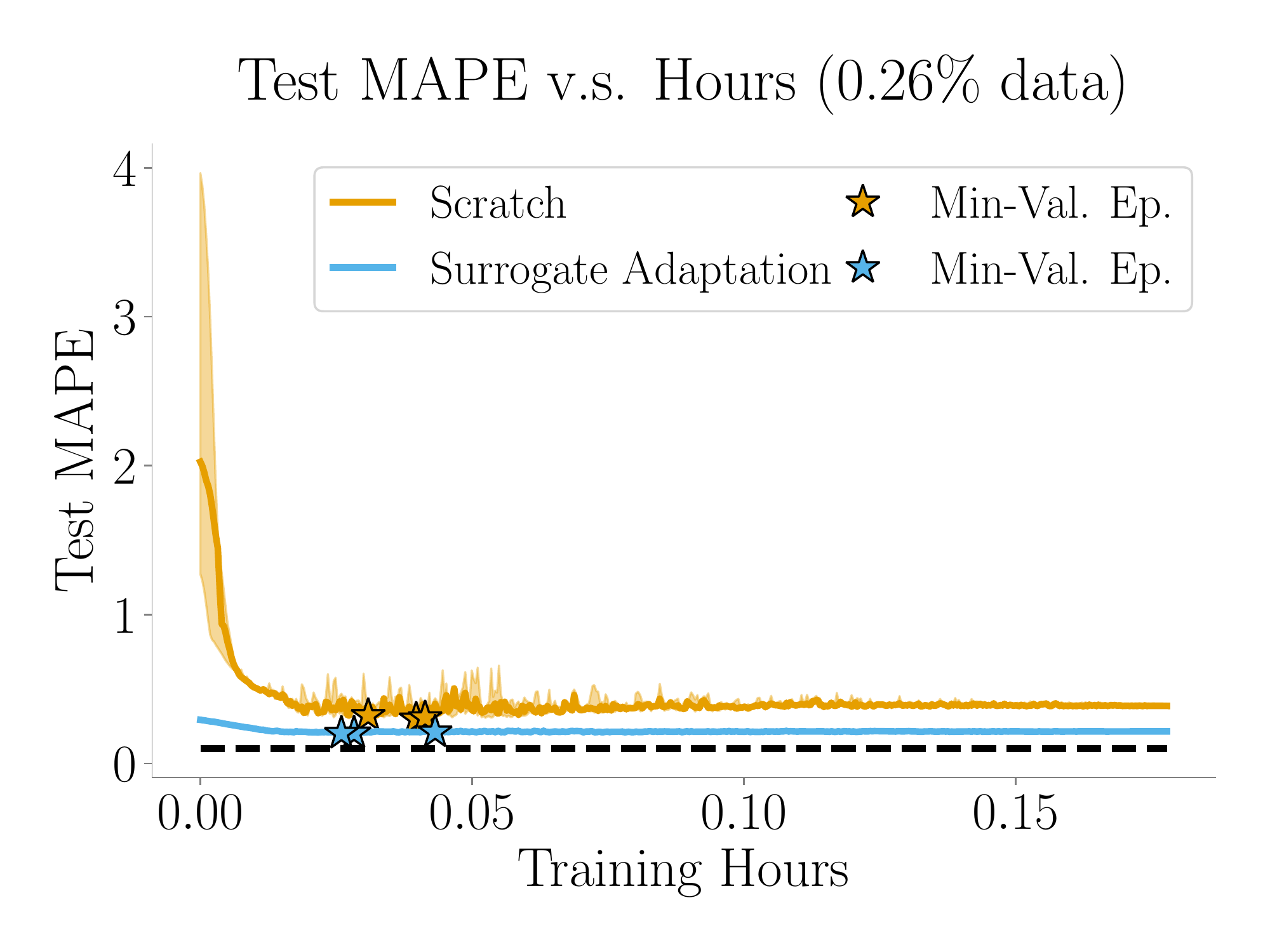}\end{minipage} \\
  \end{center}
\end{figure*}%
\begin{figure*}
  \ContinuedFloat
  \begin{center}
  {\huge\bf {\ADAPTATION{} Telemetry}}\\
  {\LARGE $0.60\%$ Data} \\
    \begin{minipage}{0.48\textwidth}\includegraphics[width=\textwidth]{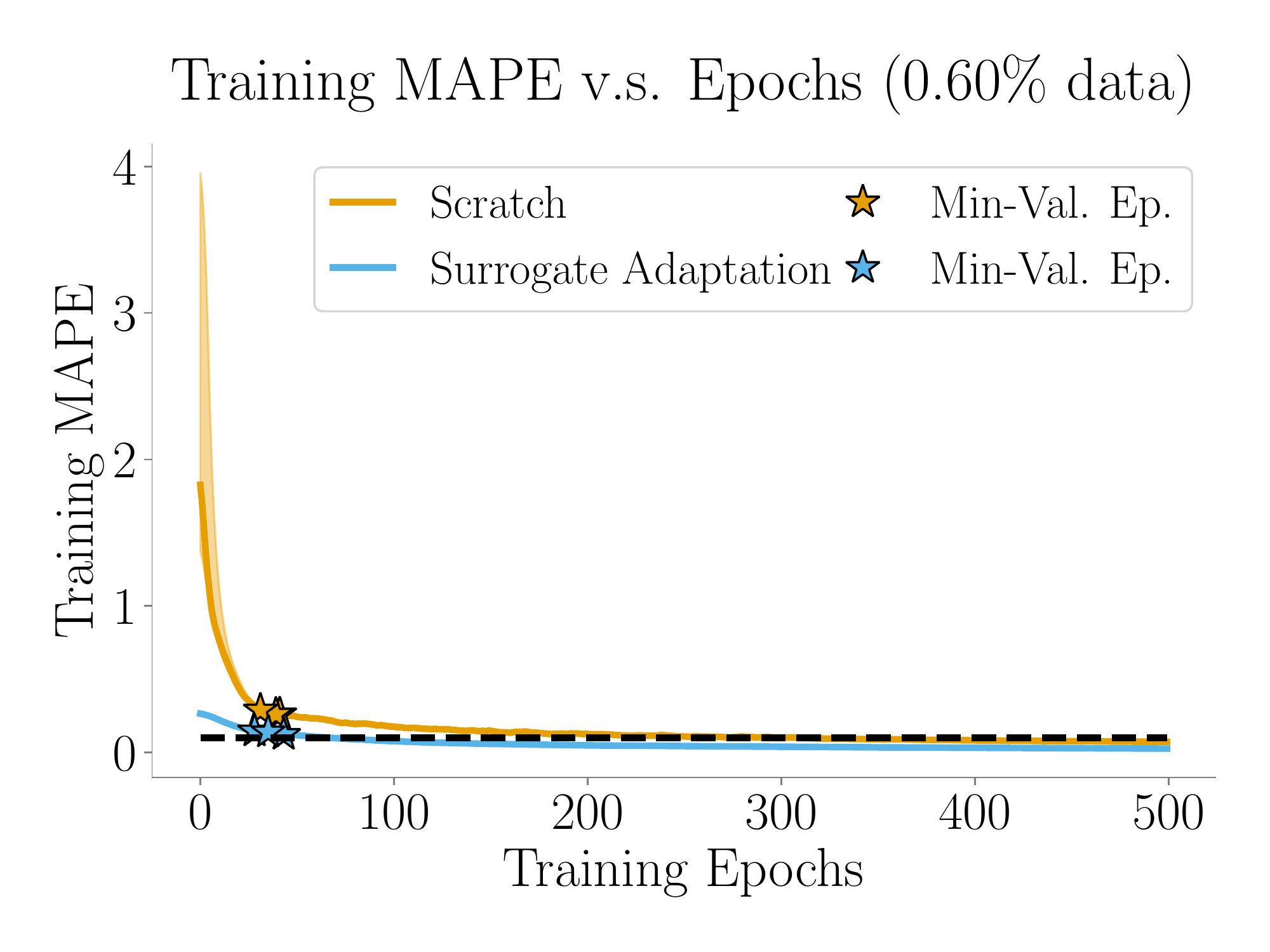}\end{minipage}
    \begin{minipage}{0.48\textwidth}\includegraphics[width=\textwidth]{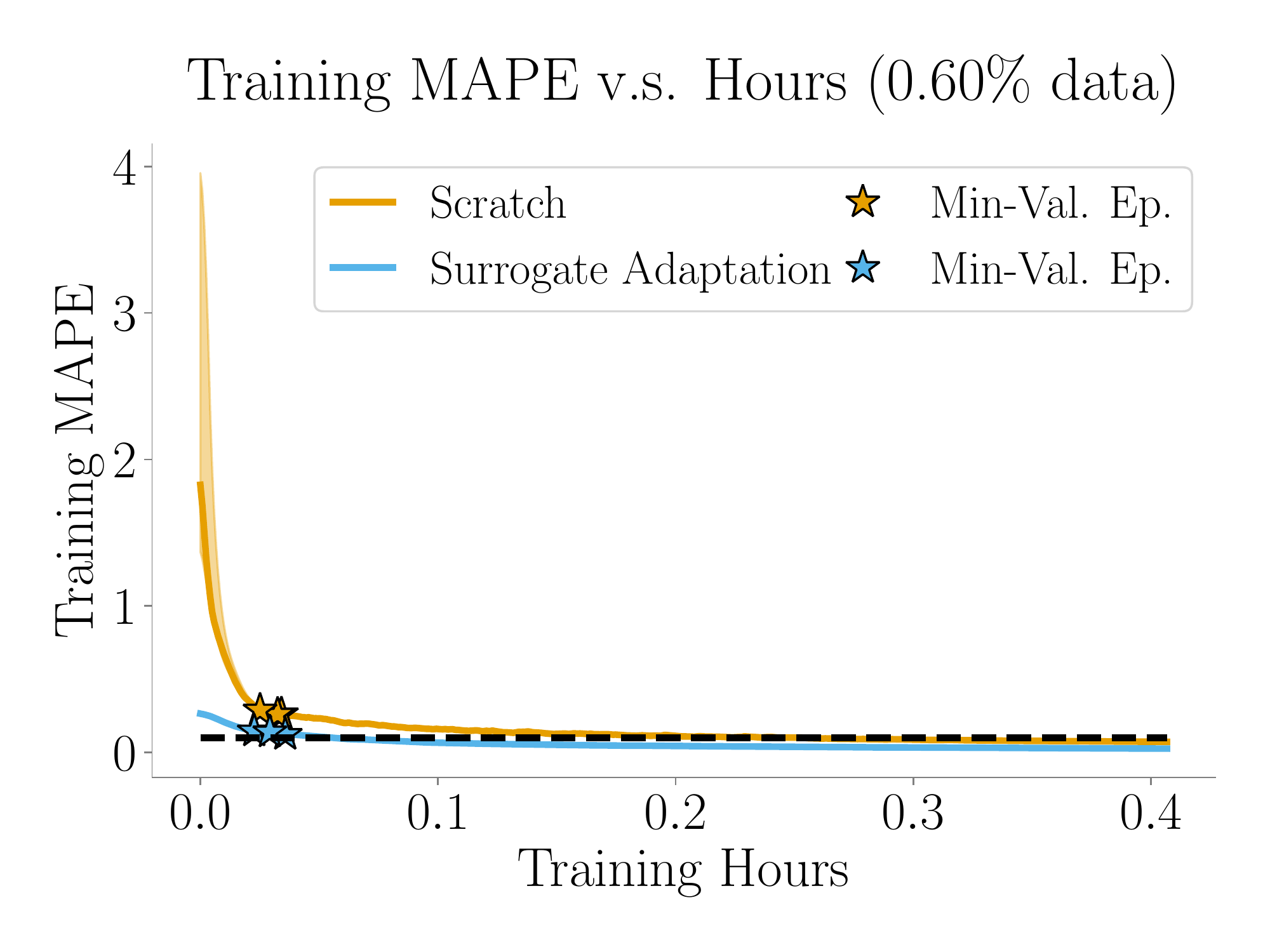}\end{minipage} \\
    \begin{minipage}{0.48\textwidth}\includegraphics[width=\textwidth]{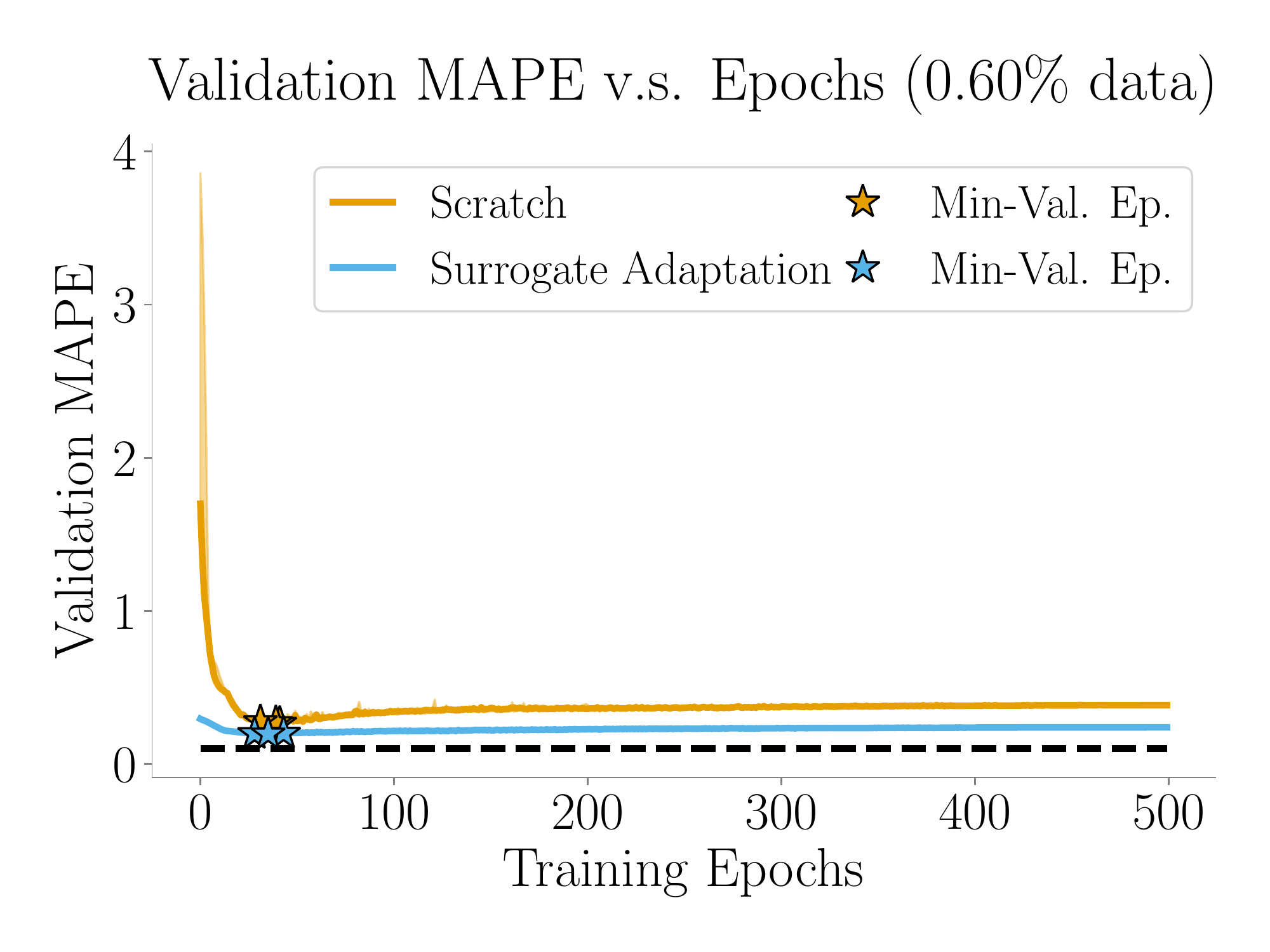}\end{minipage}
    \begin{minipage}{0.48\textwidth}\includegraphics[width=\textwidth]{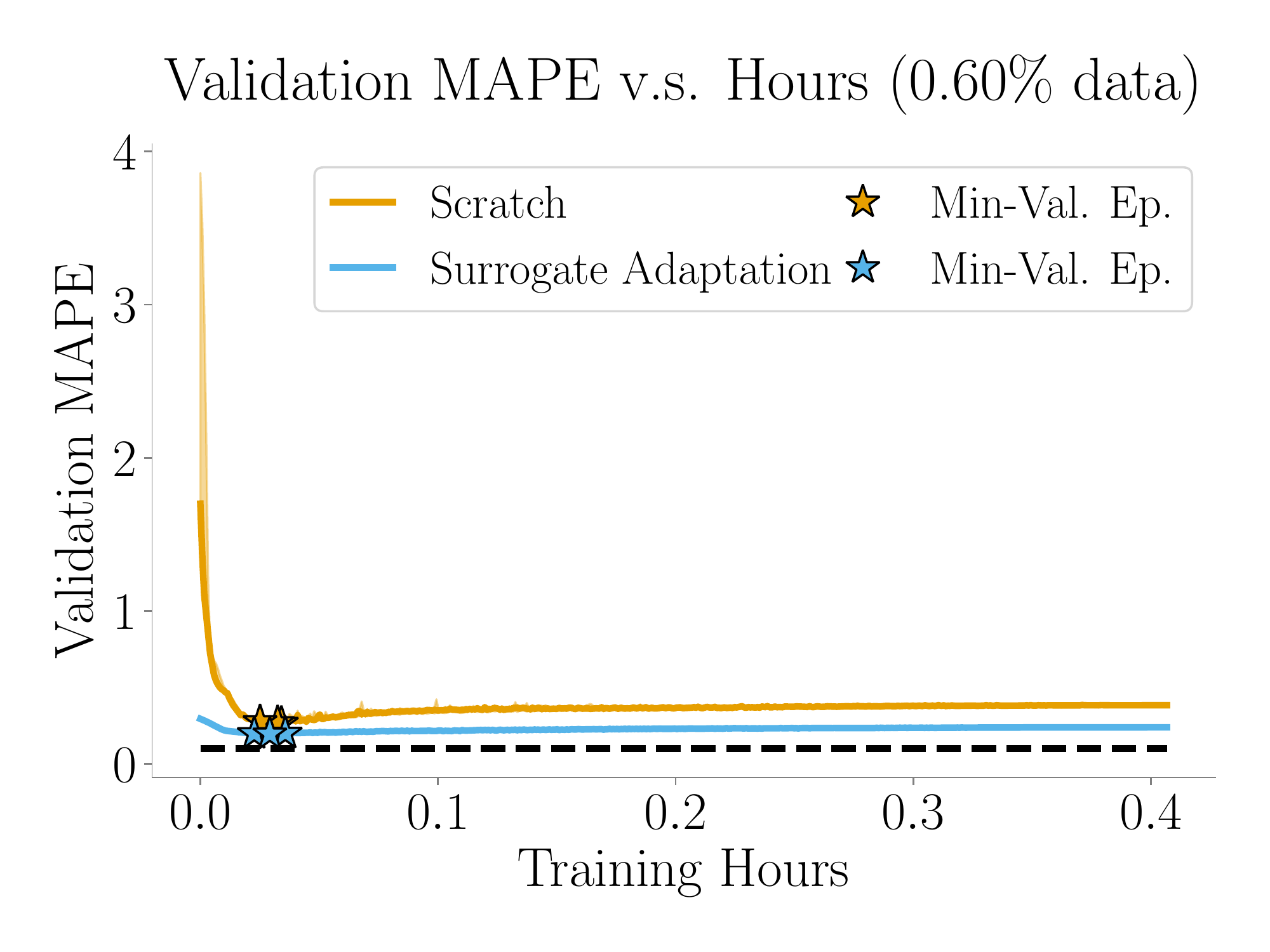}\end{minipage} \\
    \begin{minipage}{0.48\textwidth}\includegraphics[width=\textwidth]{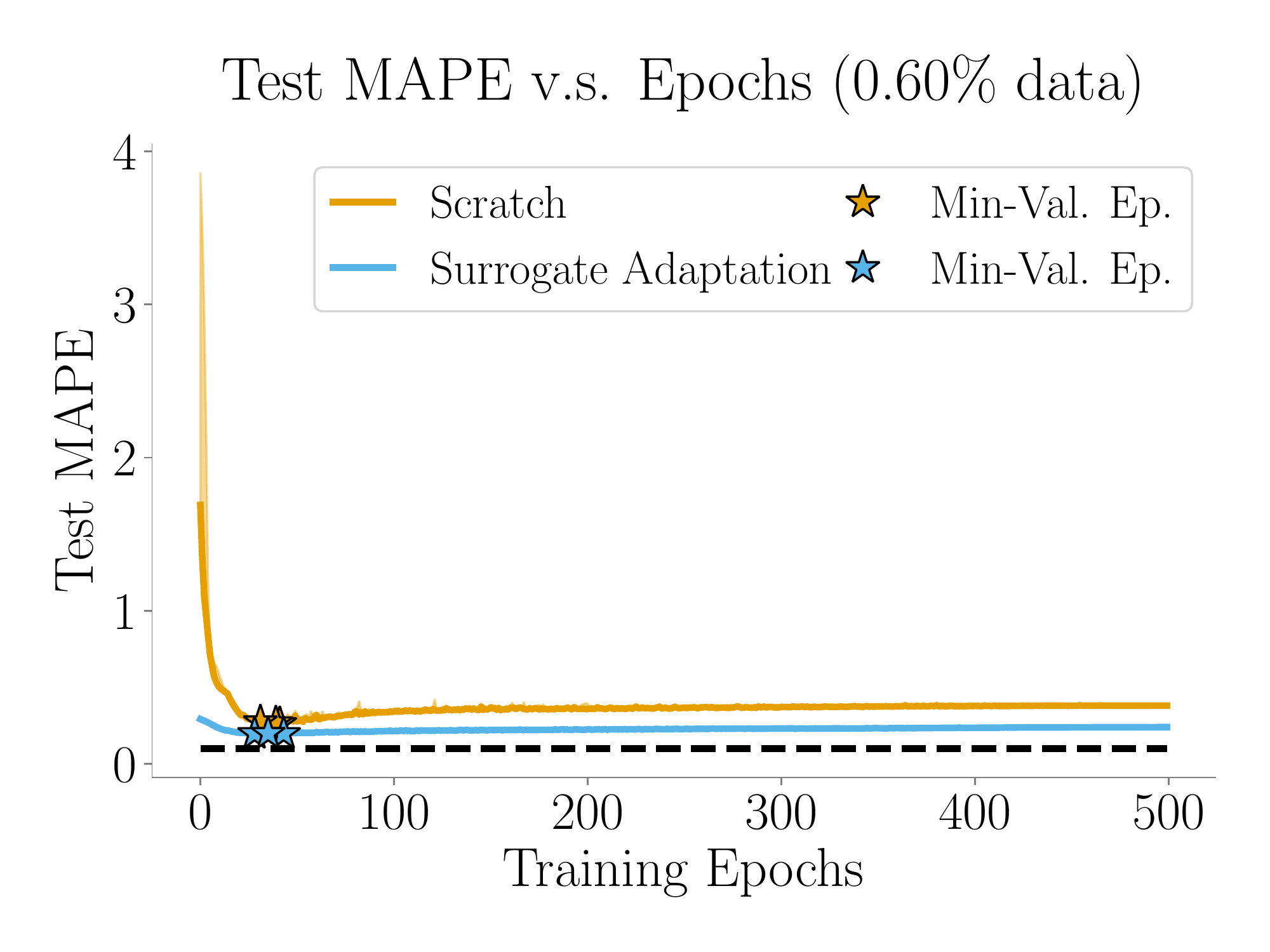}\end{minipage}
    \begin{minipage}{0.48\textwidth}\includegraphics[width=\textwidth]{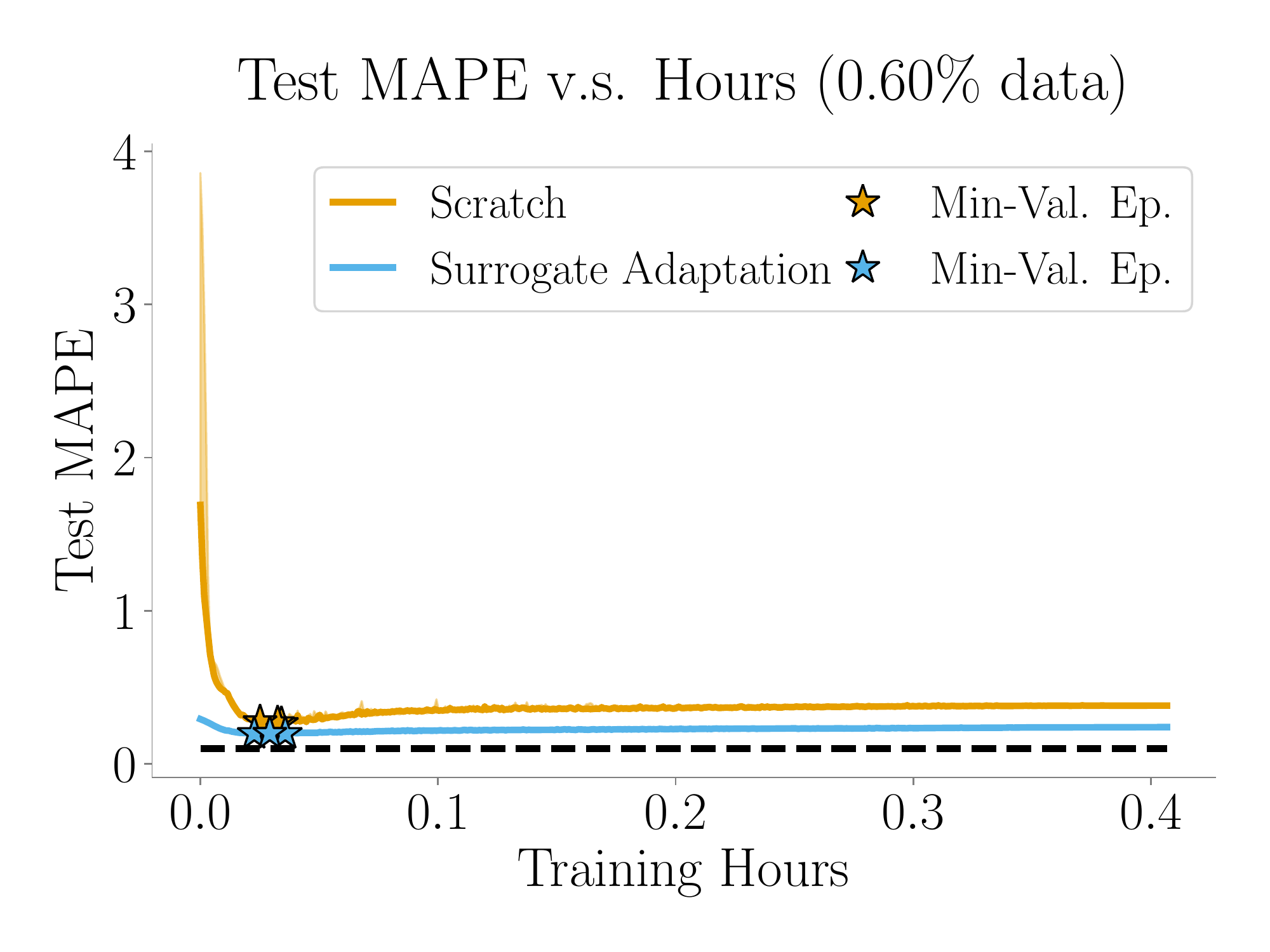}\end{minipage} \\
  \end{center}
\end{figure*}%
\begin{figure*}
  \ContinuedFloat
  \begin{center}
  {\huge\bf {\ADAPTATION{} Telemetry}}\\
  {\LARGE $1.41\%$ Data}\\
    \begin{minipage}{0.48\textwidth}\includegraphics[width=\textwidth]{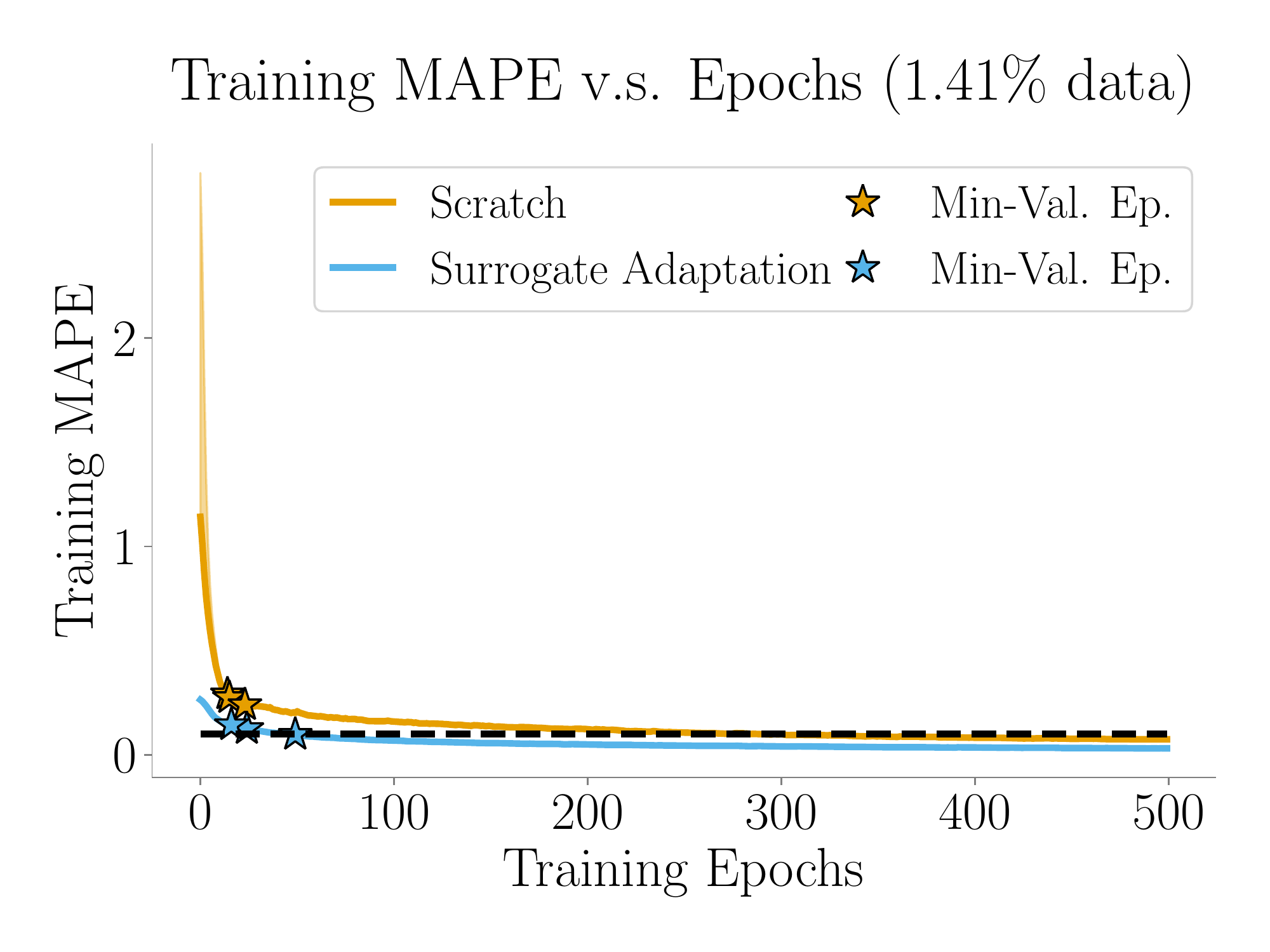}\end{minipage}
    \begin{minipage}{0.48\textwidth}\includegraphics[width=\textwidth]{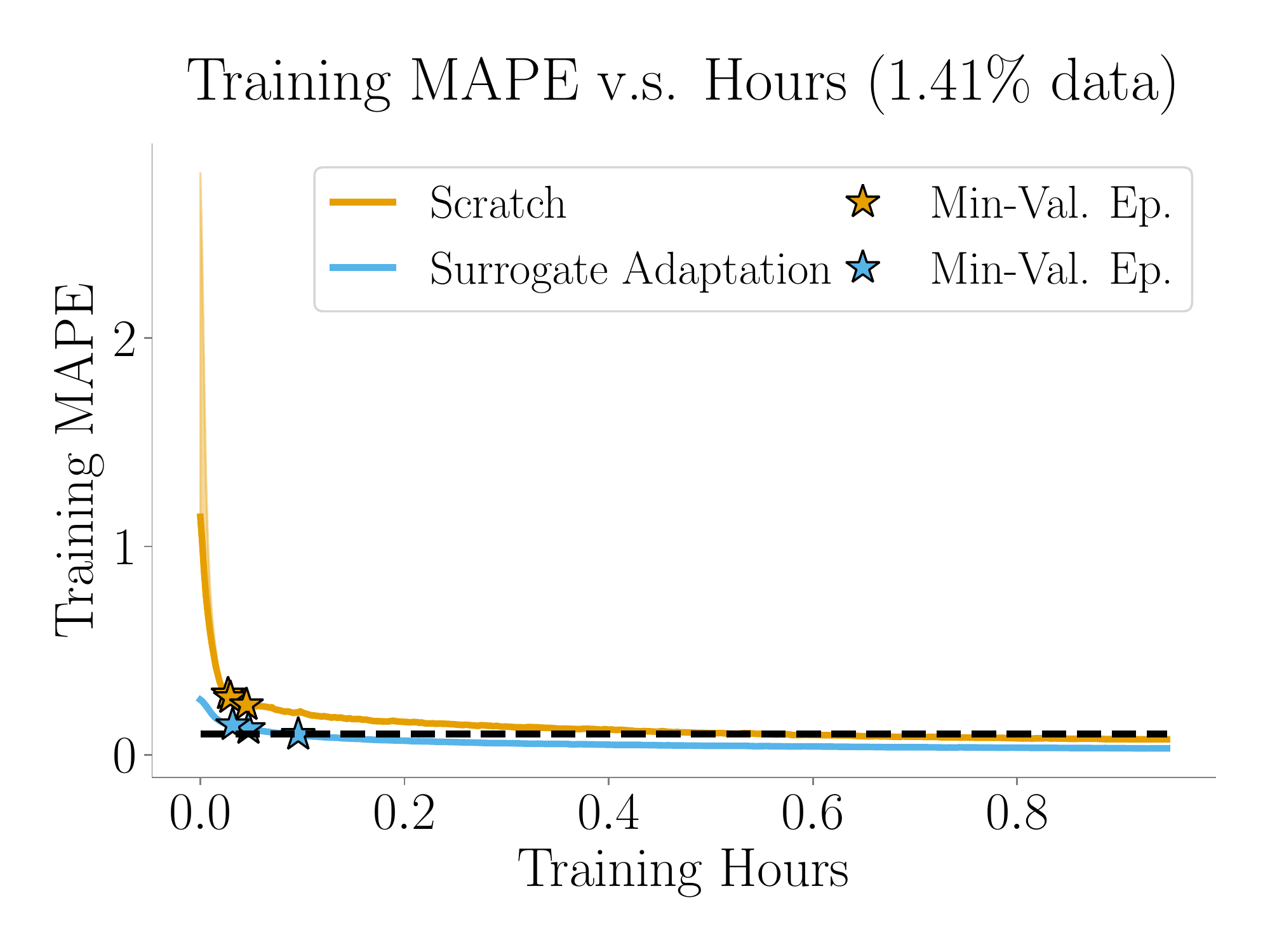}\end{minipage} \\
    \begin{minipage}{0.48\textwidth}\includegraphics[width=\textwidth]{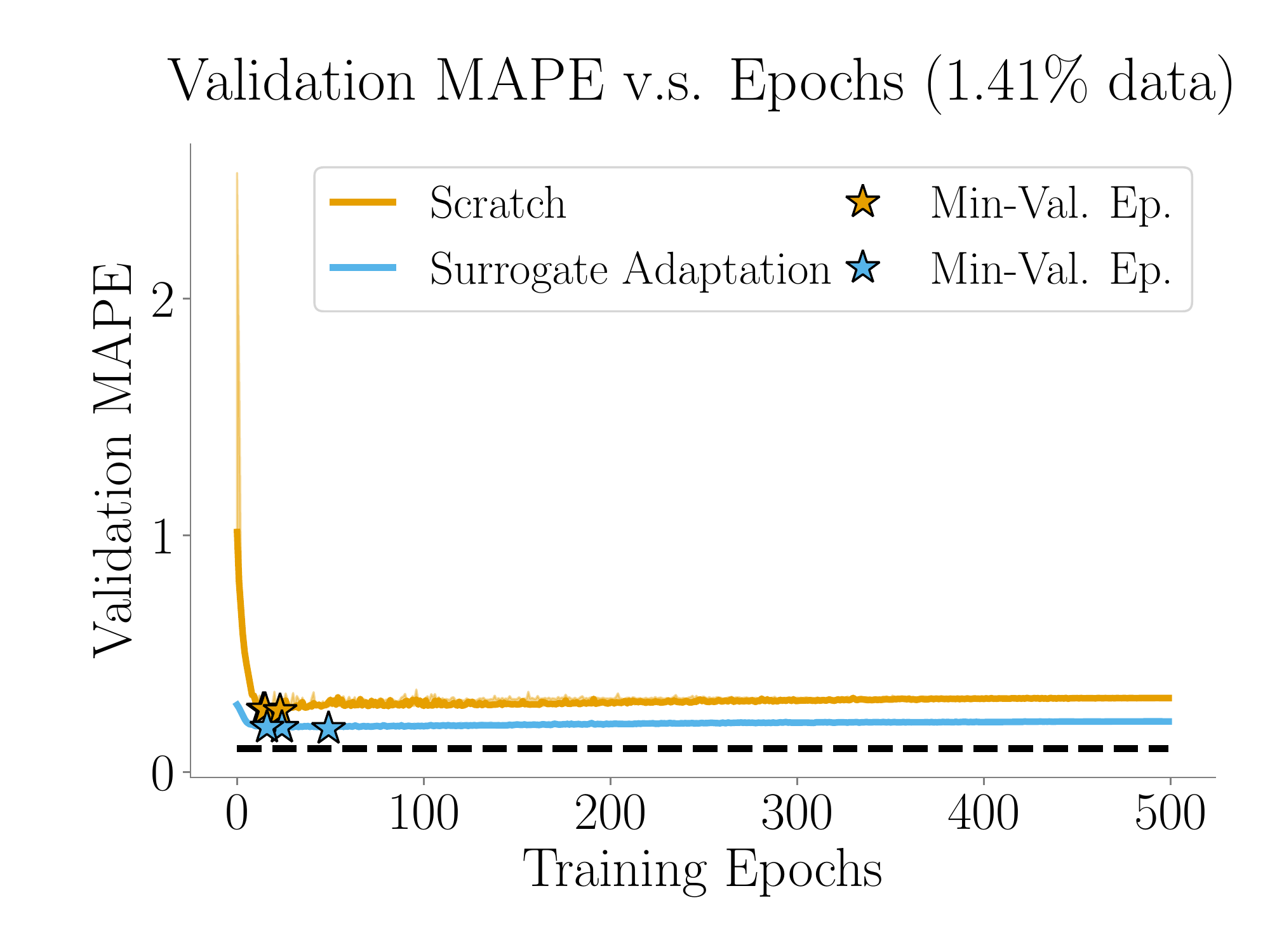}\end{minipage}
    \begin{minipage}{0.48\textwidth}\includegraphics[width=\textwidth]{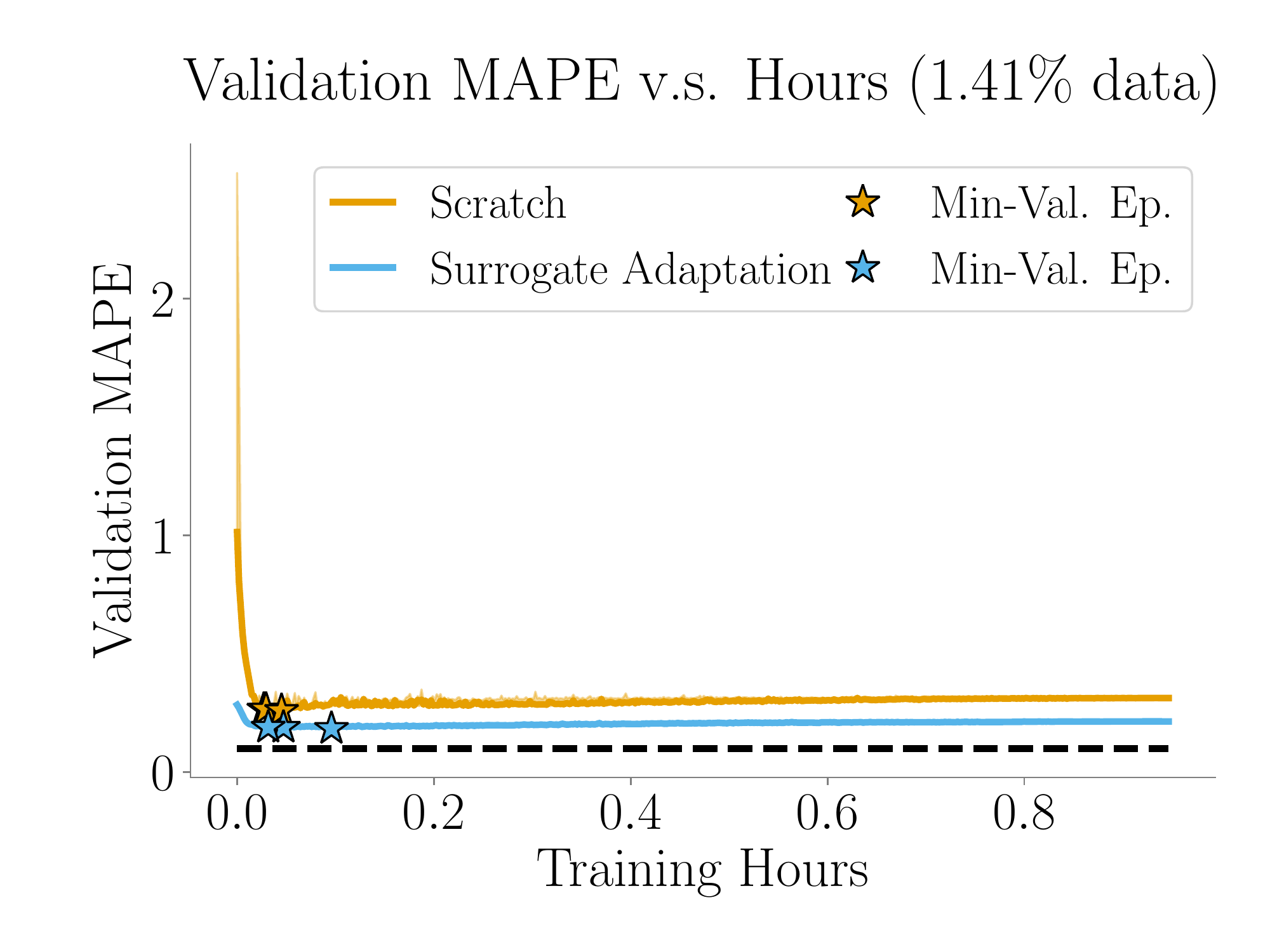}\end{minipage} \\
    \begin{minipage}{0.48\textwidth}\includegraphics[width=\textwidth]{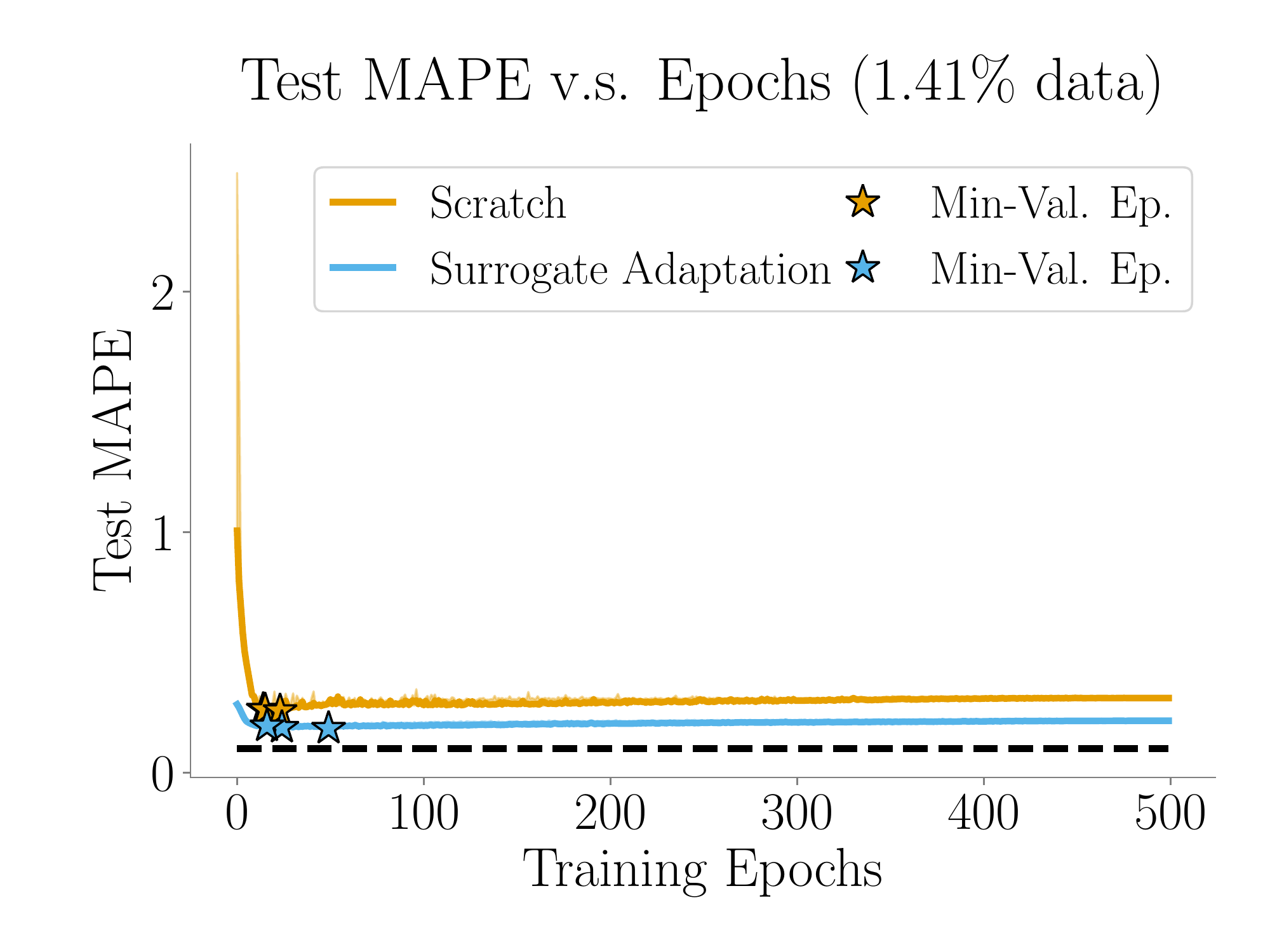}\end{minipage}
    \begin{minipage}{0.48\textwidth}\includegraphics[width=\textwidth]{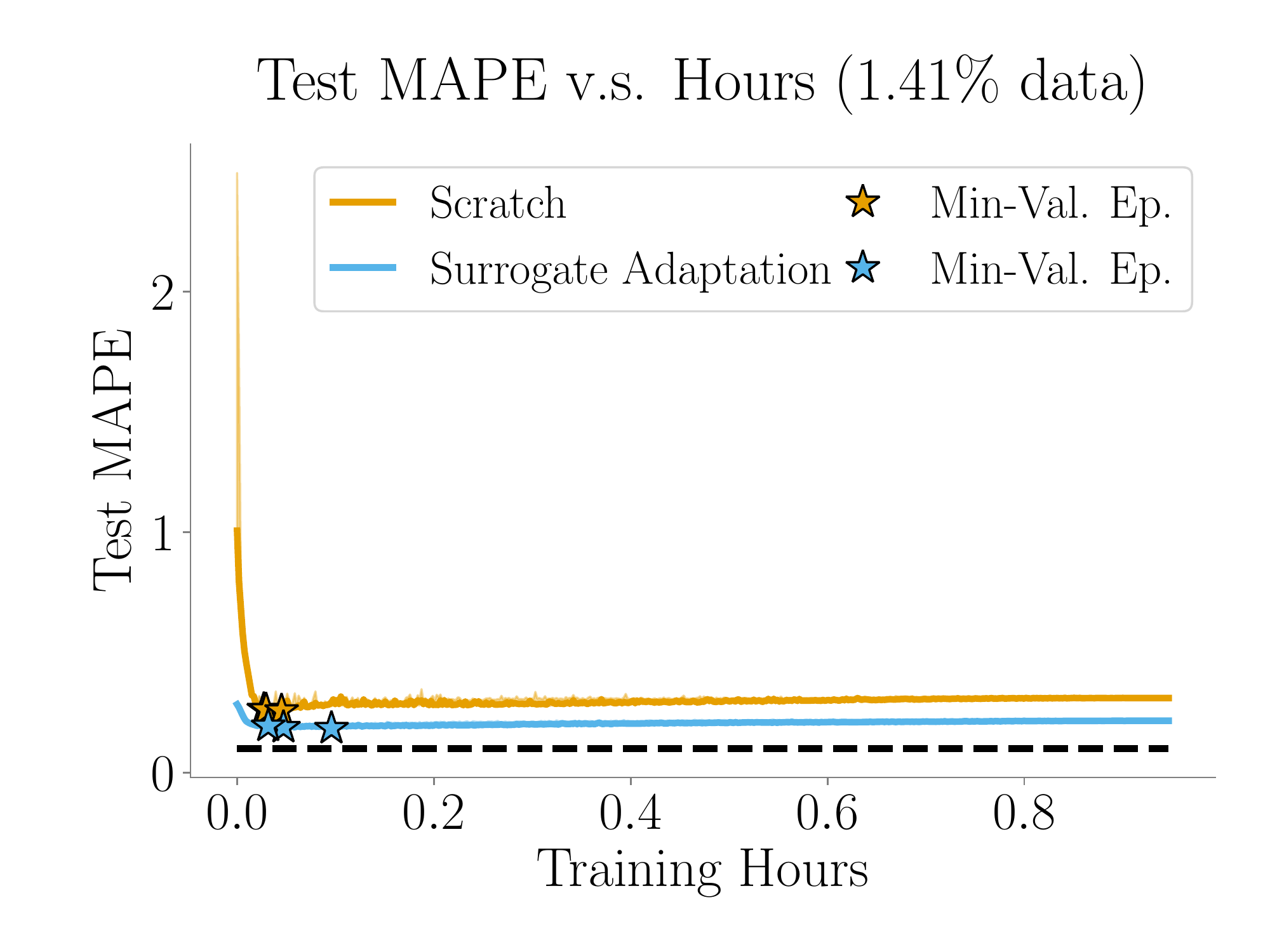}\end{minipage} \\
  \end{center}
\end{figure*}%
\begin{figure*}
  \ContinuedFloat
  \begin{center}
  {\huge\bf {\ADAPTATION{} Telemetry}}\\
  {\LARGE $3.30\%$ Data} \\
    \begin{minipage}{0.48\textwidth}\includegraphics[width=\textwidth]{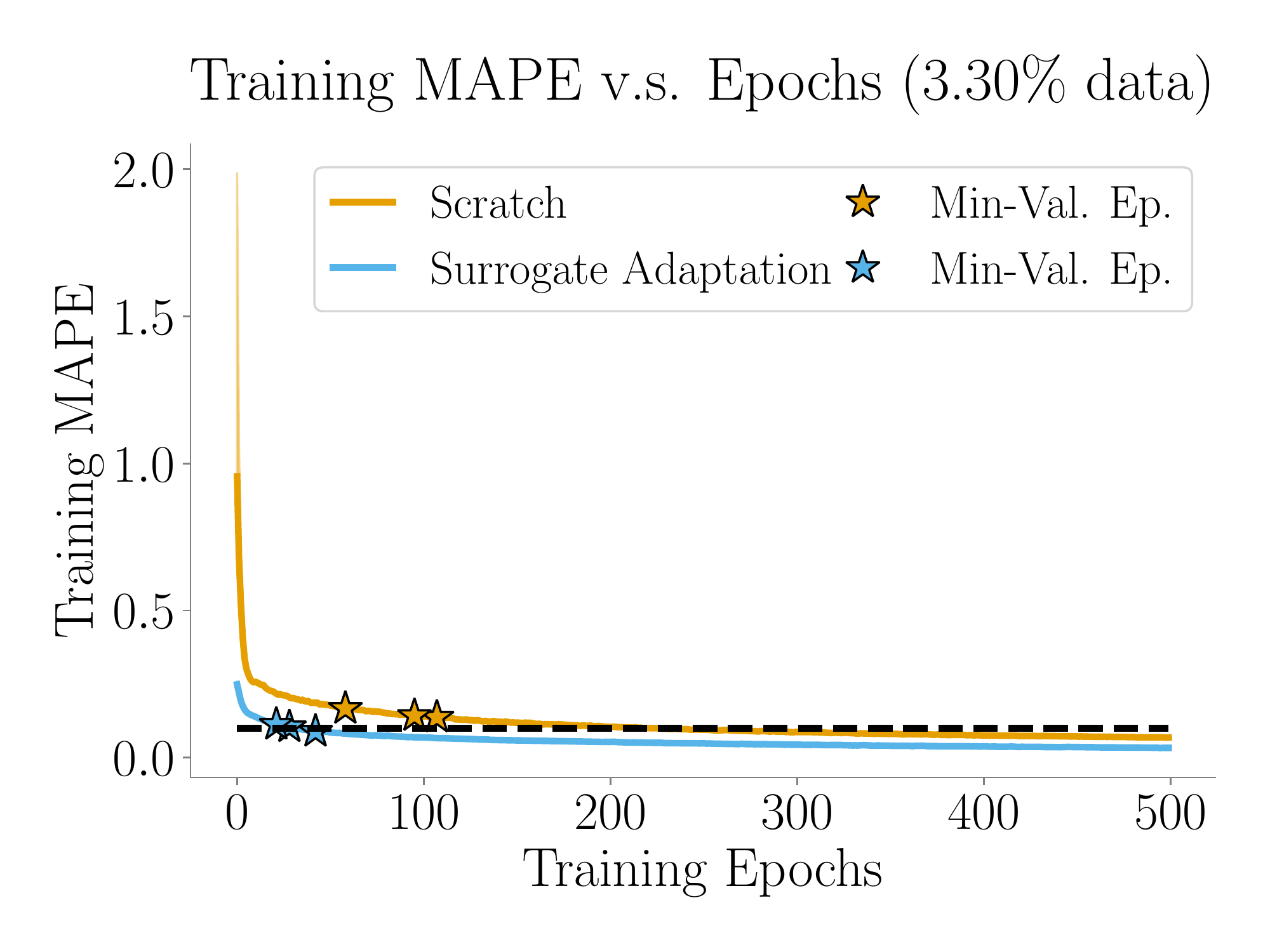}\end{minipage}
    \begin{minipage}{0.48\textwidth}\includegraphics[width=\textwidth]{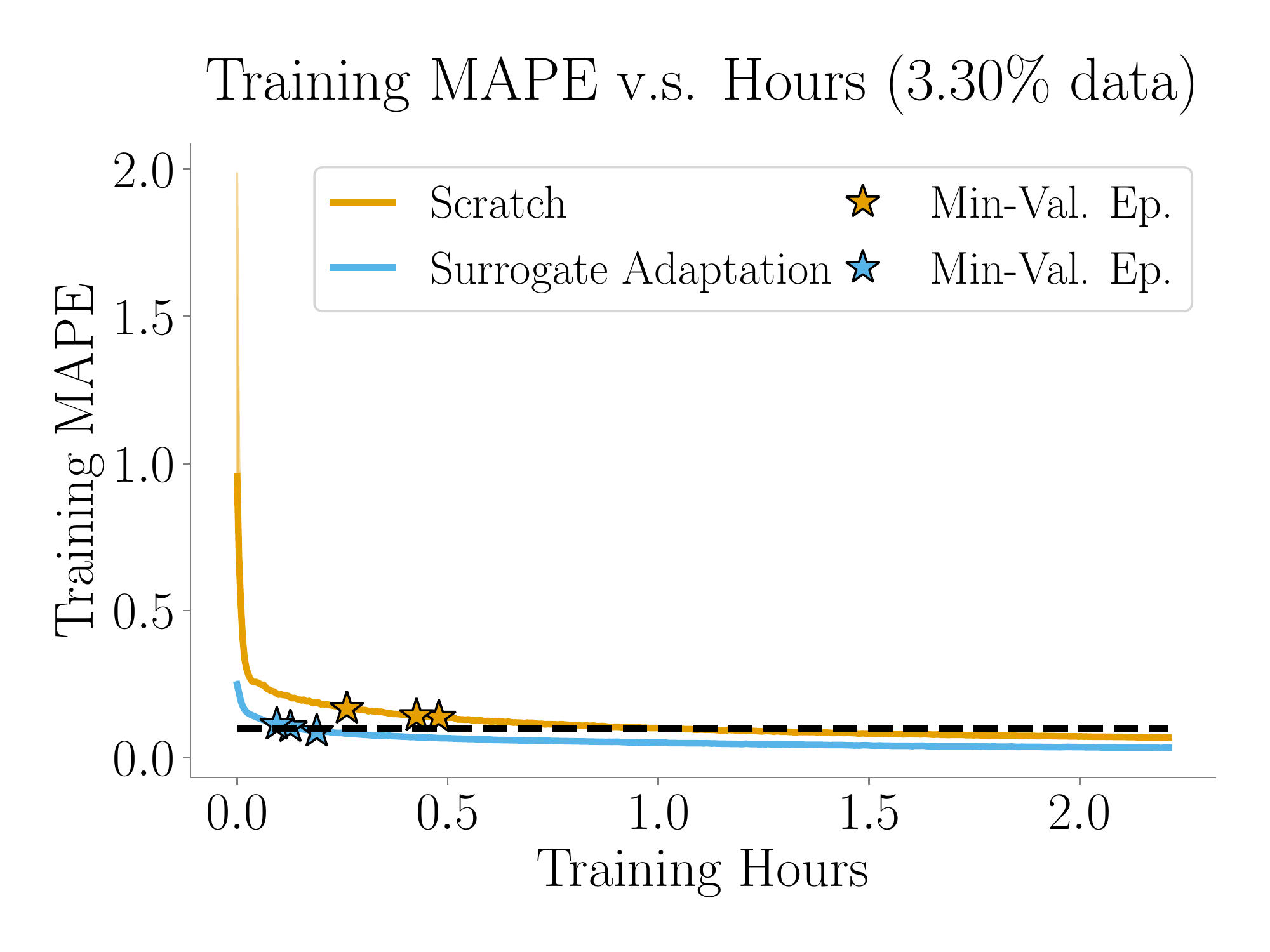}\end{minipage} \\
    \begin{minipage}{0.48\textwidth}\includegraphics[width=\textwidth]{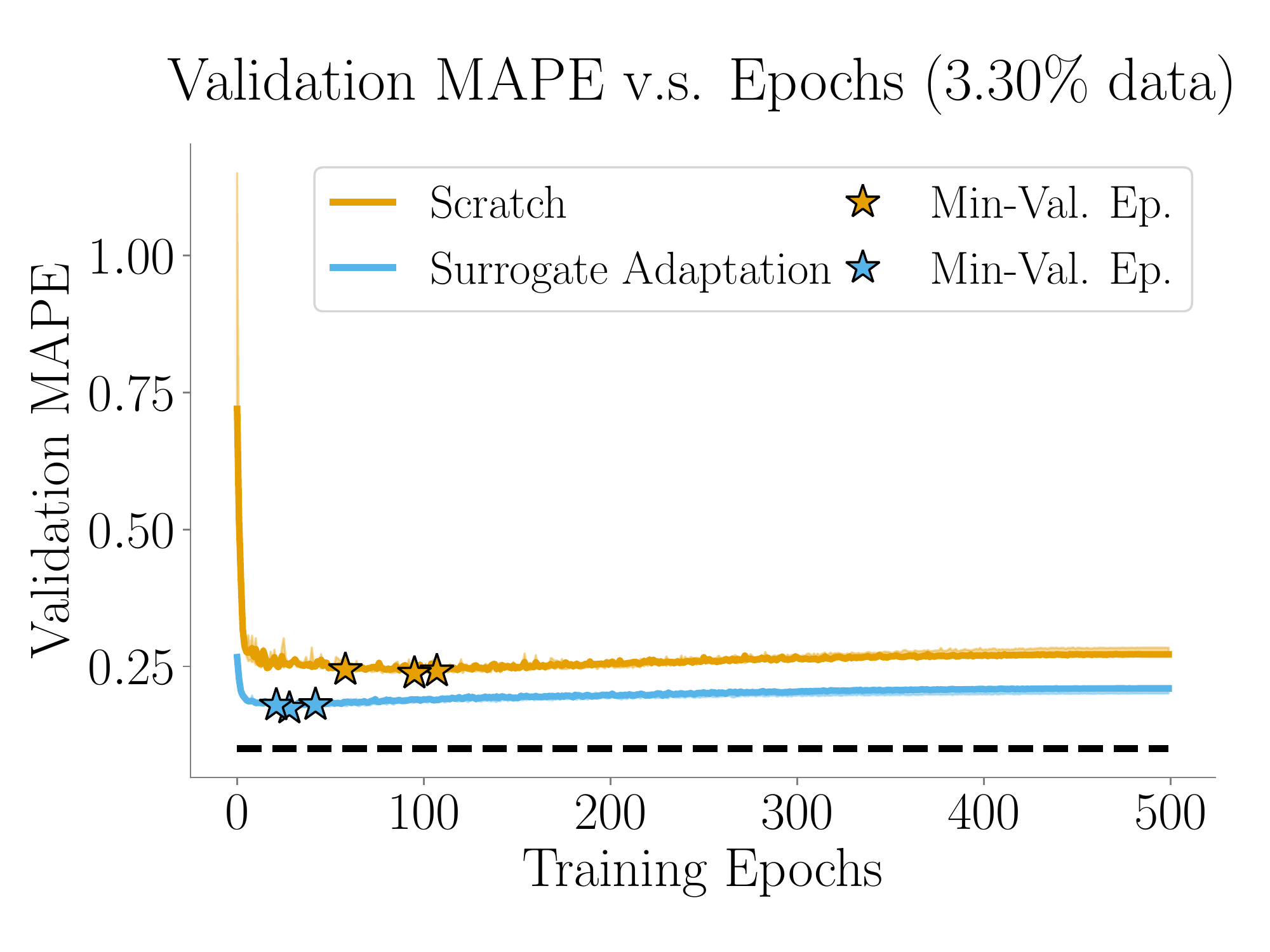}\end{minipage}
    \begin{minipage}{0.48\textwidth}\includegraphics[width=\textwidth]{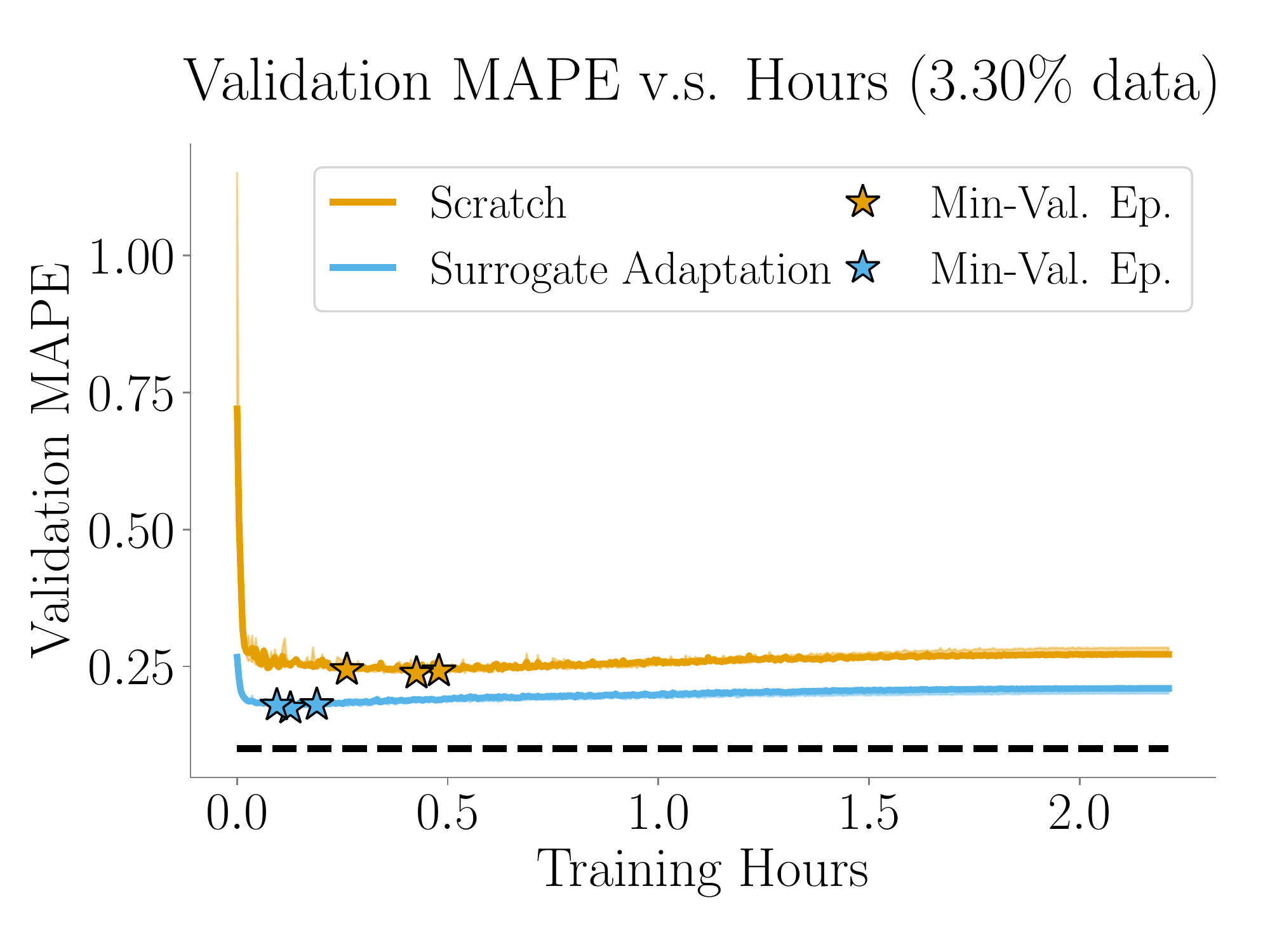}\end{minipage} \\
    \begin{minipage}{0.48\textwidth}\includegraphics[width=\textwidth]{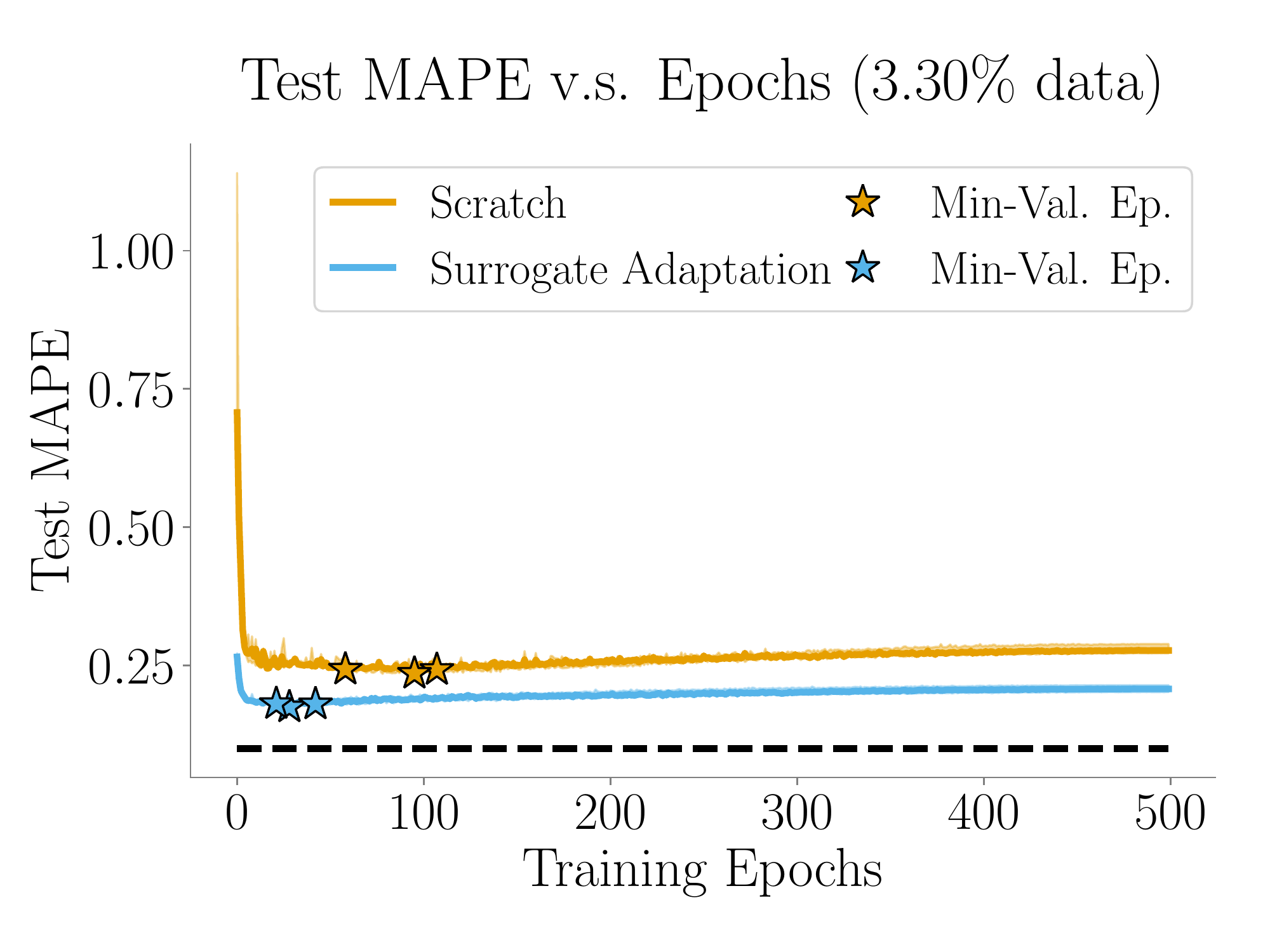}\end{minipage}
    \begin{minipage}{0.48\textwidth}\includegraphics[width=\textwidth]{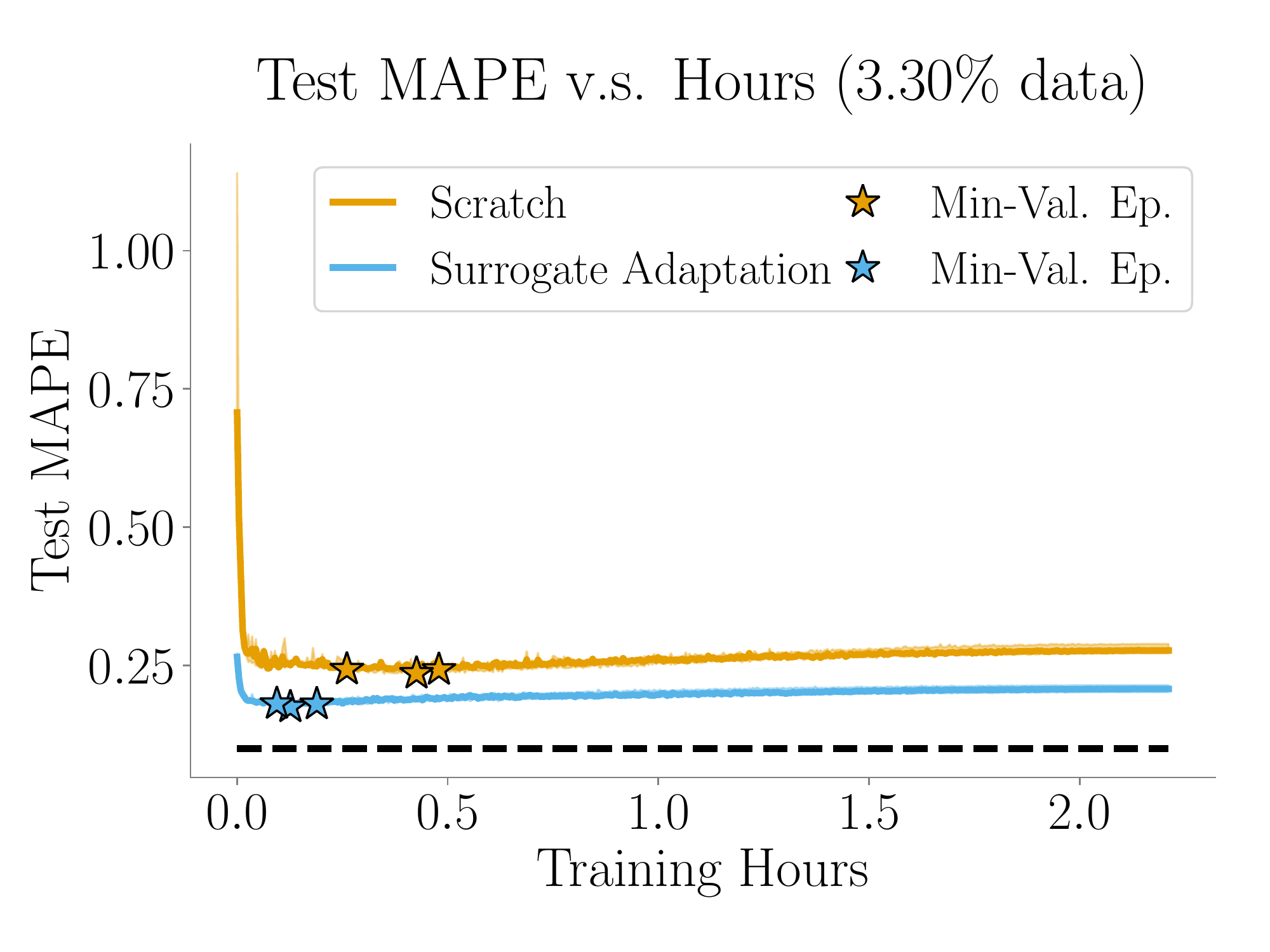}\end{minipage} \\
  \end{center}
\end{figure*}%
\begin{figure*}
  \ContinuedFloat
  \begin{center}
  {\huge\bf {\ADAPTATION{} Telemetry}}\\
  {\LARGE $7.74\%$ Data} \\
    \begin{minipage}{0.48\textwidth}\includegraphics[width=\textwidth]{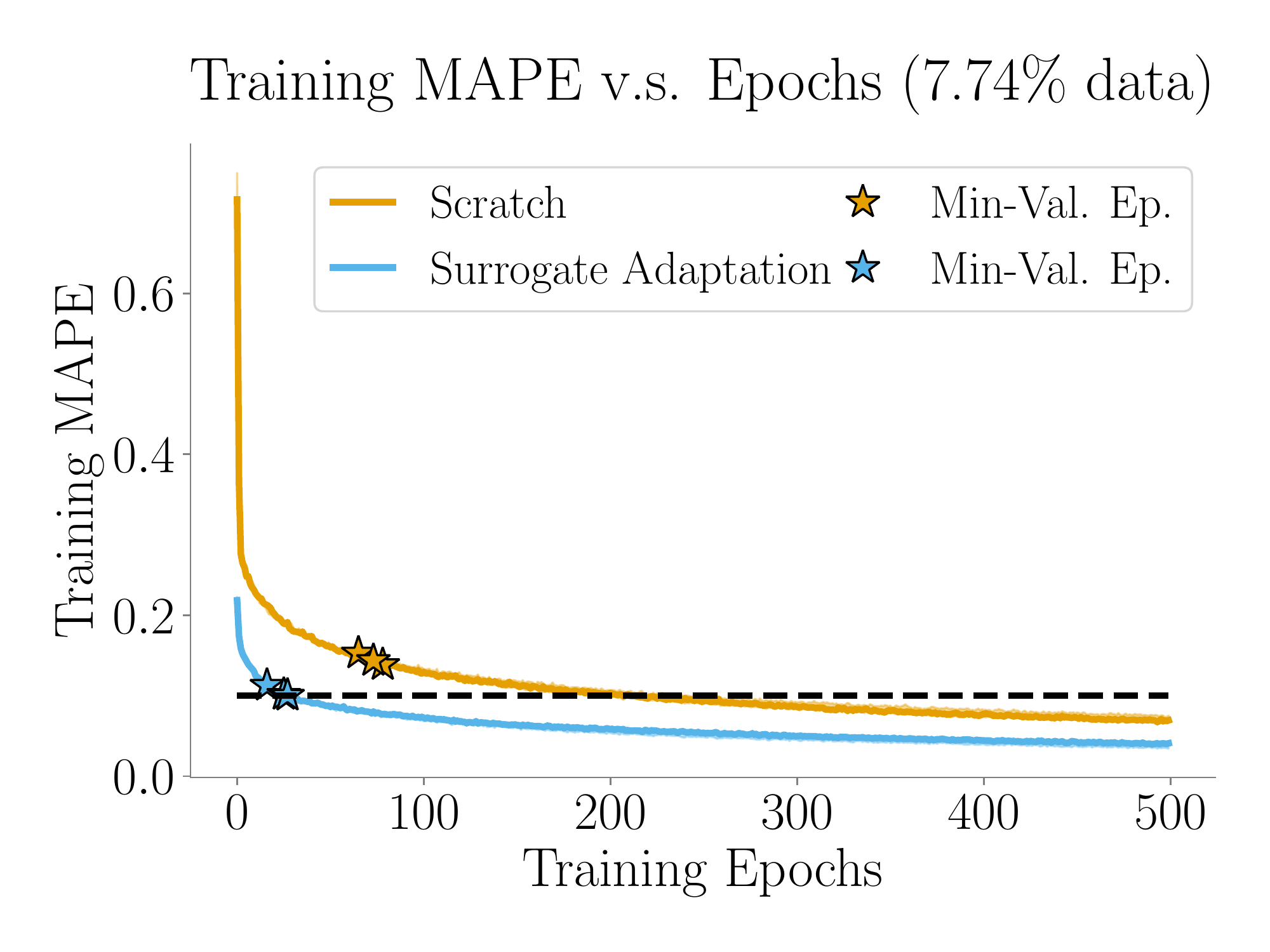}\end{minipage}
    \begin{minipage}{0.48\textwidth}\includegraphics[width=\textwidth]{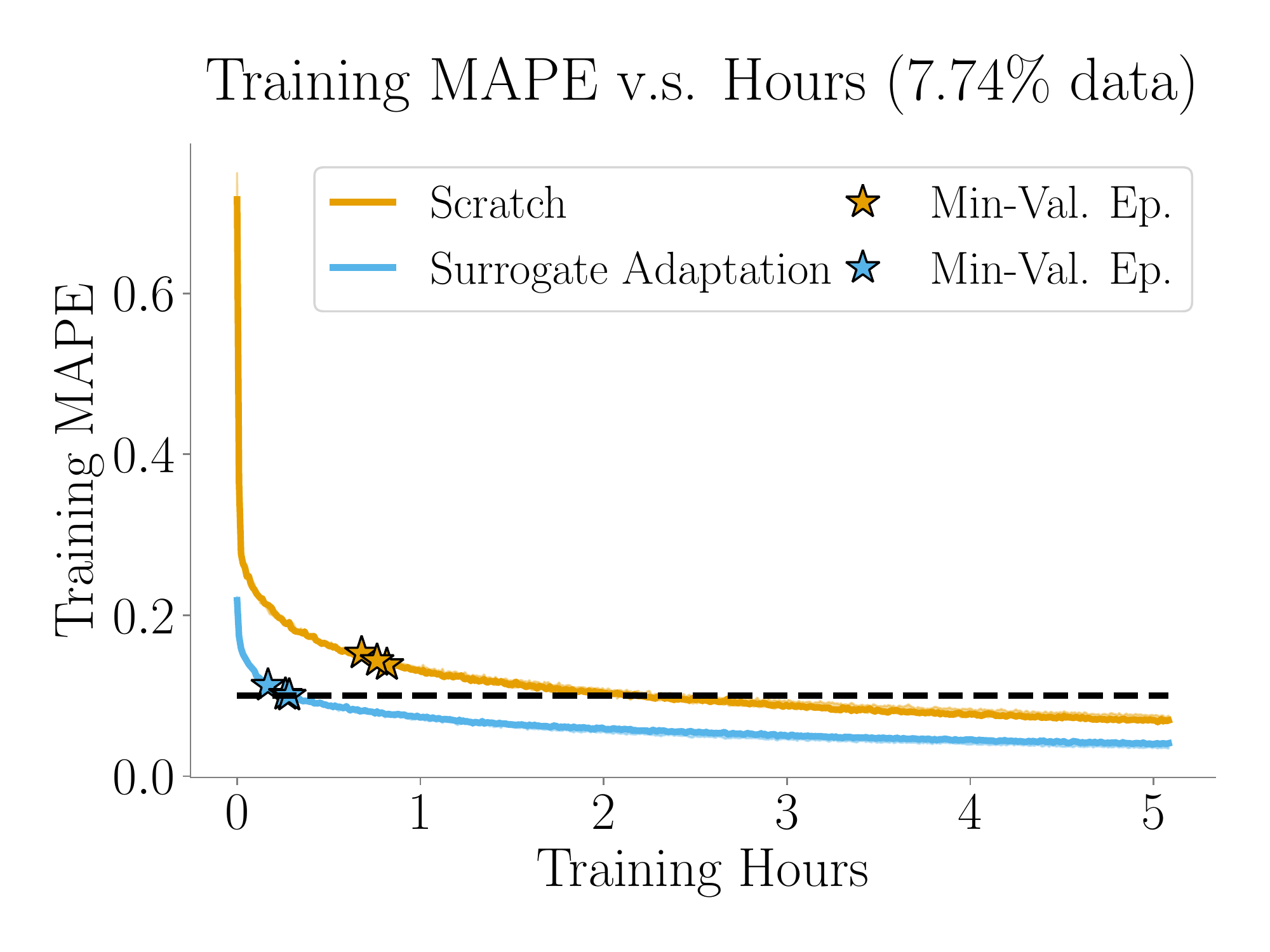}\end{minipage} \\
    \begin{minipage}{0.48\textwidth}\includegraphics[width=\textwidth]{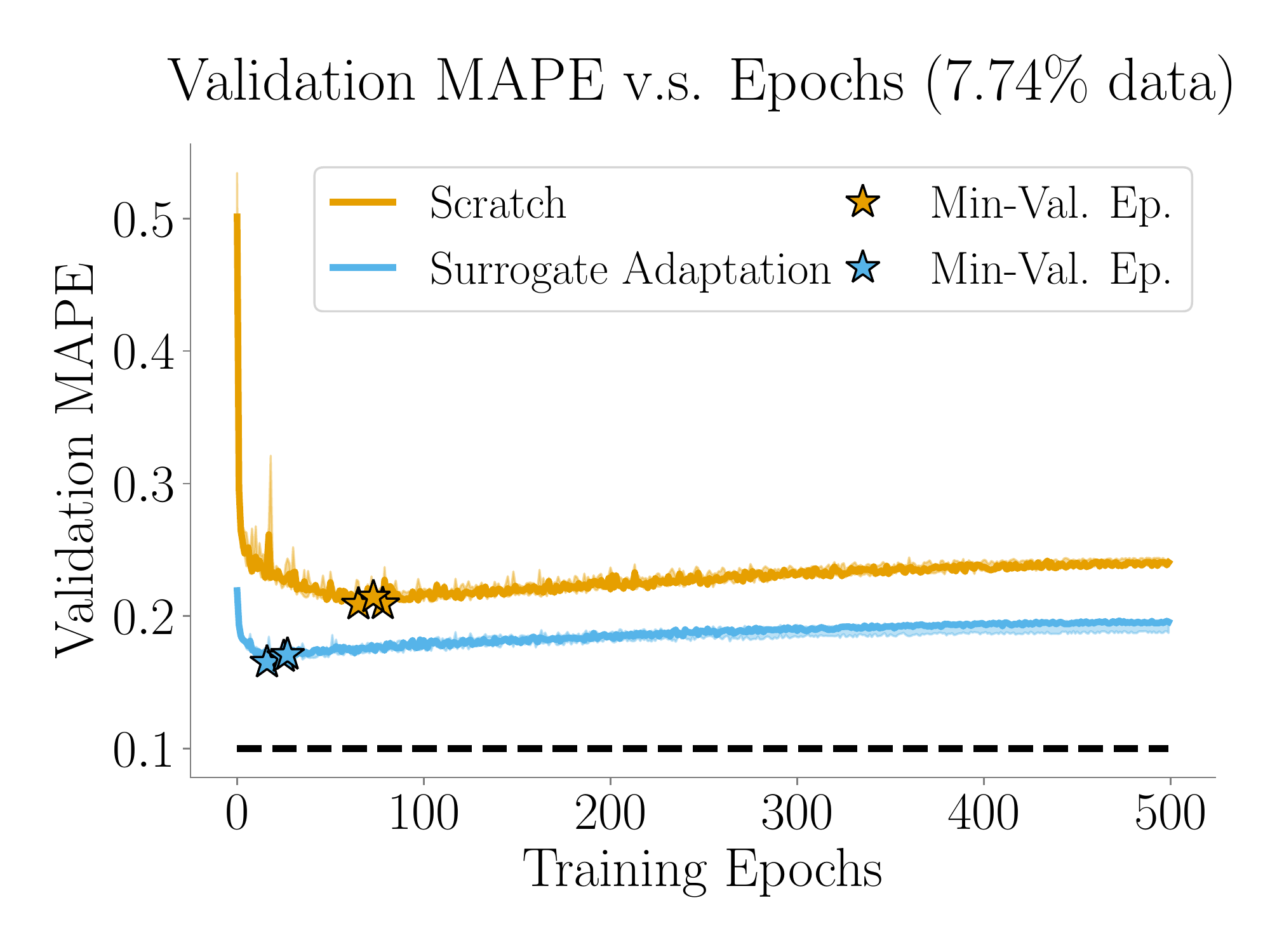}\end{minipage}
    \begin{minipage}{0.48\textwidth}\includegraphics[width=\textwidth]{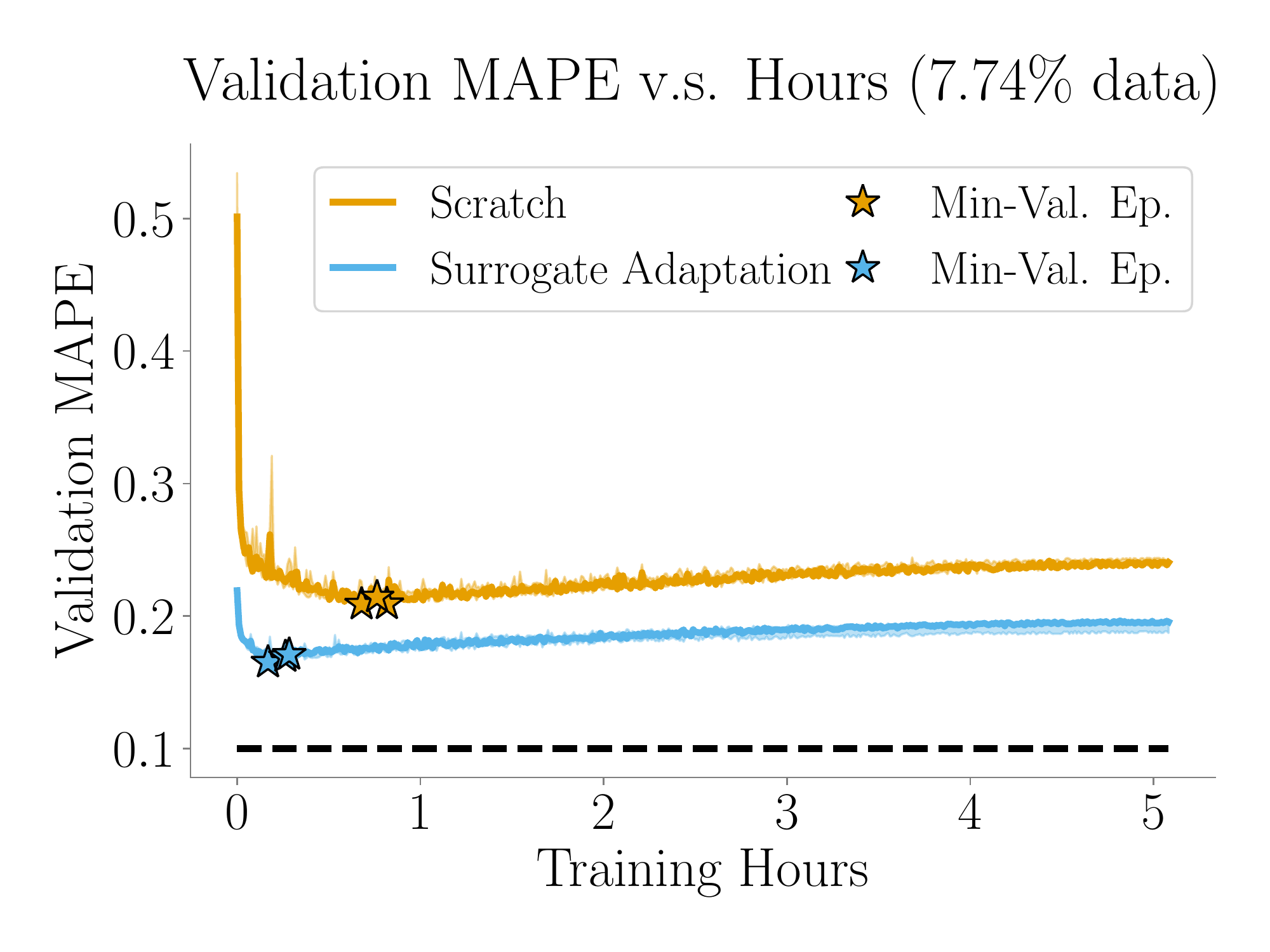}\end{minipage} \\
    \begin{minipage}{0.48\textwidth}\includegraphics[width=\textwidth]{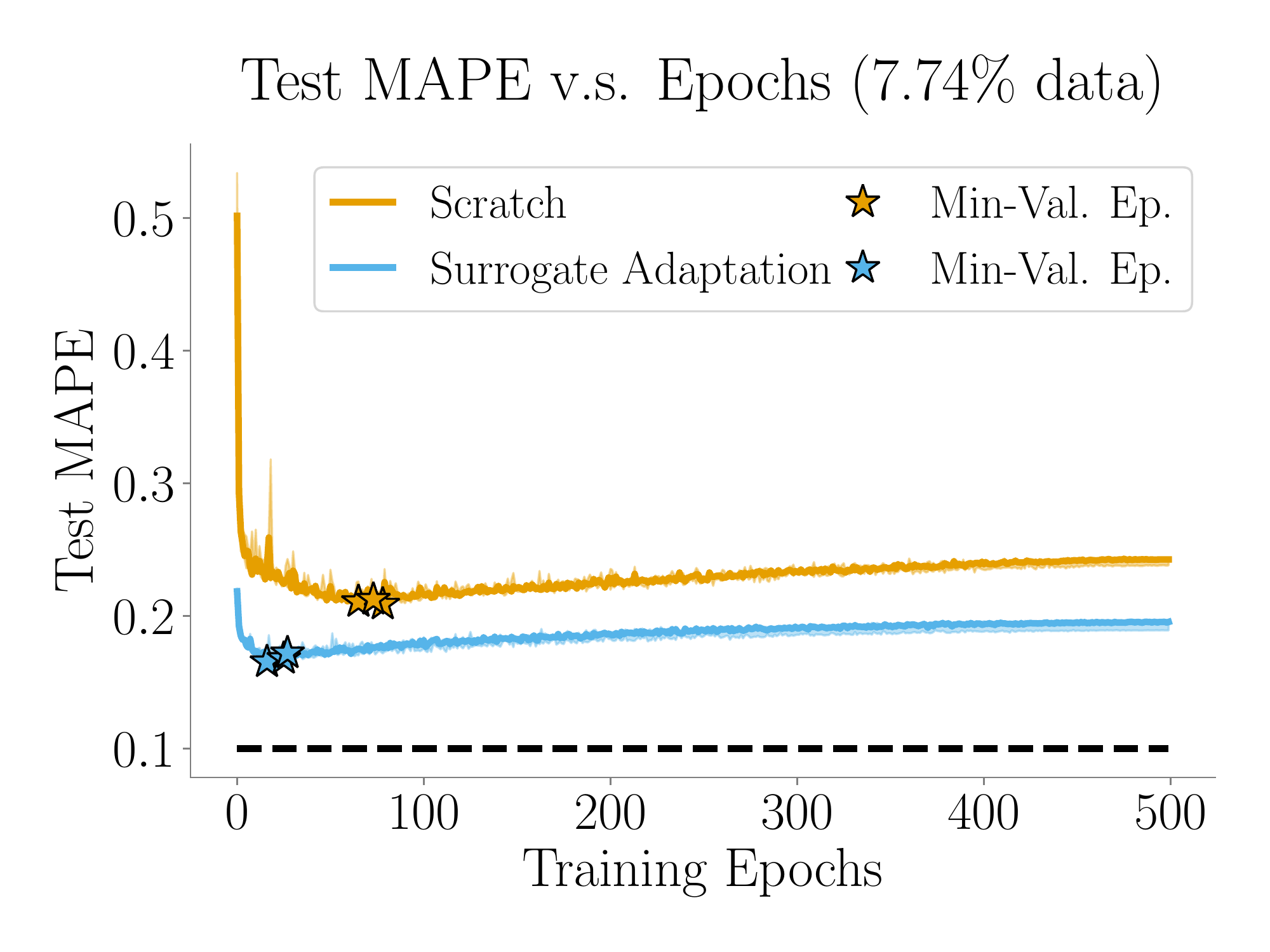}\end{minipage}
    \begin{minipage}{0.48\textwidth}\includegraphics[width=\textwidth]{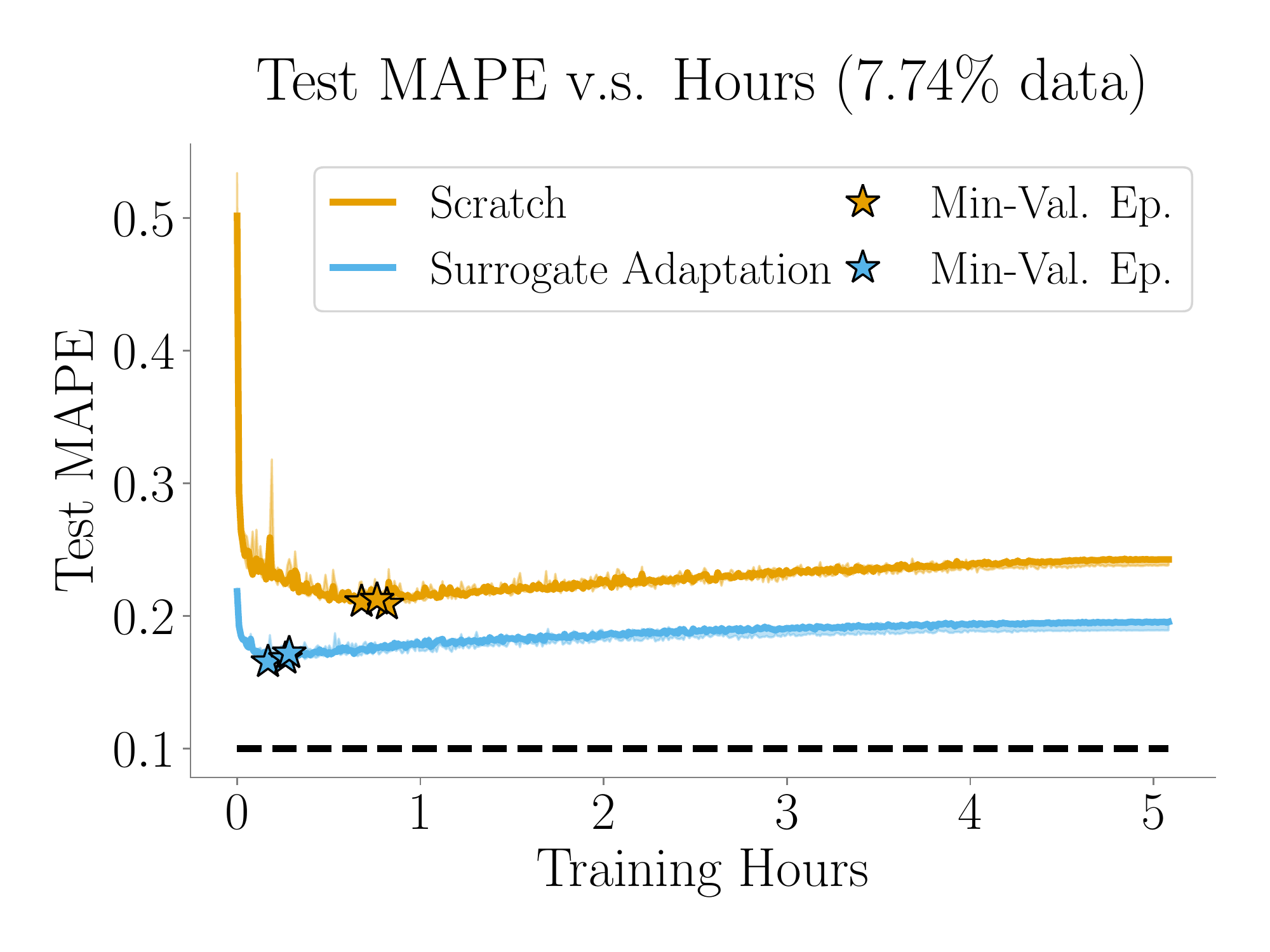}\end{minipage} \\
  \end{center}
\end{figure*}%
\begin{figure*}
  \ContinuedFloat
  \begin{center}
  {\huge\bf {\ADAPTATION{} Telemetry}}\\
  {\LARGE $18.16\%$ Data} \\
    \begin{minipage}{0.48\textwidth}\includegraphics[width=\textwidth]{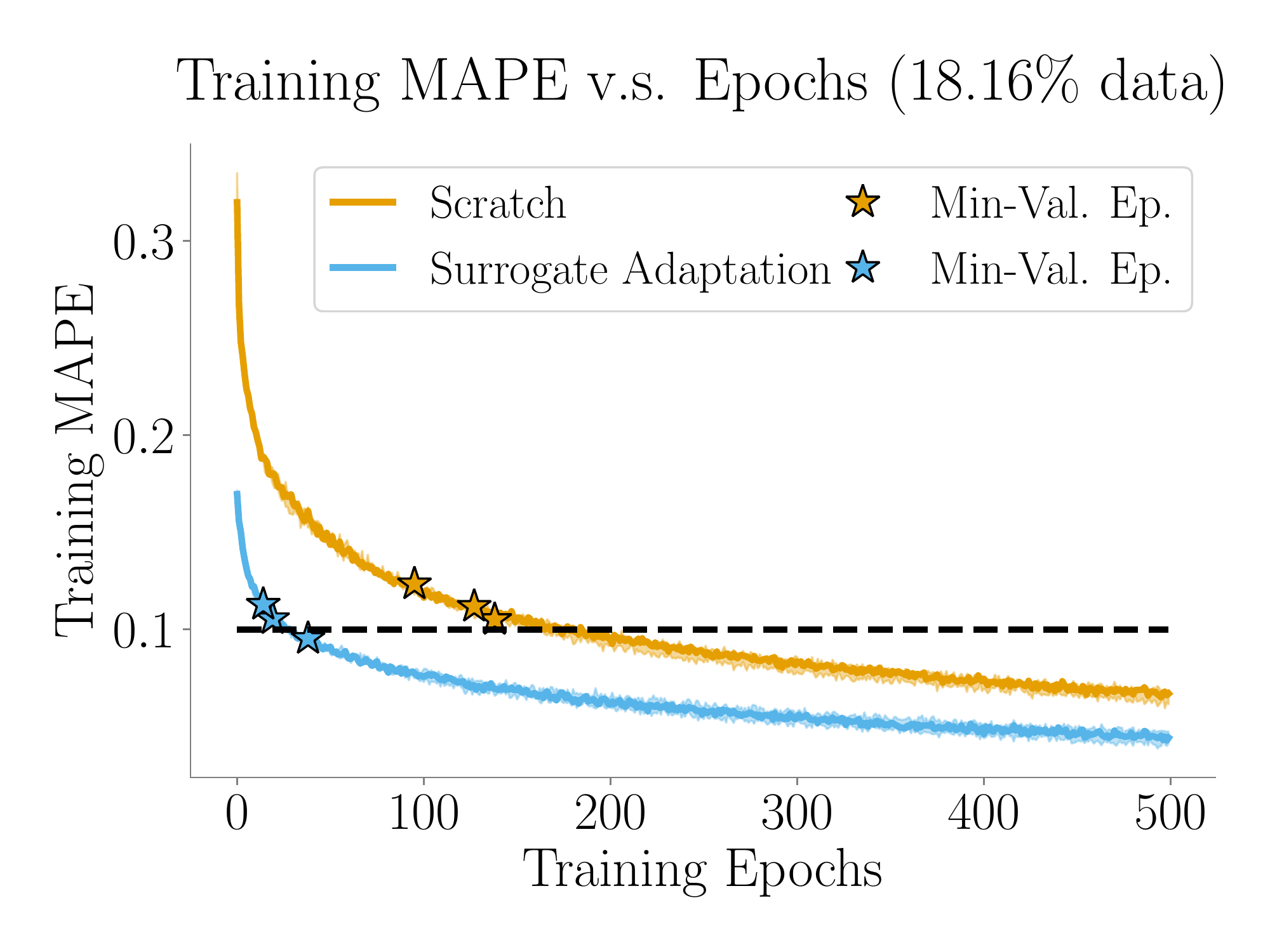}\end{minipage}
    \begin{minipage}{0.48\textwidth}\includegraphics[width=\textwidth]{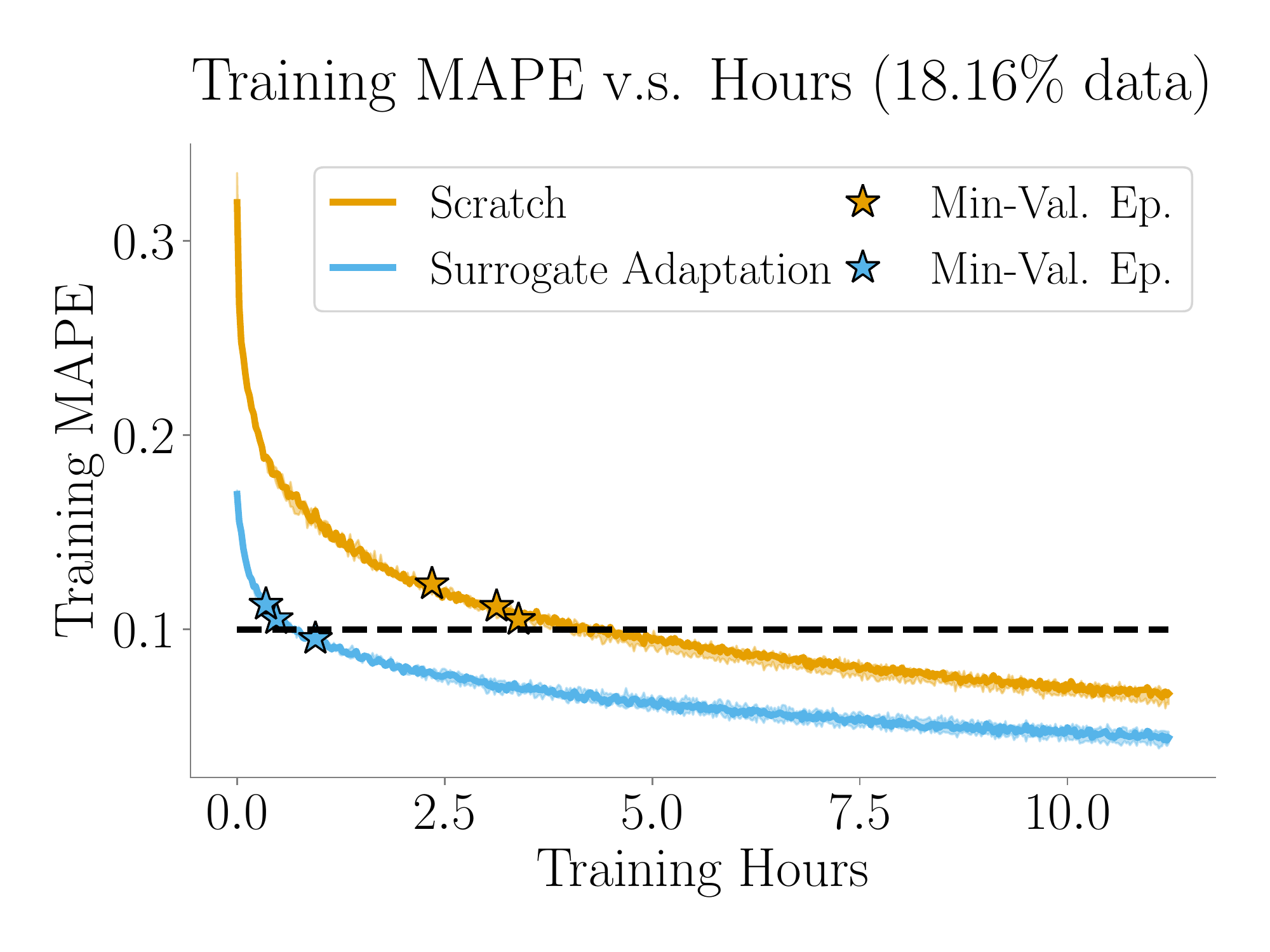}\end{minipage} \\
    \begin{minipage}{0.48\textwidth}\includegraphics[width=\textwidth]{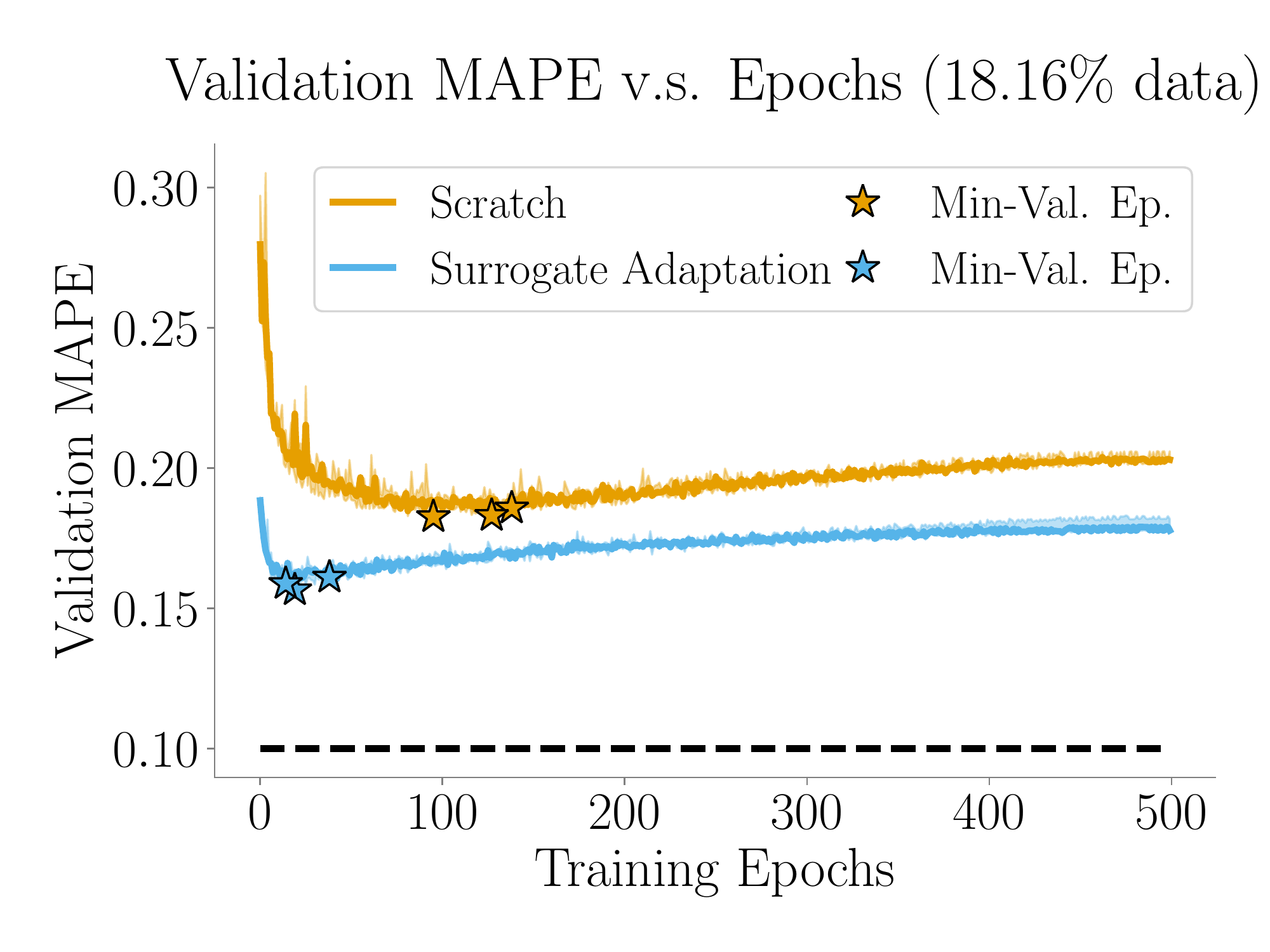}\end{minipage}
    \begin{minipage}{0.48\textwidth}\includegraphics[width=\textwidth]{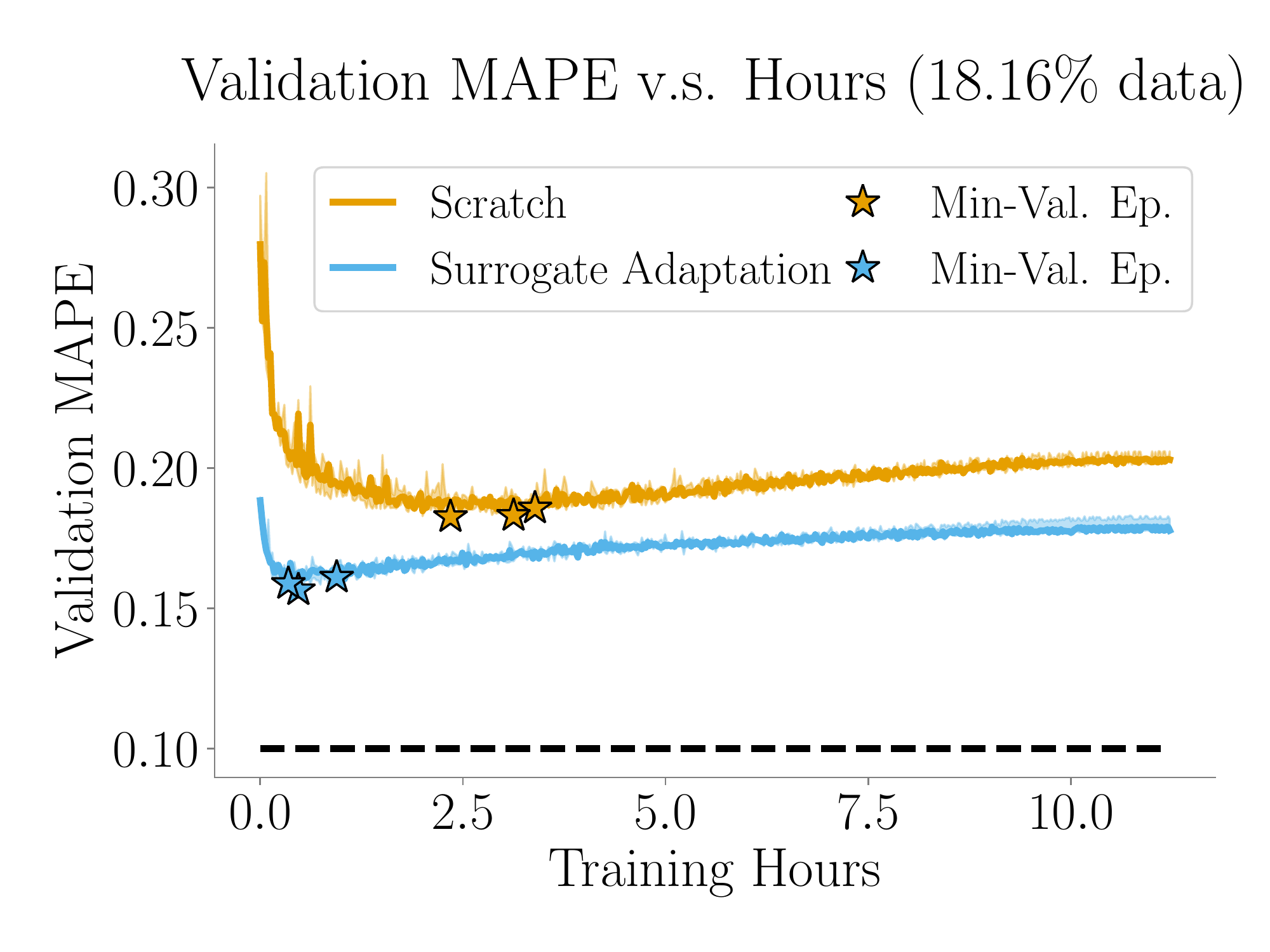}\end{minipage} \\
    \begin{minipage}{0.48\textwidth}\includegraphics[width=\textwidth]{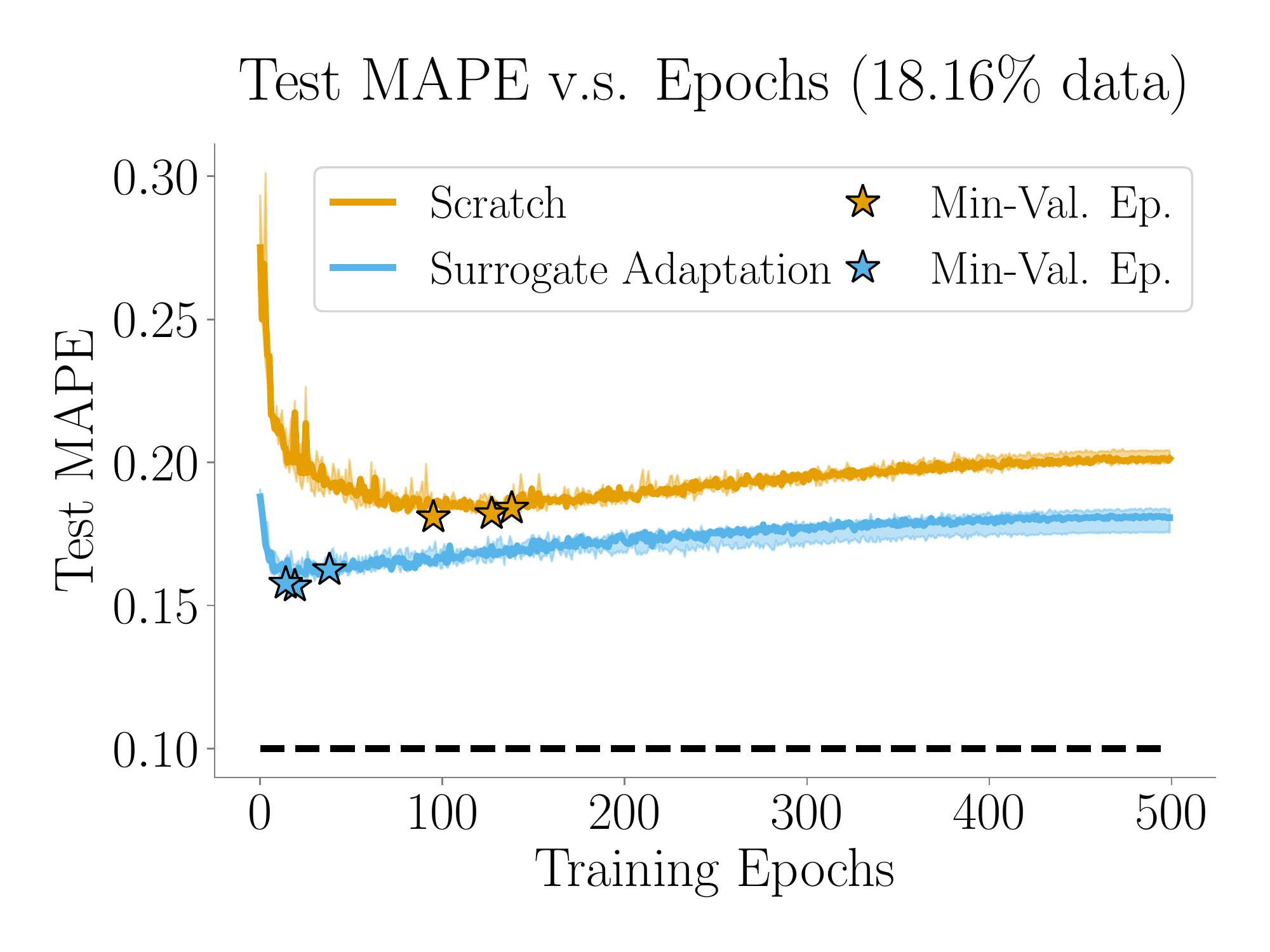}\end{minipage}
    \begin{minipage}{0.48\textwidth}\includegraphics[width=\textwidth]{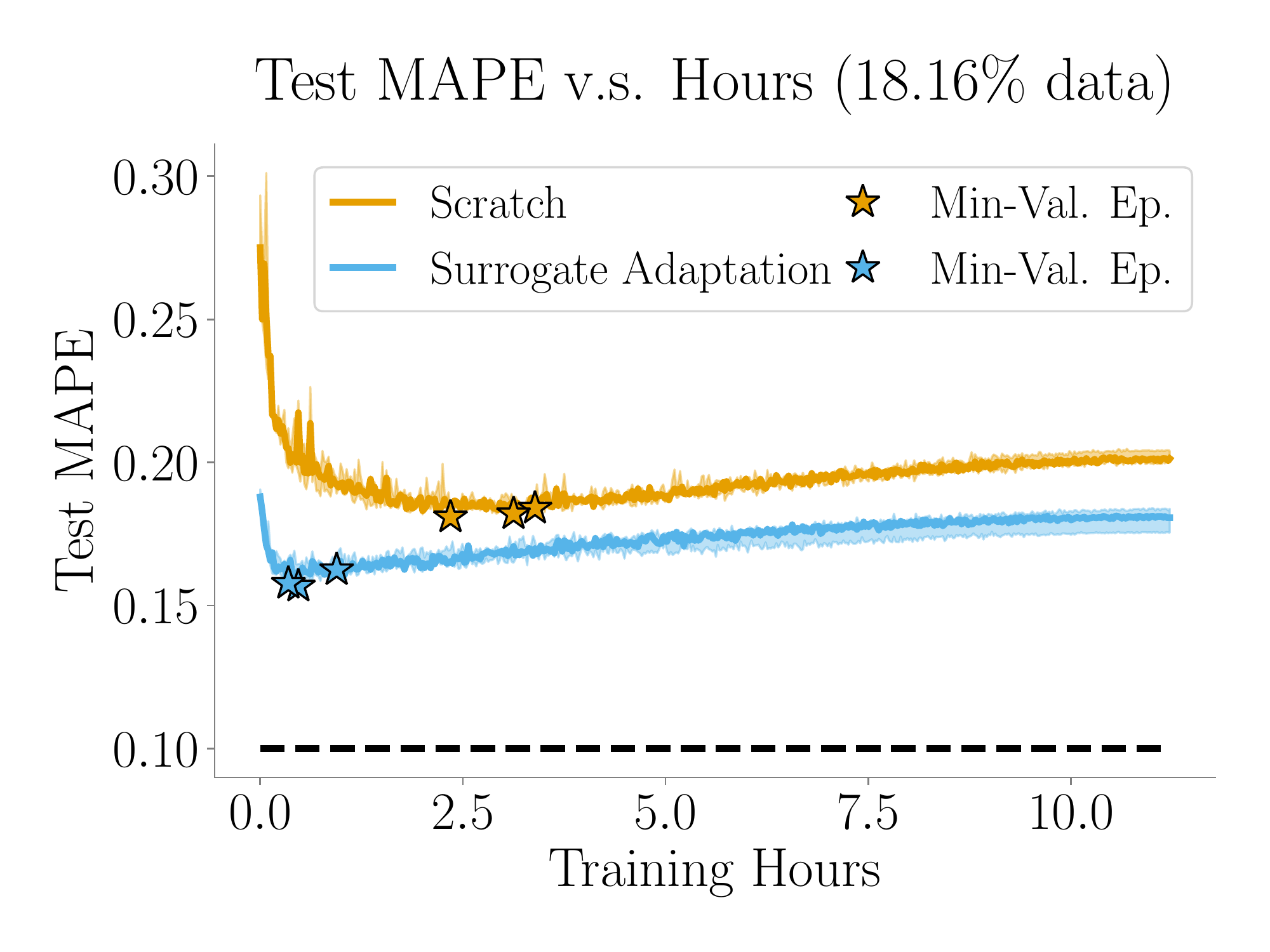}\end{minipage} \\
  \end{center}
\end{figure*}%
\begin{figure*}
  \ContinuedFloat
  \begin{center}
  {\huge\bf {\ADAPTATION{} Telemetry}}\\
  {\LARGE $42.46\%$ Data} \\
    \begin{minipage}{0.48\textwidth}\includegraphics[width=\textwidth]{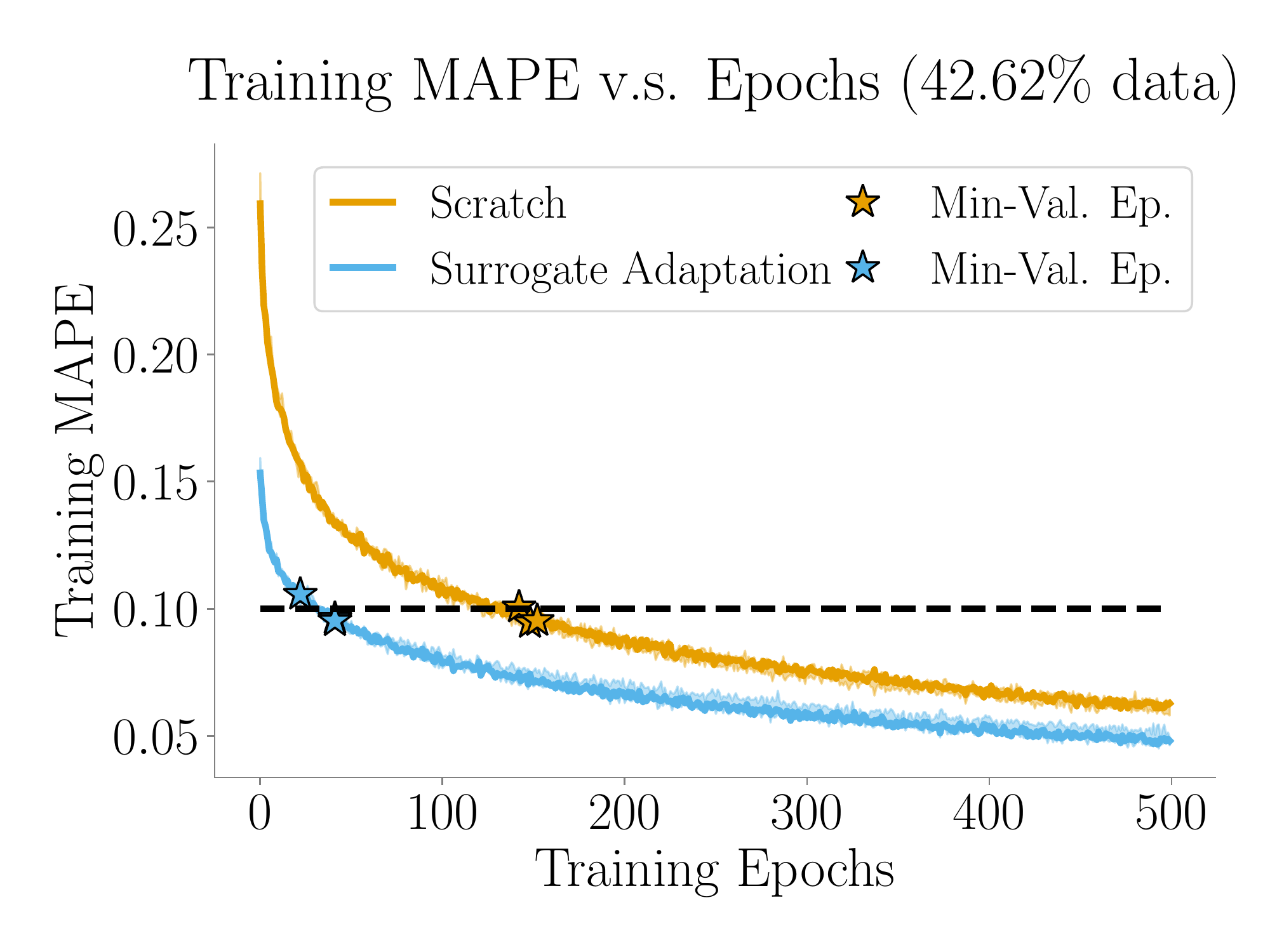}\end{minipage}
    \begin{minipage}{0.48\textwidth}\includegraphics[width=\textwidth]{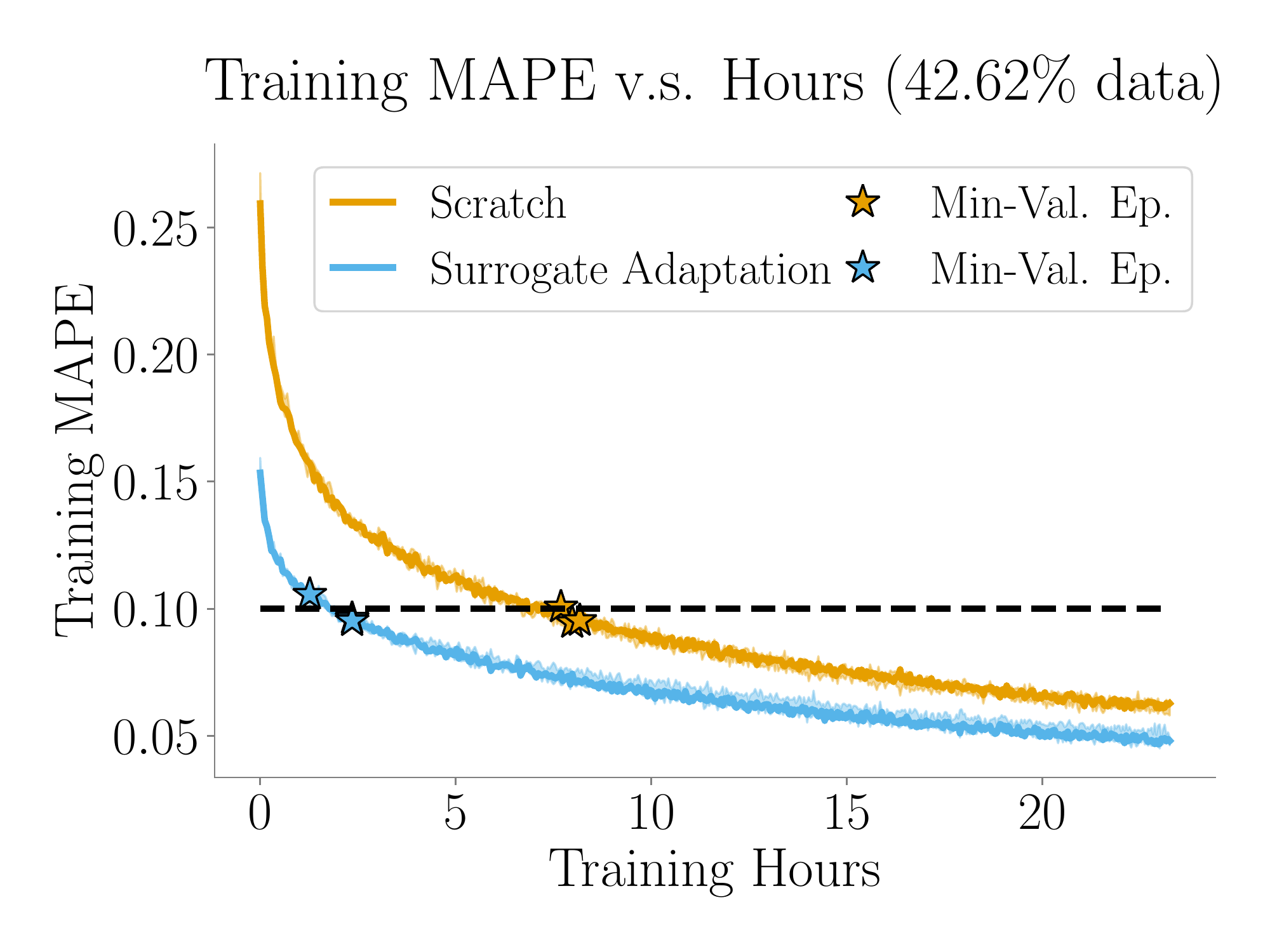}\end{minipage} \\
    \begin{minipage}{0.48\textwidth}\includegraphics[width=\textwidth]{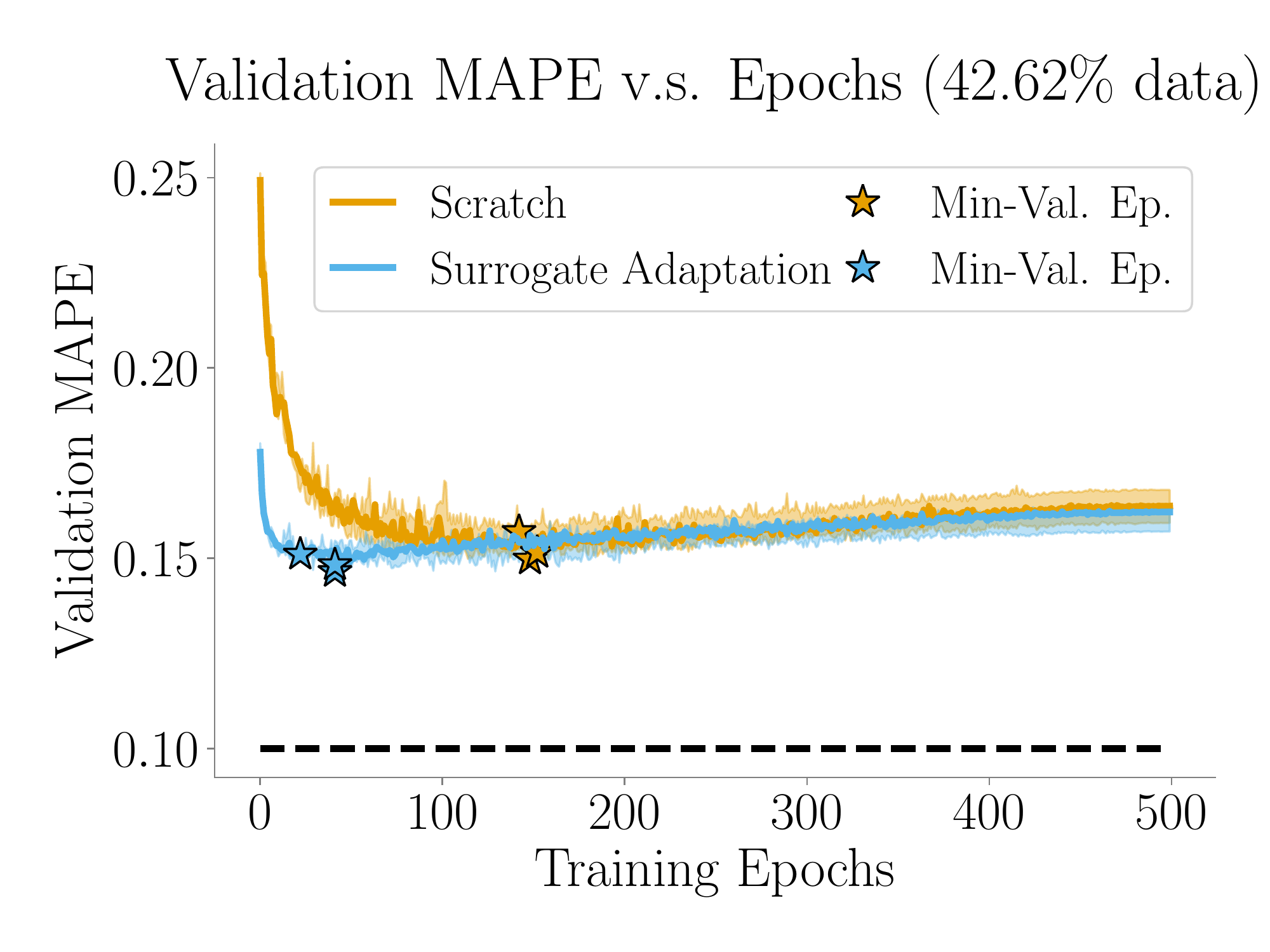}\end{minipage}
    \begin{minipage}{0.48\textwidth}\includegraphics[width=\textwidth]{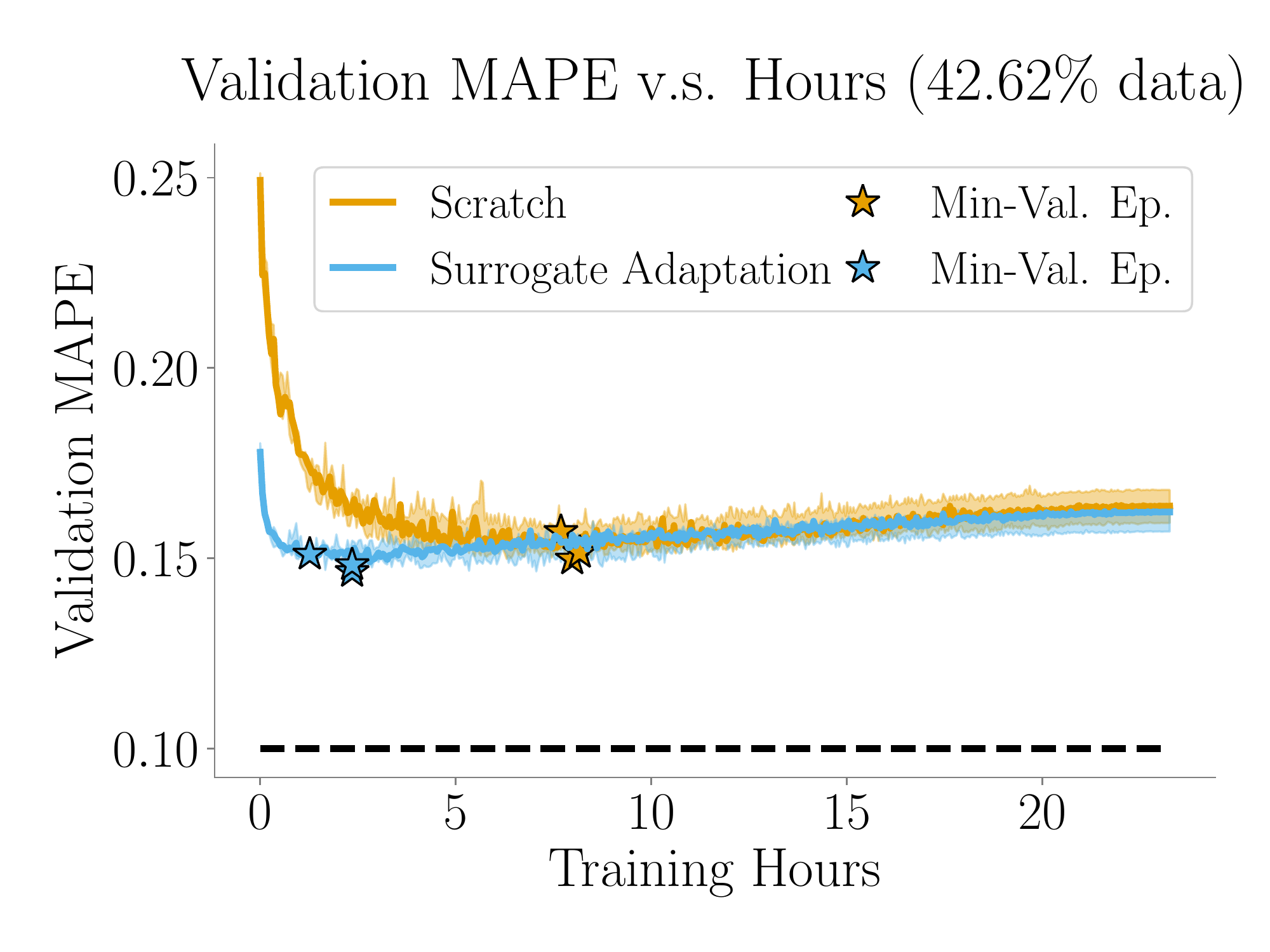}\end{minipage} \\
    \begin{minipage}{0.48\textwidth}\includegraphics[width=\textwidth]{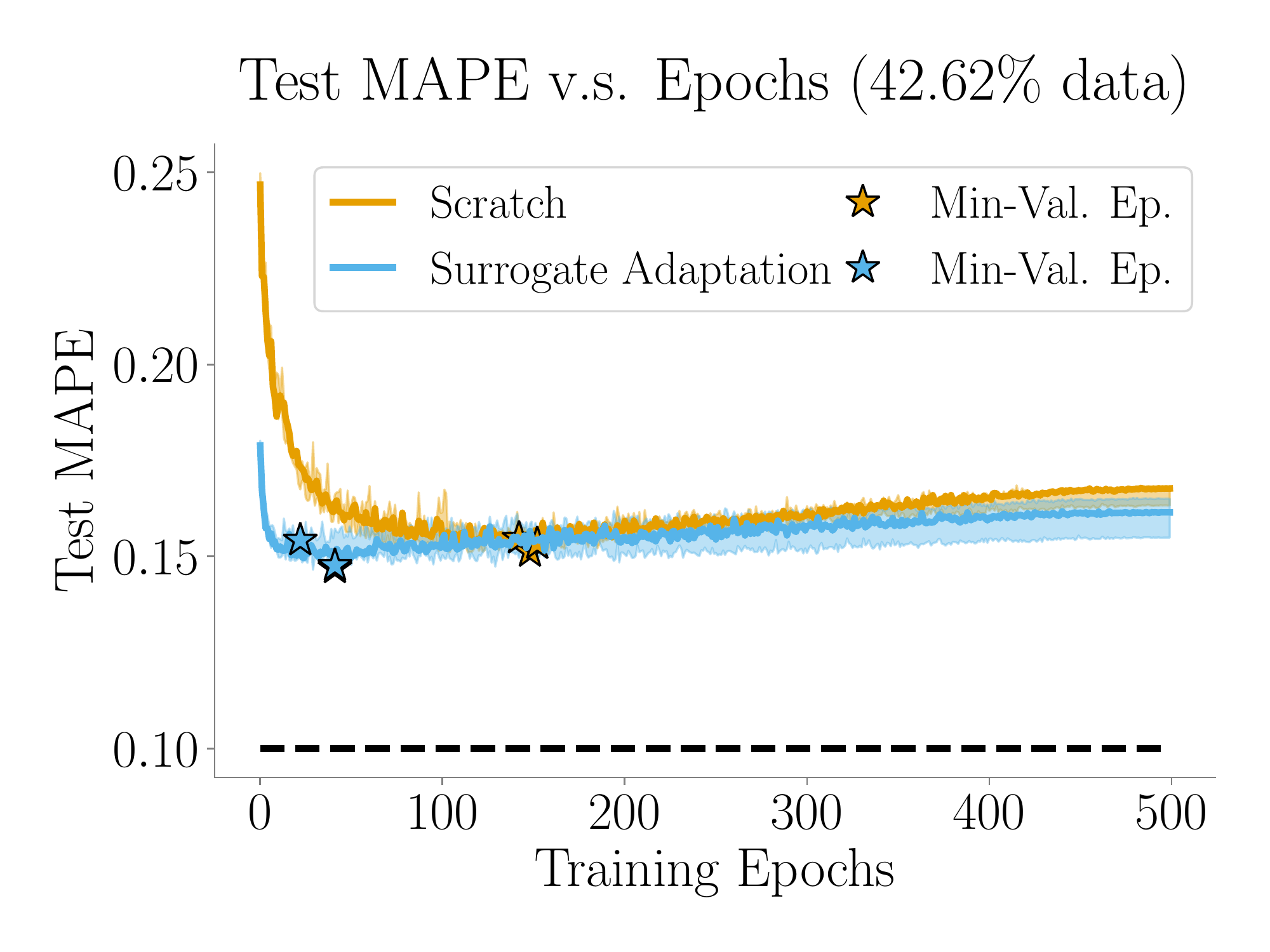}\end{minipage}
    \begin{minipage}{0.48\textwidth}\includegraphics[width=\textwidth]{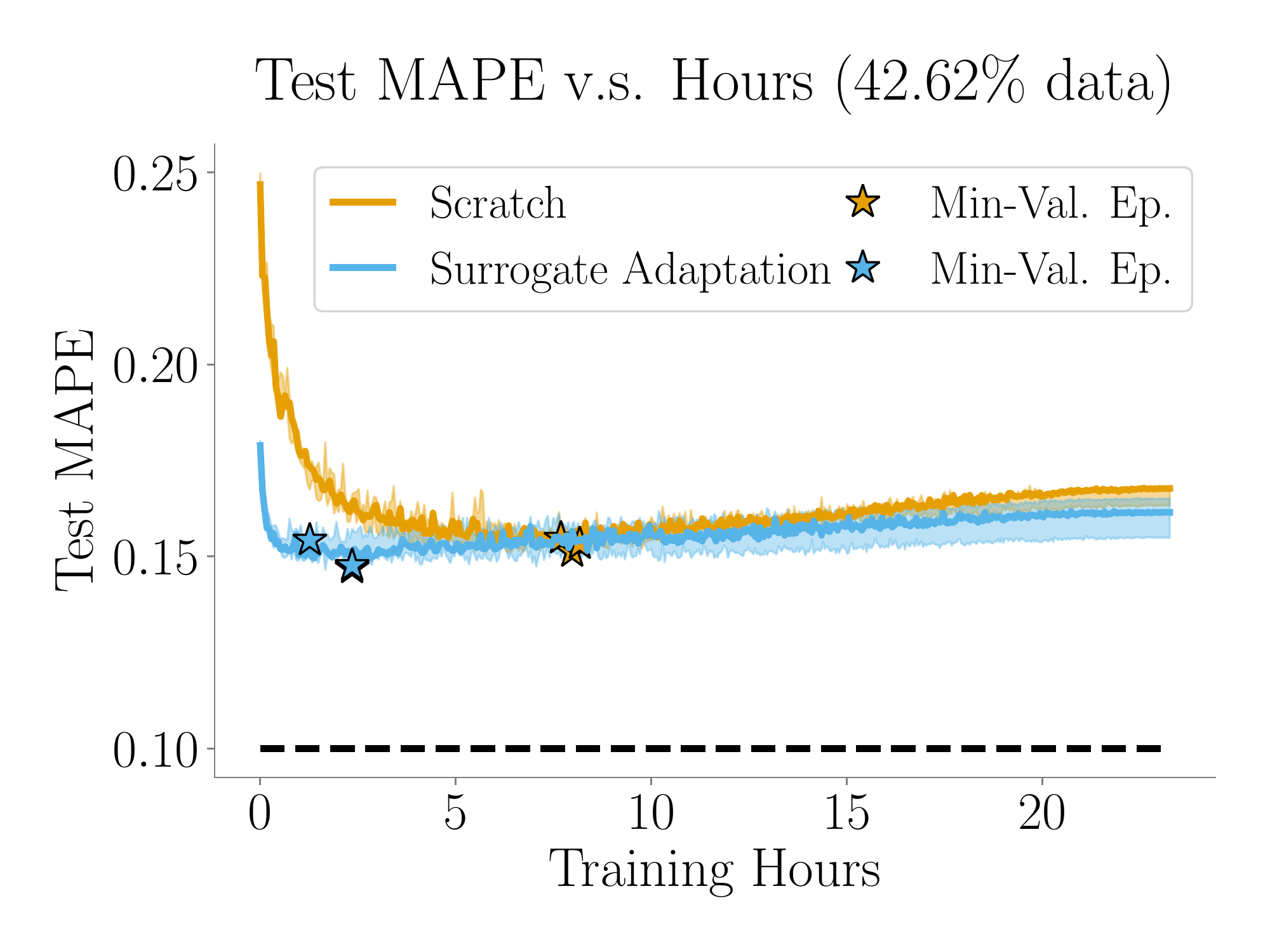}\end{minipage} \\
  \end{center}
\end{figure*}%
\begin{figure*}
  \ContinuedFloat
  \begin{center}
    {\huge\bf {\ADAPTATION{} Telemetry}}\\
    {\LARGE $100\%$ Data} \\
    \begin{minipage}{0.48\textwidth}\includegraphics[width=\textwidth]{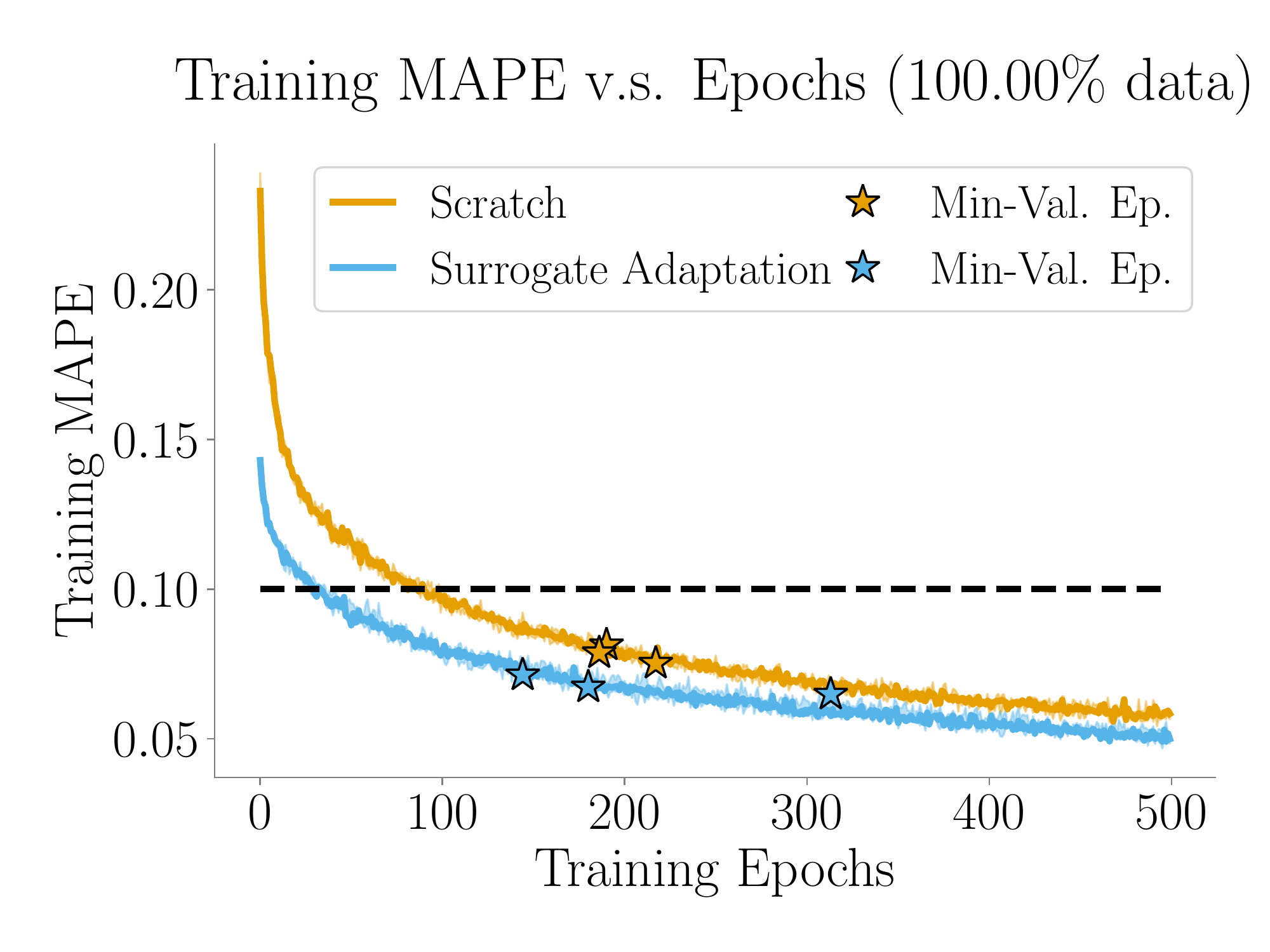}\end{minipage}
    \begin{minipage}{0.48\textwidth}\includegraphics[width=\textwidth]{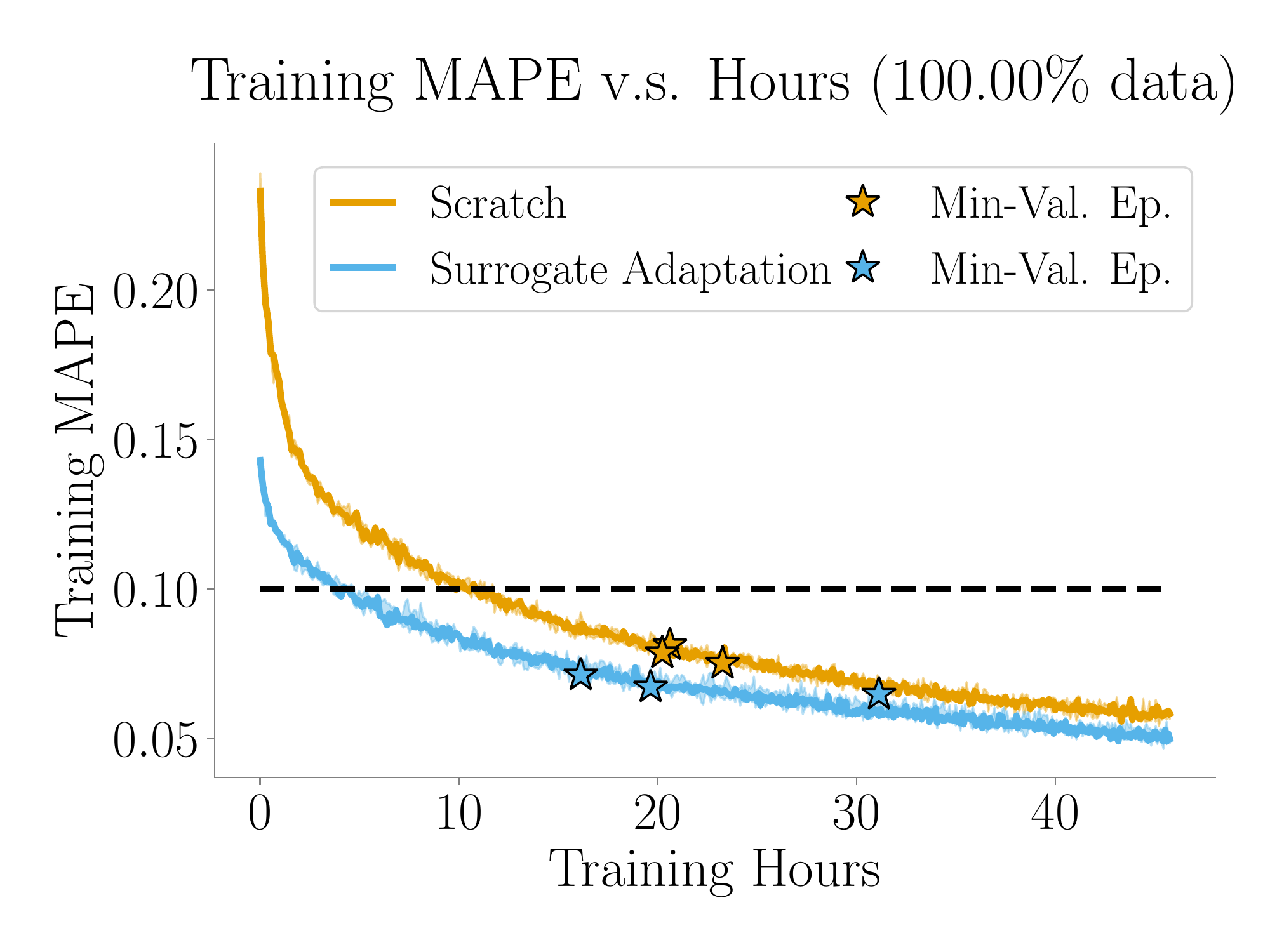}\end{minipage} \\
    \begin{minipage}{0.48\textwidth}\includegraphics[width=\textwidth]{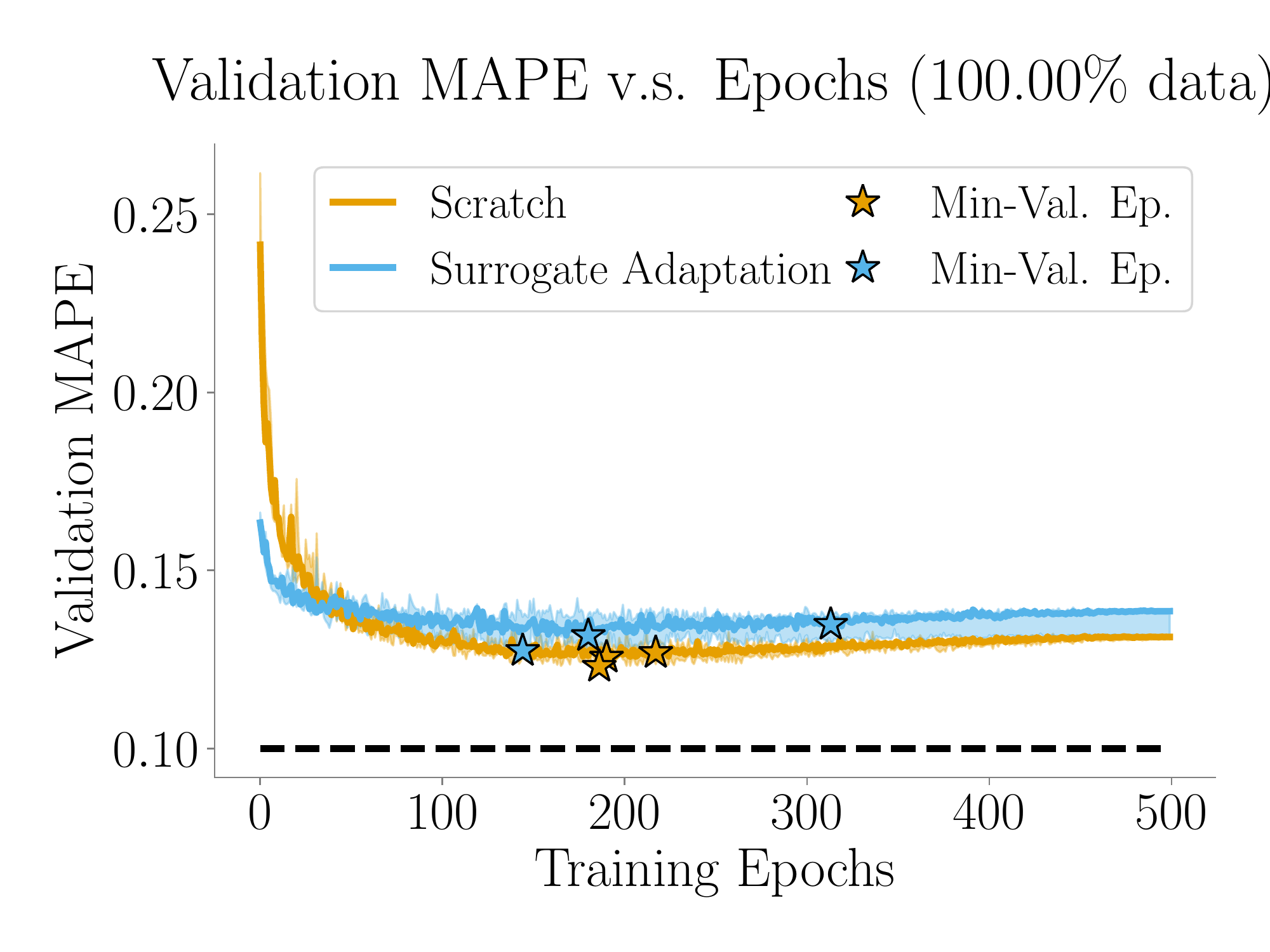}\end{minipage}
    \begin{minipage}{0.48\textwidth}\includegraphics[width=\textwidth]{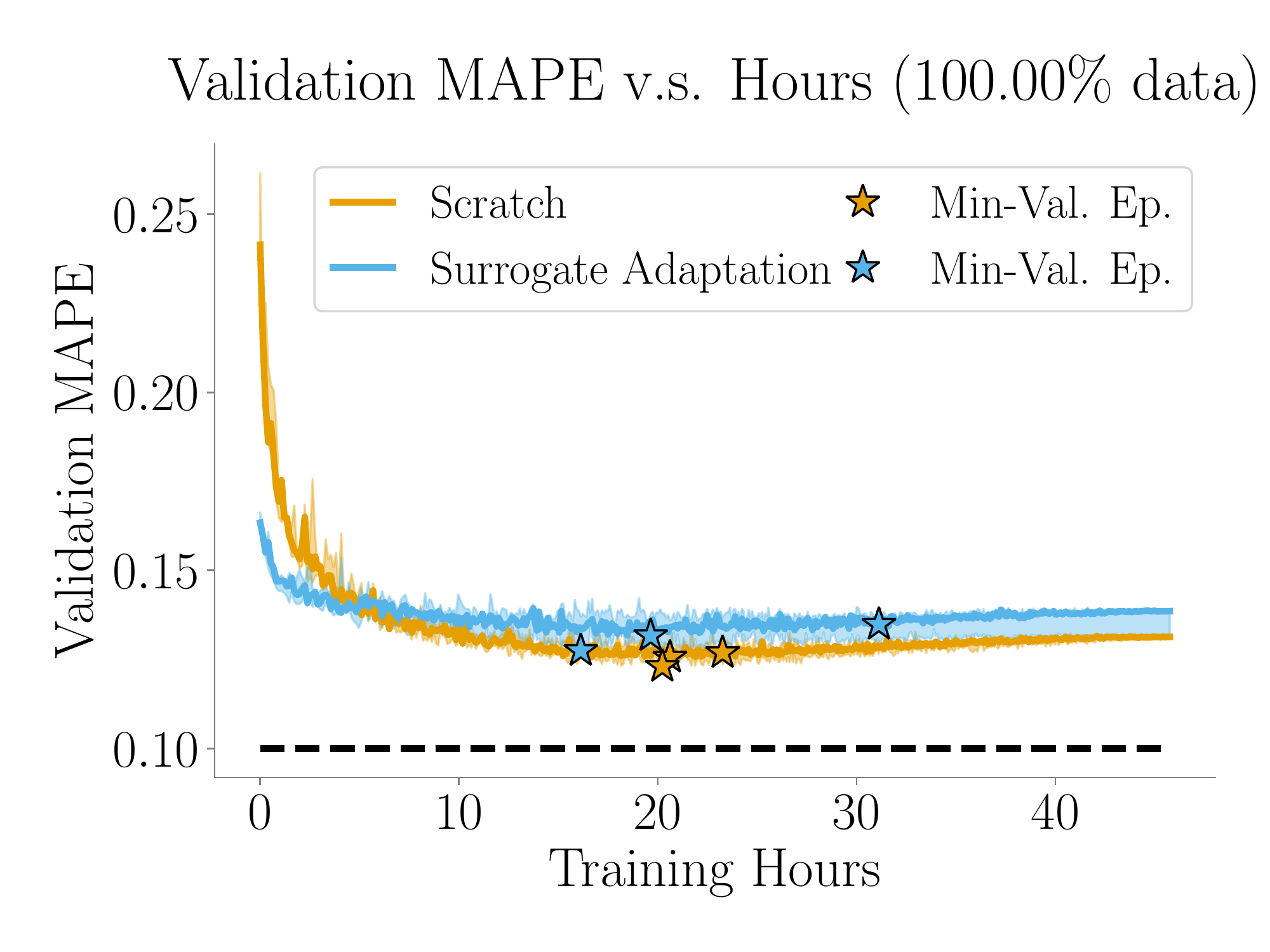}\end{minipage} \\
    \begin{minipage}{0.48\textwidth}\includegraphics[width=\textwidth]{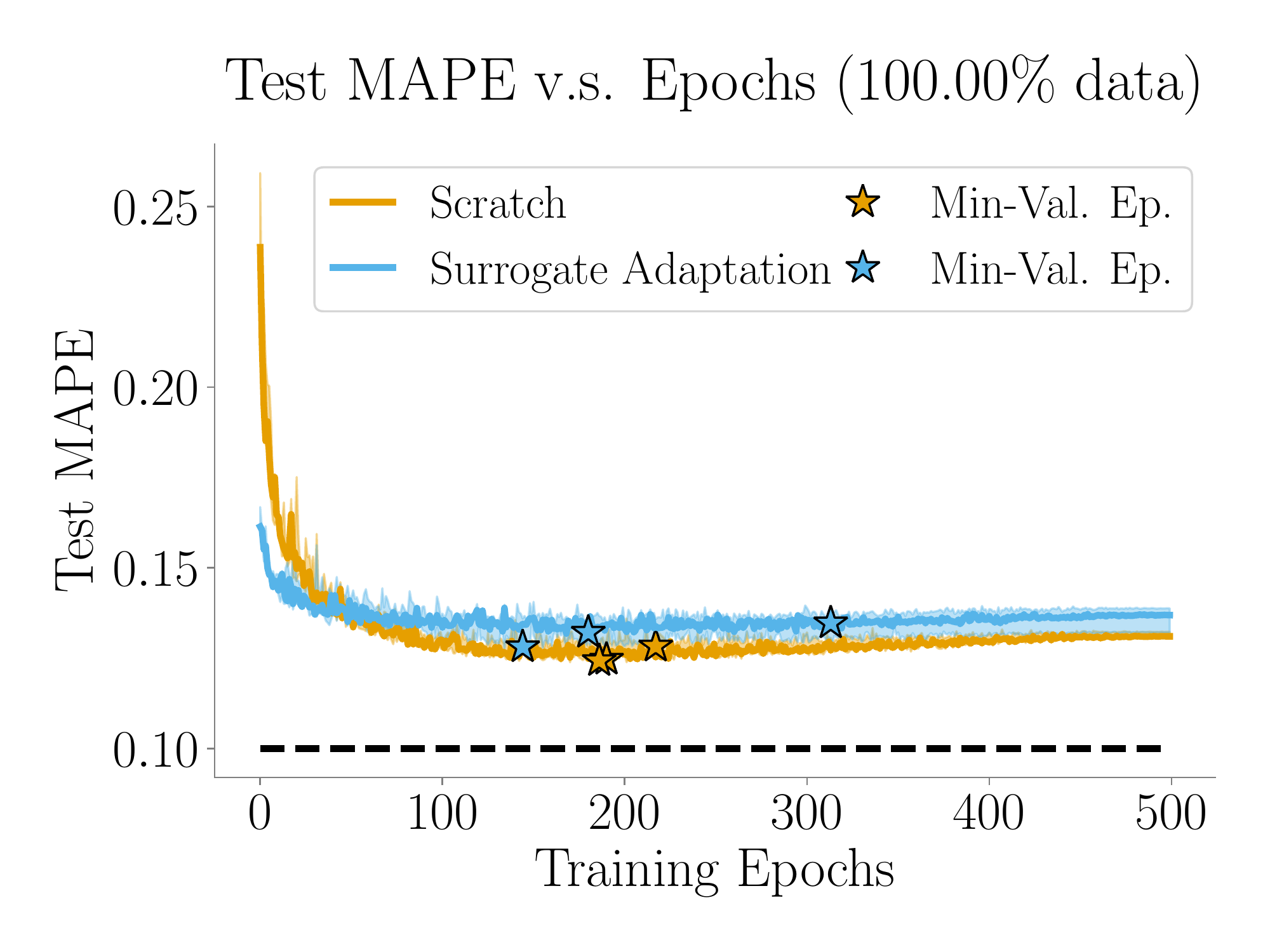}\end{minipage}
    \begin{minipage}{0.48\textwidth}\includegraphics[width=\textwidth]{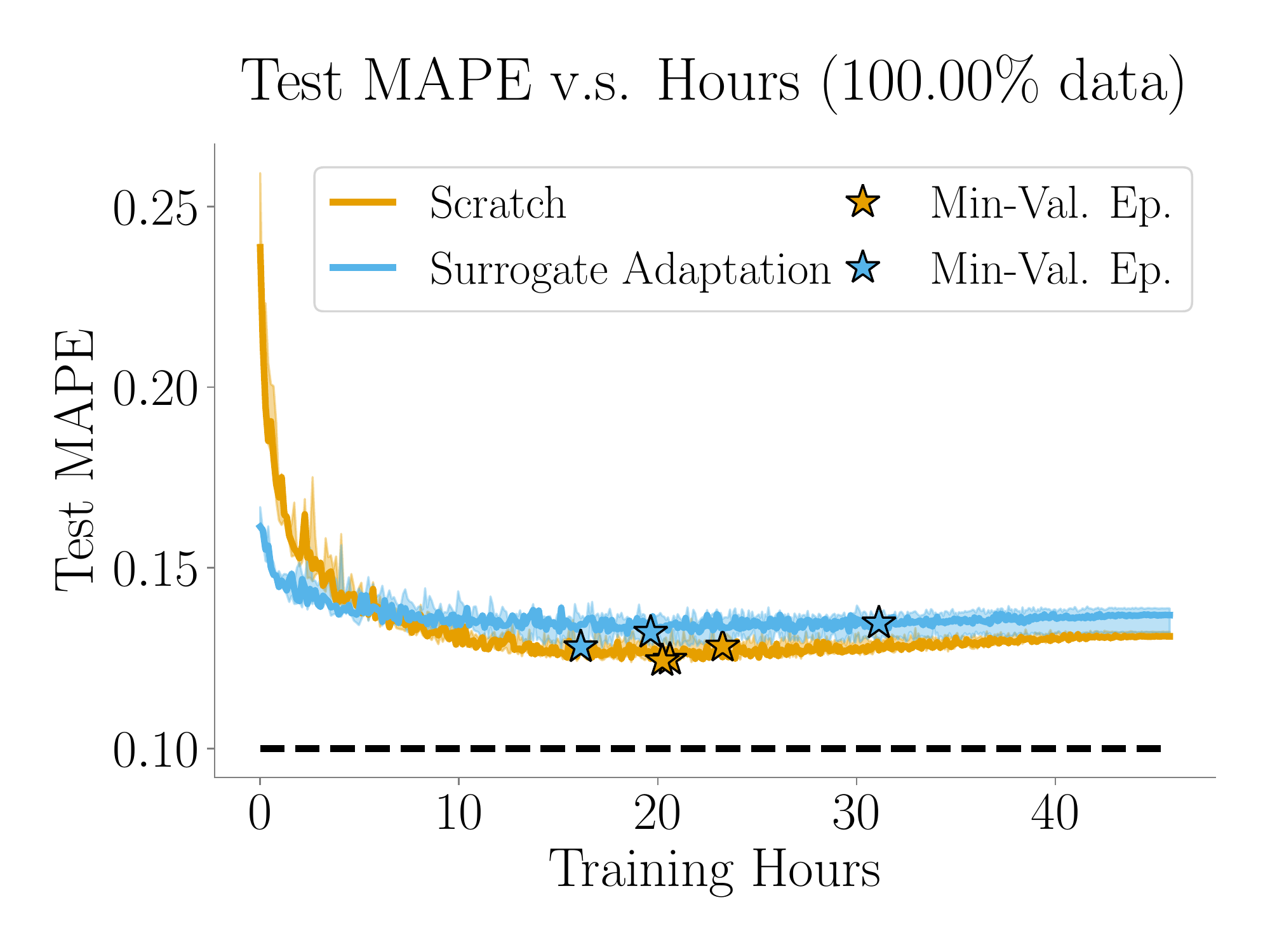}\end{minipage} \\
  \end{center}
\end{figure*}%

\end{document}